# Mesoscopic transport and quantum chaos

- **Rodolfo A. Jalabert**, Institut de Physique et Chimie des Matériaux de Strasbourg, CNRS, Université de Strasbourg, France

The field of Quantum Chaos, addressing the quantum manifestations of an underlying classically chaotic dynamics, was developed in the early eighties, mainly from a theoretical perspective. Few experimental systems were initially recognized to exhibit the versatility of being sensitive, at the same time, to their classical and quantum dynamics. Rydberg atoms (Shepelyansky, 2012-i) provided the main testing ground of Quantum Chaos concepts until the early nineties, marked by the development of microwave billiards (Stöckmann, 2010-i), ultra-cold atoms in optical lattices (Raizen, 2011-i), and low-temperature transport in mesoscopic semiconductor structures. The mesoscopic regime is attained in small condensed matter systems at sufficiently low temperatures for the electrons to propagate coherently across the sample. The quantum coherence of electrons, together with the ballistic motion characteristic of ultra-clean microstructures, motivated the proposal (Jalabert, 1990-a) of mesoscopic systems as a very special laboratory for performing measurements and testing the theoretical ideas of Quantum Chaos. Experimental realizations (Marcus, 1992-a) and many important developments, reviewed in this article, followed from such a connection.

Contents

# Introduction

## Basic concepts of Mesoscopic Physics

**Refs. (Akkermans, 1995-b; Datta, 1995-b; Imry, 2002-b)**

### Quantum coherence

The mesoscopic regime is defined by the quantum coherence of the one-electron wave-functions across the sample. In a condensed matter environment the coherence is, however, only partial, and not that of an ideal isolated quantum system. The one-electron wave-functions are well-defined over a distance $L_\Phi$ (the *phase-coherence length*) which is larger than the typical size ($a$) of the microstructure, but not infinite. Using the concept of one-electron wave-functions supposes to be away of the case of strongly correlated systems, and thus the description is actually that of weakly interacting Landau quasiparticles moving in a self-consistent field. The finite value of $L_\Phi$ arises from the residual Coulomb interaction (responsible for the quasiparticle lifetime), as well as from other elastic and inelastic phase-breaking events (coupling to the degrees of freedom of an external environment, electron-phonon scattering, etc).

### Quantum transport
**Refs. (Landauer, 1987-r; Büttiker, 1988-r; Kastner, 1992-r; Büttiker, 1993-r)**

Electronic transport, carrying an electrical current $I$, is established when a microstructure is connected to two or more electrodes (labeled by the index $l$) where electrostatic potentials $V_l$ are applied. The corresponding electrochemical potentials are $\mu_l = eV_l$, with $e$ the electron charge (see Figure 1 for a sketch of the generic case of a *two-probe* setup). The Landauer-Büttiker description of quantum transport is that of a scattering process for phase-coherent electrons traversing the microstructure in their journey between the electrodes (Landauer, 1970-a; Büttiker, 1986-a). When the time-independent (DC) potential difference $V = V_1 - V_2$ is sufficiently small, only electrons at the Fermi level of the electrodes contribute to transport, and the device operates in the linear-response regime close to equilibrium. Thus, the linear conductance (in short: the *conductance*)

$$G = \left(\frac{\partial I}{\partial V}\right)_{V=0}, \qquad (1)$$

characterizes the electronic transport. For larger $V$, the device operates in the far-from-equilibrium non-linear regime, characterized by the differential conductance $G(V) = \partial I/\partial V$. The linear regime is conceptually simpler than the non-linear one, since the knowledge of the actual electric field distribution is not required for the calculation of the conductance. It is therefore in the linear regime that the connection with Quantum Chaos has primarily been explored.

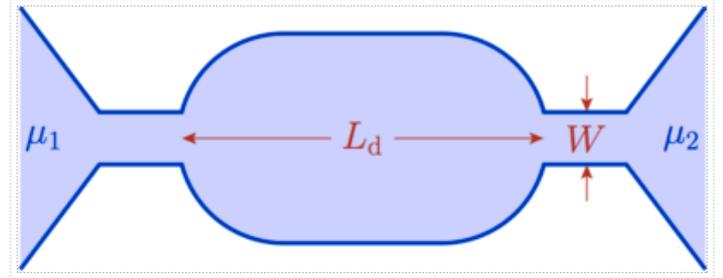

Figure 1: Typical ballistic cavity coupled to two reservoirs (characterized by electrochemical potentials $\mu_1$ and $\mu_2$) through leads of width $W$. The electrons are confined to the colored regions. $L_d$ denotes the distance between the entrance and exit leads.

When the microstructure is well-connected to the electrodes and the electron density is high enough, the electron-electron interactions are not determinant, and the mean-filed description of linear transport results in an effective single-particle approach. Alternatively, when the microstructure is weakly connected to the electrodes (i.e. through tunnel barriers like in the sketch of Figure 24) the single-electron charging effects become important and transport may be Coulomb blocked (as discussed in Sec. 5). In this last case, for not too-small microstructures the electron island thereby defined can be described through the so-called *constant-interaction model*, reducing the transport problem to a single-particle one with an additional charging energy that separates the occupied and unoccupied levels of the microstructure.

The two limiting situations of almost open and almost closed microstructures are the ones where the connection with Quantum Chaos has been further developed. But the theoretical tools used in each case are specific to the problem on hand.

## Disordered systems
**Refs. (Lee, 1985-r; Chakravarty, 1986-r; Altshuler, 1991-b; Washburn, 1992-r; Akkermans, 2007-b)**

Mesoscopic Physics was initially focused on disordered metals, where the classical motion of electrons can be thought as a random walk between the impurity sites. The phase-coherence in the multiple scattering of electrons gives rise to quantum corrections to the classical (Drude) conductance. The most studied quantum interference phenomena in disordered metals are the *Aharanov-Bohm oscillations* of the conductance in multiply connected geometries, the *weak-localization* effect (a decrease in the average conductance around zero magnetic field), and the *universal conductance fluctuations* (reproducible fluctuations in the conductance versus magnetic field or Fermi energy with a root-mean-square of the order $e^2/h$, independent of the average conductance). A perturbative treatment of disorder, followed by an average over impurity configurations, has provided the calculational tool leading to the understanding of these phenomena. The small parameter of the perturbation is $(k_F l)^{-1}$, with $k_F = 2\pi/\lambda_F$ the Fermi wave-vector, $\lambda_F$ the Fermi wave-length, and $l$ the *elastic mean-free-path* (*i.e.* the typical distance traveled by the electron between successive collisions with the impurities). Mesoscopic disordered conductors are then characterized by $\lambda_F \ll l \ll a < L_\Phi$.

## Ballistic systems
**Refs. (Beenakker, 1991-r; Davies, 1998-b; Bird, 2003-b)**

It is primarily in a "second generation" of mesoscopic systems, semiconductor microstructures, that the connection with Quantum Chaos has been developed. Extremely pure semiconductor (*GaAs/AlGaAs*) heterostructures make it possible to create a two-dimensional electron gas (2DEG) by freezing in the quantum ground state the motion perpendicular to the interface. Given the crystalline order of the interface and the fact that the dopants are away from the plane of the carriers, an electron can travel a long distance before its initial momentum is randomized. This typical distance, the *transport mean-free-path* $l_T$, is generally larger than the elastic mean-free-path (since the small-angle elastic scattering is not effective in changing the momentum direction).

Various fabrication techniques have been developed to produce a lateral confinement of the 2DEG and define one-dimensional (quantum wires) and zero-dimensional (quantum boxes or cavities) structures. Micron and sub-micron spatial resolutions allow to define, at the level of the 2DEG, mesoscopic structures smaller than the transport mean-free-path, paving the way to the *ballistic* regime. Figure 2 shows a ballistic cavity electrostatically defined by metallic gates lithographically patterned on the surface of a *GaAs/AlGaAs* heterostructure.

When $a \ll l_T$ the classical motion of the two-dimensional electrons is given by the collisions with the walls defining a cavity, with a very small lateral deflection from a straight line, due to the smooth impurity potential. The ideal case of no disorder, characterized by an infinite $l_T$, is referred to as the *clean* limit. By changing the shape of a clean cavity it is possible to go from integrable to chaotic dynamics, and then study the consequences of this transition at the quantum level. Experimentally realizable ballistic mesoscopic systems always have a finite value of $l_T$, and this fact has to be kept in mind when Quantum Chaos studies are undertaken.

The usual electronic surface densities in *GaAs/AlGaAs* 2DEG are $n_S = 1 - 3 \times 10^{11}$ $cm^{-2}$, leading to $\lambda_F = 40 - 70$ *n*m. For typical microstructures $a = 0.5 - 3$ $\mu$m. The constant improvement in fabrication techniques results in the achievement of progressively larger $l_T$, that in 2DEG as that of Figure 2 can attaint values as large as 49 $\mu$m (Kozikov, 2013-a).

Since $\lambda_F \ll a$, the devices operate far away from the extreme quantum limit, and the semiclassical approximation (presented in Sec. 2) for the electron motion inside the cavity provides a good description.

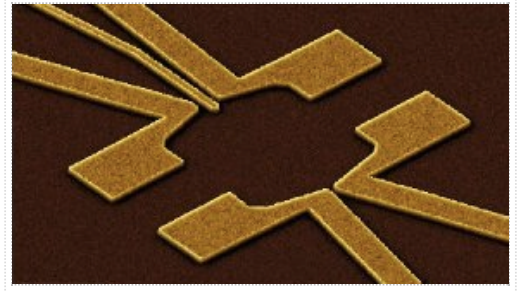

Figure 2: Image of a microstructure defined by a set of metallic gates (indicated in yellow) that can be individually addressed by different gate voltages. The 2DEG is buried a few hundred of *n*m below the surface. The unpatterned regions (brown) are accessible to electrons (Nanophysics Group, ETH-Zurich)

## Mesoscopic samples as "doubly open" quantum systems

Electrons in a mesoscopic setup are typically described by the Hamiltonian

$$\hat{H} = \hat{H}_s + \hat{H}_l + \hat{H}_{s-l} + \hat{H}_d + \hat{H}_{s-env} \ . \tag{2}$$

- $\hat{H}_s = \dfrac{\hat{\mathbf{p}}^2}{2M_e} + U(\hat{\mathbf{r}})$ is the one-particle Hamiltonian describing electrons within the sample.
- $M_e$ is the effective electron mass.
- $\hat{\mathbf{p}}$ is the momentum operator.
- $\hat{\mathbf{r}}$ is the position operator.
- $U(\mathbf{r})$ is the electrostatic energy arising from the imposed confining potential and the resulting self-consistent mean-field potential.
- $\hat{H}_l$ is the Hamiltonian of disorder-free semi-infinite leads, taken as an idealization of the electrodes.
- $\hat{H}_{s-l}$ represents the coupling between the system and the leads.
- $\hat{H}_d$ represents the electrostatic disorder arising from the impurities near or within the sample (assumed to be frozen, *i.e.* without dynamics).
- $\hat{H}_{s-env}$ represents the coupling to the one-electron environment, given mainly by the residual electron-electron interaction and the phonon bath.

As in standard open quantum systems, the coupling to the environment results in decoherence and dissipation. In addition, the coupling to the leads allows for particle exchange with the reservoirs resulting in electronic transport across the sample.

## Quantum Chaos and Mesoscopic Physics

The fruitful connection between Quantum Chaos and Mesoscopic Physics is restricted to the observables that are accessible in the laboratory. The conductance is the main physical quantity that can be inferred from transport experiments. Since the conductance is given by electron scattering, the Quantum Chaos issues accessible through mesoscopic transport are those of quantum chaotic scattering (Gaspard, 2014-i). Other central questions of Quantum Chaos, like for instance the link between the (short range) statistical properties of the spectrum of a quantum system with the nature of the underlying classical dynamics (Bohigas, 1984-a; Ullmo, 2014-i) are difficult to address in the mesoscopic regime. In well-connected microstructures it is experimentally rare to have access to single-particle energies, since the typical level spacings $\Delta$ are smaller than the thermal broadening $k_B T$ and the level-width resulting from the coupling to the electrodes. In weakly-coupled systems operating in the Coulomb blockade regime the energy spacing between resonances depends on the single-particle energy spacing, as well as on the charging energy.

It is important to keep in mind that Mesoscopic systems are not fully coherent (finite $L_\Phi/a$), nor clean (finite $l_T/a$), and do not operate in the semiclassical limit (finite $k_F a$). Therefore, Mesoscopic Physics is not an ideal laboratory for Quantum Chaos. The imperfect nature of this relationship is an essential ingredient of its interest. Mesoscopic systems are extremely useful to study the interplay between the quantum and classical worlds, and at the same time Quantum Chaos studies can be used to test fundamental questions of Condensed Matter Physics, like disorder, decoherence, dissipation and many-body effects.

# Summary of characteristic lengths

- $a$ is the typical size of the structure.
- $L_d$ is the distance between the entrance and exit leads.
- $W$ is the width of the leads.
- $\lambda_F$ is the *Fermi wave-length* of electrons within the structure.
- $L_\Phi$ is the *phase-coherence length* - distance over which the one-particle wave-functions keep their quantum coherence (determined by $\hat{H}_{\text{s-env}}$).
- $l$ is the *elastic mean-free-path* - distance between two collisions with impurities (determined by the density and the typical strength of the impurities associated to $\hat{H}_d$).
- $l_T$ is the *transport mean-free-path* - distance traveled before an initial electron momentum is randomized (determined by the spatial structure of $\hat{H}_d$).

# Geometry-dependent transport in ballistic microstructures

## Quenching of the Hall effect

The relevance of classical electron trajectories in quantum transport was first hinted in analyzing measurements of the Hall effect in restricted geometries. While the Hall resistance in a 2DEG is simply proportional to the applied transverse magnetic field, in the geometry defined by two narrow ($\sim 100$ $n$m) ballistic wires, crossing at right angles, there appeared important departures from the standard Hall effect (Roukes, 1987-a). At low temperatures ($T = 4.1$ K) and weak fields ($B \lesssim 100$ mT), the Hall resistance could be suppressed (quenched), enhanced, or even negative, depending on the details of the geometry.

The effect of geometry on transport was demonstrated by purposely designing different crosses where the ballistic electrons were scattered off the corresponding confining potentials (Ford, 1989-a). The inversion of the Hall effect was explained by the bouncing of electrons into the "wrong" probe (see Figure 3). This interpretation was supported by simulations where the transmission coefficients of the cross were identified with the probabilities obtained by a random sampling of classical trajectories (Beenakker, 1988-a).

Quantum mechanical descriptions addressing the quenching of the Hall effect pointed to the collimation of the electrons as they enter the cross region (Baranger, 1989-a). In this view, the adiabatic widening of the wires near the junction results in modes with high longitudinal momentum that are preferentially populated, inducing the quenching of the Hall effect. Such a quantum description is in line with the relevance of geometry and classical trajectories since, in a semiclassical description, collimation corresponds to electron trajectories continuing straight ahead with a small angular spread.

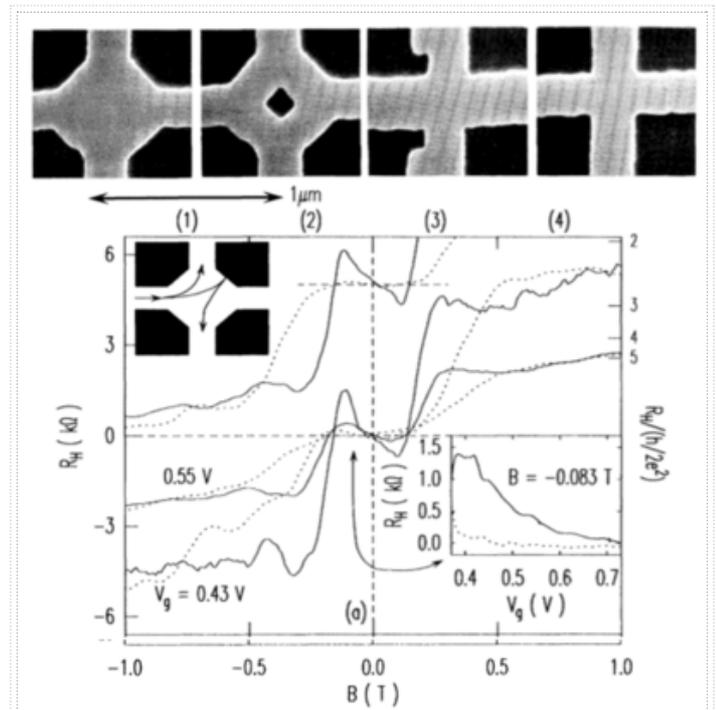

Figure 3: (Top) Electron micrographs of the following devices: (1) a widened cross, (2) a widened cross with a central dot, (3) a cross with narrow probes, and (4) a "normal" (nominally perfect) cross. Main figure: Hall resistance versus magnetic field for a sample consisting of devices (1) (solid lines) and (4) (dotted lines), for several values of the gate voltage $V_g$ controlling the channel width and the electron density. The traces vertically offset by 5 k$\Omega$ present results for a different, nominally identical, sample. Upper inset: The electron paths for the widened cross. Lower inset: A gate-voltage sweep showing that Hall resistance has the "wrong" sign for all values of $V_g$. (Adapted from Ref. (Ford, 1989-a), copyright 1989, American Physical Society.)

## Conductance fluctuations in the ballistic regime

For temperatures $T \lesssim 100$ mK the quenching of the Hall effect was still observable, but the Hall resistance exhibited fluctuations as a function of the external magnetic field or the gate voltage defining the cross geometry (Ford, 1988-a). Such fluctuations were reproducible under the cycling of the control parameter, similarly to the conductance fluctuations of disordered mesoscopic systems.

The purely classical approaches describing electronic transport at Helium temperatures could not account for the fluctuations encountered at ultra-low temperatures. Thus, the ballistic conductance fluctuations were proposed as arising from the quantum interference between the multiple paths that electrons can undertake traversing the microstructure (Jalabert, 1990-a). In a semiclassical approach these paths are classical trajectories, which might exhibit a chaotic character in a sufficiently complex geometry. Such a connection provides the link between ballistic conductance fluctuations and quantum chaotic scattering.

# The scattering approach to the conductance

**Refs. (Imry, 1986-r; Büttiker, 1988-r; Büttiker, 1993-r; Stone, 1995-r; Baranger, 1999-r; Jalabert, 2000-r; Mello, 2004-b)**

## The basic components: sample, leads and reservoirs

In the scattering (Landauer-Büttiker) approach to quantum transport, the electrical resistance arises from the elastic scattering that electrons suffer while traversing a mesoscopic structure connected to electrodes with fixed electrochemical potentials $\mu_l$ (that do not vary while giving and accepting electrons). In this idealized view the mesosocpic *sample* (microstructure) is connected to *reservoirs* (electrodes and measuring devices) through *leads* (ideal contacts). The role of the reservoirs is crucial for dealing with an infinite total system and a continuous spectrum.

The simplest setup is the two-probe configuration of Figure 1, within a two-dimensional space spanned by vectors $\mathbf{r} = (x, y)$, operating in the linear regime with an applied voltage $V$ which is very small ($\mu_1 - \mu_2 = eV \ll \mu_1$). The multi-probe case (Büttiker, 1986-a) does not pose new fundamental problems, but the theoretical description becomes more complicated since a matrix of conductance coefficients must be introduced. Even though the scattering theory is applicable to an arbitrary number of spatial dimensions, the restriction to two degrees of freedom is motivated in view of the application to 2DEG and for the simplicity of the notation. The restriction to the linear regime is a crucial approximation within the scattering approach. Going beyond the linear regime poses considerable difficulty due to the necessity to describe the self-consistent electrostatic potential resulting from the imposed voltages and the electron-electron interactions in the sample (Christen, 1996-a).

In the sketch of Figure 1 the leads have a finite length, while in actual microstructures the entrance to the ballistic cavity is often done through quantum point contacts (QPCs) tuned to the conductance plateaus. The choice of collinear leads in not crucial, and it is adopted in order to simplify the description.

The spin-degenerate case has been the main focus for investigating the connection between Mesoscopic Transport and Quantum Chaos. It applies to the description of well-connected microstructures, as well as to weakly-coupled quantum dots (QDs) which are not very small (far away from the few-electron limit).

The possibility of having superconducting contacts (not contemplated in this review) opens an interesting area for Quantum Chaos studies (Engl, 2010-a).

## Reservoir and lead states

A scattering approach is built from asymptotically-free quantum states, which in the case of quantum transport are those of the reservoir and the leads. The electrons in the reservoirs have the dispersion relation of a free electron gas

$$\varepsilon = \frac{\hbar^2 k^2}{2 M_\mathrm{e}} \ . \tag{3}$$

- $k$ is the magnitude of the wave-vector defining the state (assumed to be two-dimensional).
- $M_\mathrm{e}$ is the effective electron mass.

The contacts between the reservoir and the sample are idealized as semi-infinite, quasi-one dimensional, disorder-free leads providing the set of asymptotic states necessary for the scattering description. Taking the $x$-direction as the longitudinal one, the incoming ($-$) and outgoing ($+$) *modes* in lead 1 (left) and 2 (right) with energy $\varepsilon$ are, respectively,

$$\varphi^{(\mp)}_{1,\varepsilon,a}(\mathbf{r}) = \frac{c}{\sqrt{k_a}} \ \exp\left[\pm i k_a^\mp x\right] \phi_a(y) \, , \quad x < 0 \, , \tag{4}$$

$$\varphi^{(\mp)}_{2,\varepsilon,a}(\mathbf{r}) = \frac{c}{\sqrt{k_a}} \ \exp\left[\mp i k_a^\mp x\right] \phi_a(y) \, , \quad x > 0 \, .$$

- $\varepsilon = \varepsilon_a^{(\mathrm{t})} + \varepsilon_a^{(\mathrm{l})}$ .

- $\varepsilon_a^{(t)}$ is the quantized energy of the $a^{th}$ transverse *channel* with wave-function $\phi_a(y)$.
- $\varepsilon_a^{(l)} = \hbar^2 k_a^2 / 2 M_e$ is the longitudinal energy.
- The lowest modes satisfying $k_a^2 > 0$ are propagating ($N$ in each lead), the modes with $k_a^2 < 0$ are evanescent.
- $k_a$ is the longitudinal wave-vector ($k_a > 0$ is always chosen for the propagating modes).
- $v_a = \hbar k_a / M_e$ is the longitudinal velocity for the propagating modes.
- $k_a^{\mp}$ stands for an infinitesimal negative (positive) imaginary part given to $k_a$ for incoming (outgoing) modes, needed to define the proper time-ordering.
- $c = \sqrt{M_e / 2\pi\hbar^2}$ is a normalization constant.

The choice of $c$ corresponds, *up to a numerical factor*, to the widely used unit-flux normalization condition. The current density, per spin and unit energy, in the $x$-direction associated with the right (left)-moving mode $1(2), \varepsilon, a$ is given by $\pm e/h |\phi_a(y)|^2$. The electrical current, per spin and unit energy, is $\pm e/h$. The overall signs result from the convention of taking as positive the current of positive charges moving from left to right.

A general separable confining potential in the leads yields an $x$-dependent $k_a$. The restriction to confining potentials that are $x$-independent in the asymptotic regions eliminates this dependence, thus simplifying the description. Furthermore, the choice of a hard wall confinement in the $y$ direction (by taking leads of width $W$), leads to the transverse energies

$$\varepsilon_a^{(t)} = \frac{\hbar^2 q_a^2}{2 M_e} \quad, \tag{5}$$

and channel wave-functions

$$\phi_a(y) = \sqrt{\frac{2}{W}} \sin[q_a(y-W)] . \tag{6}$$

- $q_a = \dfrac{\pi a}{W}$ is the $a^{th}$ transverse wave-vector satisfying $k_a = \sqrt{k^2 - q_a^2}$.

## Scattering states and scattering matrix

Once a quantum-coherent scatterer (of linear extension $L_d$ in the $x$ direction) is placed at the coordinate origin, the incoming modes $\varphi_{1(2),\varepsilon,a}^{(-)}$ give rise to *outgoing scattering states* (defined for all $x$) that in the asymptotic regions are, respectively,

$$\Psi_{1,\varepsilon,a}^{(+)}(\mathbf{r}) = \begin{cases} \varphi_{1,\varepsilon,a}^{(-)}(\mathbf{r}) + \sum_{b=1}^{N} r_{ba}\, \varphi_{1,\varepsilon,b}^{(+)}(\mathbf{r}), & x \ll -L_d/2 \\ \sum_{b=1}^{N} t_{ba}\, \varphi_{2,\varepsilon,b}^{(+)}(\mathbf{r}), & x \gg L_d/2 \end{cases} \tag{7}$$

$$\Psi_{2,\varepsilon,a}^{(+)}(\mathbf{r}) = \begin{cases} \varphi_{2,\varepsilon,a}^{(-)}(\mathbf{r}) + \sum_{b=1}^{N} r'_{ba}\, \varphi_{2,\varepsilon,b}^{(+)}(\mathbf{r}), & x \gg L_d/2 \\ \sum_{b=1}^{N} t'_{ba}\, \varphi_{1,\varepsilon,b}^{(+)}(\mathbf{r}), & x \ll -L_d/2 \end{cases}$$

The $N \times N$ matrices $r$ ($r'$) and $t$ ($t'$) characterize, respectively, the *reflection and transmission matrices* from lead $l=1$ ($l=2$). The choice of having an equal number of $N$ propagating modes in each of the two leads is nonessential. The normalization chosen for the modes (4) ensures that the outgoing scattering states constitute an orthonormal basis verifying

$$\int d\mathbf{r}\; \Psi_{l,\varepsilon,a}^{(+)}(\mathbf{r})^* \,\Psi_{\bar{l},\bar{\varepsilon},\bar{a}}^{(+)}(\mathbf{r}) = \delta_{l\bar{l}}\; \delta(\varepsilon - \bar{\varepsilon})\; \delta_{a\bar{a}} . \tag{8}$$

The $2N \times 2N$ *scattering matrix* $S$, relating incoming and outgoing modes, is given by

$$S = \begin{pmatrix} r & t' \\ t & r' \end{pmatrix} . \tag{9}$$

Current conservation dictates that the incoming and outgoing electron fluxes should be equal, implying that $S$ is a *unitary* matrix ($SS^\dagger = I$). In terms of the total transmission ($T = \sum_{a,b} |t_{ba}|^2$) and reflection ($R = \sum_{a,b} |r_{ba}|^2$) coefficients, the unitarity condition is expressed as $T + R = N$. Also, unitarity dictates that $T = T'$ and $R = R'$. In the absence of magnetic field, the time-reversal symmetry implies that $S$ is a *symmetric* matrix ($S^{\mathrm{T}} = S$). For simplicity the energy dependence of the various components of the scattering matrix is not explicitly written. The special cases of spatially symmetric cavities (invariant under reflection with respect to a longitudinal or transverse axis) result in matrices $S$ presenting additional symmetries, with a block structure (Baranger, 1996-a).

## Current-density and electrical current

The current-density operator is defined as

$$\hat{\mathbf{j}}(\mathbf{r}) = \frac{e}{2M_{\mathrm{e}}} \left[ \left( \hat{\mathbf{p}} - \frac{e}{c} \mathbf{A}(\hat{\mathbf{r}}) \right) \delta(\hat{\mathbf{r}} - \mathbf{r}) + \delta(\hat{\mathbf{r}} - \mathbf{r}) \left( \hat{\mathbf{p}} - \frac{e}{c} \mathbf{A}(\hat{\mathbf{r}}) \right) \right] . \tag{10}$$

- $e$ is the absolute value of the electronic charge.
- c is the speed of light.
- $\mathbf{A}$ is the vector potential.

The matrix elements of the $x$-component of $\hat{\mathbf{j}}(\mathbf{r})$ in the basis of the scattering states read

$$\left[ j^x(\mathbf{r}) \right]_{\bar{a}a}^{\bar{l}l} (\bar{\varepsilon}, \varepsilon) = \frac{e\hbar}{2iM_{\mathrm{e}}} \left[ \Psi_{\bar{l},\bar{\varepsilon},\bar{a}}^{(+)}(\mathbf{r})^* \frac{\partial}{\partial x} \Psi_{l,\varepsilon,a}^{(+)}(\mathbf{r}) - \Psi_{l,\varepsilon,a}^{(+)}(\mathbf{r}) \frac{\partial}{\partial x} \Psi_{\bar{l},\bar{\varepsilon},\bar{a}}^{(+)}(\mathbf{r})^* \right] - \frac{e^2}{M_{\mathrm{e}}c} \mathbf{A}(\mathbf{r}) \Psi_{\bar{l},\bar{\varepsilon},\bar{a}}^{(+)}(\mathbf{r})^* \Psi_{l,\varepsilon,a}^{(+)}(\mathbf{r}) . \tag{11}$$

The diagonal matrix element represents the current-density per spin and unit energy associated with the state $\Psi_{l,\varepsilon,a}^{(+)}$. Assuming a magnetic field free incoming lead $l$, it is useful to define an $N \times N$ current operator $\mathcal{I}_{l,\varepsilon}$ at energy $\varepsilon$, whose matrix elements in the subspace of scattering states $l, \varepsilon$ are

$$\left[ \mathcal{I}_{l,\varepsilon} \right]_{\bar{a}a} = \int_{S_x} \mathrm{d}y \, \left[ j^x(\mathbf{r}) \right]_{\bar{a}a}^{ll} (\varepsilon, \varepsilon) . \tag{12}$$

Current conservation implies that the definition is independent of the cross section $S_x$ chosen for the integration over the transverse coordinate. Given the one-to-one correspondence between incoming modes and outgoing scattering states, the current matrix elements (involving scattering states) can be identified with those of $t^\dagger t$ (involving lead modes), through

$$\left[ \mathcal{I}_{1,\varepsilon} \right]_{\bar{a}a} = \frac{e}{h} \left[ t^\dagger t \right]_{\bar{a}a} . \tag{13}$$

The diagonal matrix element $\left[ \mathcal{I}_{1,\varepsilon} \right]_{aa}$ is the current (per spin and unit energy) associated with the scattering state $1, \varepsilon, a$

$$I_{1,\varepsilon,a} = \frac{e}{h} \sum_{b=1}^{N} |t_{ba}|^2 = \frac{e}{h} \left( 1 - \sum_{b=1}^{N} |r_{ba}|^2 \right) . \tag{14}$$

Summing the current associated with each propagating mode amounts to an *incoherent* superposition of modes, that in the zero-temperature limit leads to a total current from left to right given by

$$I = \int_{\mu_2}^{\mu_1} \mathrm{d}\varepsilon \sum_{a=1}^{N} 2\pi\hbar v_a \rho_a(\varepsilon) \, I_{1,\varepsilon,a} \tag{15}$$

- $\rho_a(\varepsilon) = (\pi\hbar v_a)^{-1}$ is the one-dimensional density of lead modes per unit length (including the spin degeneracy factor).
- The index identifying the incoming lead is no longer needed, since the total current is the same in both leads.
- The factor $2\pi\hbar v_a$ arises from the change of the flux normalization used in Eq. (4) into a flux measured in particles arriving to the scatterer per unit time.

## Conductance is transmission

In the linear-response regime the energy integral in (15) is dominated by the contribution at the Fermi energy of the reservoirs $\varepsilon_{\mathrm{F}} \simeq \mu_1, \mu_2$. The two-probe Landauer-Büttiker formula for the linear conductance reads

$$G = \frac{I}{V} = \frac{2e^2}{h} T = G_0 \, g . \tag{16}$$

- $T = \sum_{a,b} T_{ba} = \mathrm{Tr}[t^\dagger t]$ is the total transmission coefficient (in short: the *transmission*).
- $T_{ba} = |t_{ba}|^2$ is the transmission coefficient between modes $a$ and $b$.
- The trace is taken over the incoming, right-moving modes $a$, the sum over $b$ corresponds to the outgoing, right-moving modes.
- $G_0 = 2e^2/h$ is the *quantum of conductance*.

- The dimensionless conductance $g$ is equal, in the two-probe case, to transmission coefficient.

The remarkably simple-looking form of the Landauer-Büttiker formula (16) hides some subtle issues that are thoroughly discussed in the corresponding literature of Mesoscopic Phyiscis (Imry, 1986-r). Prominent among them are the contact resistance (responsible for the non-zero resistance of perfectly transmitting samples) and the energy dissipation mechanisms (taking place in the reservoirs).

The above-sketched counting argument leading to Eq. (16) can be put on a rigorous framework by using the the linear response formalism of the conductivity (Kubo formula) within a wave-guide geometry (Fisher, 1981-a, Szafer, 1988-a). The extension to finite magnetic fields (Baranger, 1989b-a, Shepard, 1991-a, Nöckel, 1993-a) presents some subtleties, but the final form is still the simple-looking Eq. (16).

## Shot noise
### Refs. (Blanter, 2000-r)

The DC linear conductance involves stationary quantum states and therefore it does not account for time-dependent processes present in the transport problem. Time-dependent current fluctuations caused by the discreteness of the electronic charge, known as *shot noise*, have a zero-frequency power spectrum given by

$$P = 4 \int_0^\infty \mathrm{d}\tau \langle \delta I(\tau + \tau_0) \, \delta I(\tau) \rangle . \tag{17}$$

- The fluctuations are taken with respect to the stationary value $I$ of the current.
- The average is taken with respect to the initial time $\tau_0$.

Uncorrelated carriers are characterized by

$$P = P_{\mathrm{Poisson}} = 2eI = gP_0 . \tag{18}$$

- $P_0 = 2e \, G_0 V$.
- $V$ is the applied voltage.

When the electrons arrive from reservoirs containing degenerate electron gases, the correlations in the electron transmission imposed by the Pauli principle result in (Büttiker, 1990-a)

$$P = P_0 \, \mathrm{Tr}[tt^\dagger (\mathbf{1} - tt^\dagger)] , \tag{19}$$

which is in general smaller than $P_{\mathrm{Poisson}}$. Despite their similar structure, Eq. (19) for the shot noise contains temporal information not present in the expression (16) of the conductance.

## Polar decomposition of the scattering matrix
### Refs. (Stone, 1991-r)

The scattering matrix $S$, defined in (9), when studied within a random-matrix approach (presented in Sec. 3) is conveniently parametrized in the so-called polar decomposition as

$$S = \begin{pmatrix} u_3 & 0 \\ 0 & u_4 \end{pmatrix} \begin{pmatrix} -\mathcal{R} & \mathcal{T} \\ \mathcal{T} & \mathcal{R} \end{pmatrix} \begin{pmatrix} u_1 & 0 \\ 0 & u_2 \end{pmatrix} . \tag{20}$$

- $u_l$ ($l = 1, \ldots, 4$) are $N{\times}N$ unitary matrices.
- $\mathcal{R}$ is a diagonal matrix with non-zero elements $\mathcal{R}_n = \left[\dfrac{\lambda_n}{1 + \lambda_n}\right]^{1/2}$.
- $r^\dagger r = u_1^\dagger \mathcal{R}^2 u_1$, and then $R_n = \mathcal{R}_n^2$ are called *reflection eigenvalues*.
- $\mathcal{T}$ is a diagonal matrix with non-zero elements $\mathcal{T}_n = \left[\dfrac{1}{1 + \lambda_n}\right]^{1/2}$.
- $t^\dagger t = u_1^\dagger \mathcal{T}^2 u_1$, and then $T_n = \mathcal{T}_n^2$ are called *transmission eigenvalues*.
- $\lambda_n$ are real positive parameters.

In the *unitary* case without symmetries the matrix $S$ has $4N^2$ independent real parameters, and the polar decomposition (20) is not unique since it introduces $N$ extra parameters. In the *orthogonal* case, the time-reversal and spin rotation symmetries dictate that $S^\mathrm{T} = S$. Thus, $u_3 = u_1^\mathrm{T}$ and $u_4 = u_2^\mathrm{T}$. The number of independent parameters in the polar decomposition is then reduced to $2N^2 + N$ (where $N^2$ parameters appear for each of the two $N{\times}N$ unitary matrices, together with the $N$ parameters $\lambda_n$). In the *symplectic* case where the spin degeneracy is broken (for instance by spin-orbit scattering in the sample) and no magnetic field is applied, the size of $S$ has to be doubled in order to account for the spin indices. The matrices $u_3$ and $u_4$ are also given in terms of $u_1$ and $u_3$, and the $\lambda_n$ parameters have a twofold (Kramers) degeneracy (Mello, 1991-a).

The transmission coefficient is given by the sum of the transmission eigenvalues

$$T = \sum_{n=1}^{N} T_n = \sum_{n=1}^{N} \mathcal{T}_n^2 \,. \tag{21}$$

## Transmission eigenmodes and scattering eigenstates

The transmission eigenvectors (of the matrices $t^\dagger t$ and $t'^\dagger t'$) are given by the columns of the matrices $u_1$ and $u_2$. Similarly, the *transmission eigenmodes* are of the form

$$\varrho_{1,\varepsilon,n}^{(-)}(\mathbf{r}) = \sum_{a=1}^{N} [u_1]_{na}^* \, \varphi_{1,\varepsilon,a}^{(-)}(\mathbf{r}) \,, \quad x < 0 \,, \tag{22}$$

$$\varrho_{2,\varepsilon,n}^{(-)}(\mathbf{r}) = \sum_{a=1}^{N} [u_2]_{na}^* \, \varphi_{2,\varepsilon,a}^{(-)}(\mathbf{r}) \,, \quad x > 0 \,.$$

In the same way as the incoming modes (4) generate the outgoing scattering states (7), the transmission eigenmodes (22) give rise to *scattering eigenstates* $\chi_{l,\varepsilon,n}$ that are eigenfunctions of the current operator $\mathcal{I}_{l,\varepsilon}$. The latter can be written as linear combinations of the scattering states

$$\chi_{l,\varepsilon,n}^{(+)}(\mathbf{r}) = \sum_a c_{l,\varepsilon,a}^{(n)} \, \Psi_{l,\varepsilon,a}^{(+)}(\mathbf{r}) \,. \tag{23}$$

The coefficient $c_{1(2),\varepsilon,a}^{(n)}$ coincides with the matrix element $\left[u_{1(2)}\right]_{na}^*$ of Eq. (22) up to an overall $n$-dependent phase.

The incoherent superposition of scattering eigenstates provides an alternative path to formulate the scattering approach to the conductance. Working in this basis, $\sum_n I_{1,\varepsilon,n} = (e/h) \sum_n \mathcal{T}_n^2 = (e/h) \sum_{b,a} |t_{ba}|^2$ and thus Eq. (14) follows

## Scattering amplitudes in terms of the Green function

The formal theory of scattering adapted to a wave-guide (lead) geometry allows to relate the *retarded Green function* $\mathcal{G}(\mathbf{r}, \bar{\mathbf{r}}, \varepsilon)$ to the matrix elements of $S$. Writing the spectral decomposition of $\mathcal{G}$ in the basis of the scattering states leads to transmission and reflection amplitudes between modes $a$ and $b$ given by (Fisher, 1981-a)

$$t_{ba} = i\hbar (v_a v_b)^{1/2} \, \exp\left[-i(k_b^+ x - k_a^+ \bar{x})\right] \int_{S_x} \mathrm{d}y \int_{S_{\bar{x}}} \mathrm{d}\bar{y} \, \phi_b^*(y) \, \mathcal{G}(\mathbf{r}, \bar{\mathbf{r}}, \varepsilon) \, \phi_a(\bar{y}) \,, \tag{24}$$

$$r_{ba} = -\delta_{ab} \, \exp\left[i(k_b^+ x + k_a^+ \bar{x})\right] \exp\left[ik_b^+ |x - \bar{x}|\right] + i\hbar (v_a v_b)^{1/2} \, \exp\left[-i(k_b^+ x + k_a^+ \bar{x})\right] \int_{S_x} \mathrm{d}y \int_{S_{\bar{x}}} \mathrm{d}\bar{y} \, \phi_b^*(y) \, \mathcal{G}(\mathbf{r}, \bar{\mathbf{r}}, \varepsilon) \, \phi_a(\bar{y}) \,.$$

The integrations take place at transverse cross sections $S_{\bar{x}}$ on the left lead and $S_x$ on the right (left) lead for the transmission (reflection) amplitudes. The physical observables are obtained from the transmission and reflection coefficients ($T_{ba} = |t_{ba}|^2$ and $R_{ba} = |r_{ba}|^2$) between modes, which, by current conservation, do not depend on the choice of the transverse cross sections. Expressing the scattering amplitudes in terms of Green functions is extremely useful for analytical and numerical computations. Diagrammatic perturbation theory, as well as semiclassical expansions, are conveniently built in terms of Green functions.

## *R*-matrix formalism for a quantum dot connected to leads
### Refs. (Alhassid, 2000-r)

In the case of weak coupling between the dot and the leads the conductance is signed by the resonance states, that develop from the eigenstates of the closed dot. In this limit it is therefore useful to express the scattering matrix in terms of the eigenvalues and eigenvectors of the isolated dot. Such a connection can be established by adopting the *R*-matrix theory, originally developed in the context of Nuclear Physics, to the dot-lead setup (Jalabert, 1992-a). In this approach, the scattering matrix (9) is written as

$$S = \frac{\mathbf{1} + iK}{\mathbf{1} - iK} \,. \tag{25}$$

The Hermitian matrix $K$ is related to the $R$-matrix through $K = (kP)^{1/2} R (kP)^{1/2}$, and can be defined by its matrix elements

$$K_{c'c}(\varepsilon) = \frac{1}{2} \sum_\nu \frac{\gamma_{\nu,c'} \gamma_{\nu,c}^*}{\varepsilon_\nu - \varepsilon} \,. \tag{26}$$

- The index $c = (l, a)$ stands for the lead $l = 1, 2$ and the mode number $a = 1 \ldots N$.
- $\varepsilon_\nu$ is the energy of the $\nu$-th state of the isolated dot.
- $\gamma_{\nu,c} = \sqrt{\dfrac{\hbar^2 k_c P_c}{M_e}} \int_0^W dy\, \psi_\nu(x^{(l)}, y)\, \phi_a(y)$ is the *partial-width amplitude* for the decay of level $\nu$ into channel $c$.
- $\psi_\nu(x, y)$ is the $\nu$-th eigenfunction of the dot defined by the boundary conditions $\psi_\nu = 0$ at the walls of the cavity and $(\partial \psi_\nu / \partial n) - h_l \psi_\nu = 0$ at the dot-lead interfaces ($\hat{n} = \pm \hat{x}$ sets the normal direction to each interface and $h_l$ is an arbitrary constant that might depend on the interface $l = 1, 2$).
- $\phi_a(y)$ is the $a$-th channel wave-function defined in Eq. (6).
- $x^{(l)}$ is the value of the $x$ coordinate at the lead entrances. In Figure 24 $x^{(1)} = 0$ and $x^{(2)} = L_d$.
- $W$ is the width of the leads.
- $kP$ is a diagonal matrix.
- $k_c$ is the longitudinal wave-vector associated with mode $a$ of lead $l$.
- $P_c$ is the *penetration factor* to tunnel through the barrier for the channel $c$ (given by the decay of the scattering wave-functions across the barrier).

The matrix $K$ can be expressed in an arbitrary basis $\Psi_\mu$ of wave-functions of the dot

$$K(\varepsilon) = \pi\, \mathcal{W}^\dagger \left( \hat{H}_s - \varepsilon \right)^{-1} \mathcal{W} \,. \tag{27}$$

- $\hat{H}_s$ is the Hamiltonian of the dot, considered as a single-particle system (see Eq. (2)).
- The rectangular matrix $\mathcal{W}$ is defined by the couplings $\mathcal{W}_{\mu,c} = \sqrt{\dfrac{\hbar^2 k_c P_c}{2\pi M_e}} \int_0^W dy\, \Psi_\mu(x^{(l)}, y)\, \phi_a(y)$.

The scattering matrix can then be written as

$$S = \frac{\mathbf{1} + i\pi\, \mathcal{W}^\dagger \left( \hat{H}_s - \varepsilon \right)^{-1} \mathcal{W}}{\mathbf{1} - i\pi\, \mathcal{W}^\dagger \left( \hat{H}_s - \varepsilon \right)^{-1} \mathcal{W}} = \mathbf{1} - 2i\pi\, \mathcal{W}^\dagger \left( \hat{H}_\text{eff} - \varepsilon \right)^{-1} \mathcal{W} \,. \tag{28}$$

- The effective Hamiltonian $\hat{H}_\text{eff} = \hat{H}_s - i\pi\, \mathcal{W}\mathcal{W}^\dagger$ is not Hermitian, as it contains an imaginary self-energy describing the decay of the dot states into the leads.

When the typical level spacing $\Delta$ of the dot states is larger than the level widths, the energy-dependence of the scattering amplitudes close to the $\nu$-th eigen-level is dominated by the corresponding contribution to the matrix $K$ in Eq. (26). In this simple case, the transmission amplitude between modes $a$ and $b$ take, for $|\varepsilon - \tilde{\varepsilon}_\nu| \ll \Delta$, the Breit-Wigner form

$$t_{ba} = t_{ba}^{(0)} + i\, \frac{\gamma_{\nu,(1,a)}\, \gamma_{\nu,(2,b)}^*}{\varepsilon - \tilde{\varepsilon}_\nu + i\Gamma_\nu/2} \,. \tag{29}$$

- The weakly $\varepsilon$-dependent term $t_{ba}^{(0)}$ arises from the contribution of levels different than $\nu$.
- $\tilde{\varepsilon}_\nu$ is a renormalized energy, correcting $\varepsilon_\nu$ by the shift due to the barriers and the effect of far-away levels.
- $\Gamma_\nu = \sum_c |\gamma_{\nu,c}|^2$ is the level-width of the state $\nu$.

## Scattering phase and Friedel sum-rule

Since the scattering matrix $S$ is unitary its eigenvalues are pure phases $e^{i\varphi_c}$, with $c = 1, \ldots, 2N$. The Wigner time can then be written as

$$\tau(\varepsilon) = \frac{\hbar}{2iN} \text{Tr}\left[S^\dagger \frac{dS}{d\varepsilon}\right] = \frac{\hbar}{2N} \sum_{c=1}^{2N} \frac{d\varphi_c}{d\varepsilon}, \qquad (30)$$

and be related to the average (smoothed) density of states of the dot through

$$\langle d(\varepsilon) \rangle = \frac{1}{\pi\hbar} \tau(\varepsilon) . \qquad (31)$$

The case of leads supporting only one mode is particularly important, specially in problems concerning Coulomb blockade (see Sec. 5). The $S$-matrix is then $2 \times 2$, and can be parametrized with 4 angles as

$$S = e^{i\zeta} \begin{pmatrix} ie^{i\xi} \cos\theta & e^{i\eta} \sin\theta \\ e^{-i\eta} \sin\theta & ie^{-i\xi} \cos\theta \end{pmatrix}. \qquad (32)$$

- $\zeta \in [0, \pi)$ is the *scattering phase*.
- $\theta \in [0, \pi)$ sets the value of the transmission.
- $\eta \in [0, 2\pi)$. For time-reversal symmetric systems $\eta = 0, \pi$.
- $\xi \in [0, 2\pi)$. For systems that are left-right symmetric $\xi = 0, \pi$.
- $t = |t| e^{i\alpha} = \sin\theta \, e^{i(\zeta - \eta)}$.
- $\alpha = \zeta - \eta$ is the *transmission phase*.

The eigenphases of a $2 \times 2$ scattering matrix are given, for the case where $\cos\theta \cos\xi > 0$, by $\varphi_1 = \zeta + \text{Arcsin}(\cos\theta \cos\xi)$ and $\varphi_2 = \zeta + \pi - \text{Arcsin}(\cos\theta \cos\xi)$. Thus,

$$\tau(\varepsilon) = \hbar \, \frac{d\zeta}{d\varepsilon} , \qquad (33)$$

and the Friedel sum-rule relating the changes of the number of particles $N_s$ added to the dot with the corresponding change in the scattering phase can be written as

$$\Delta\zeta = \pi \, \Delta N_s . \qquad (34)$$

# Chaotic scattering

**Refs. (Smilansky, 1989-r; Tél, 1990-r; Gaspard, 2014-i)**

## Transient chaos

The study of a physical system from the Quantum Chaos point of view usually starts with the analysis of its classical dynamics. In the case of mesoscopic transport the classical scattering problem has to be considered. The concept of chaos, developed for closed systems and related to the long-time properties of the trajectories, has to be re-examined in open systems since the trajectories exit the scattering region after a finite amount of time.

The *transient chaos* of a scattering problem is characterized by the infinite set of trajectories which stay forever in the scattering region. This set is constituted by the periodic unstable orbits staying within the scattering region (the *strange repeller*) and their stable manifold (the open trajectories that converge to the previous ones in the infinite-time limit). Chaotic scattering is obtained when the dynamics in the neighborhood of the repeller is chaotic in the usual sense, and this set has a fractal dimension in the space of classical trajectories. When an incoming particle enters the scattering region, it approaches the strange repeller, bounces around close to this set for a while, and it is eventually ejected from the scattering region (if it did not have the right initial conditions to be trapped).

# Time-delay function

Fixing a point $\bar{y}$ at the entrance of the cavity (see Figure 7), and varying the injection angle $\bar{\theta}$ with which the classical trajectories impinge, allows to define the time-delay function $\tau_{\bar{y}}(\bar{\theta})$ as the time that the trajectory with initial conditions $(\bar{y},\bar{\theta})$ spends inside the cavity. Similarly, the function $\tau_{\bar{\theta}}(\bar{y})$ is defined if the fixed and scanned variables are switched. In the case of a chaotic cavity the curve $\tau_{\bar{y}}(\bar{\theta})$ has a fractal character. The infinitely trapped trajectories give the divergences of $\tau_{\bar{y}}(\bar{\theta})$ and determine the self-similar structure. These signatures are evident in Figure 4, presenting the time-delay function for a stadium billiard, similarly to what is obtained in other examples of chaotic scattering (see Figure 4 of Ref. (Gaspard, 2014-i) for the paradigmatic case of the three-disk problem).

# Escape rate and dwell time

The rate at which particles escape from a scattering region with chaotic dynamics results from a balance between the rate in which nearby trajectories diverge away from the repeller (characterized by their largest *Lyapunov exponent* $\lambda$) and the rate at which the chaotic escaping trajectories are folded back into the scattering region (depending on the density of the repeller, that is measured by its *fractal dimension d*). If particles are randomly injected in the scattering region, the survival probability at time $\tau$ will be $P(\tau) = e^{-\gamma\tau}$, with the *escape rate* $\gamma = \lambda(1-d)$ (Gaspard, 1989-a). The inverse of the escape rate is the typical time spent by the particles in the scattering region, and it is usually referred as the *dwell time*.

# Length-distribution in billiards

In ideal billiards the potential is completely flat within the sample. Thus, the total length $L$ of a trajectory inside the cavity (from entrance to exit) and the corresponding time $\tau$ are simply related by $L = v\tau$, where $v$ is the constant velocity of the scattering particles. As shown in Figure 5.a, the length-distribution (equivalent to the distribution of escape times) for a cavity with the shape of a stadium follows an exponential law (solid line) $P(L) = e^{-\gamma_{cl}L}$ (with $\gamma_{cl} = \gamma/v$). Such a distribution is independent on the chosen set of initial conditions for sampling the trajectories. The numerically obtained curve $P(L)$ becomes ragged for large $L$, due to the finite number of trajectories taken into account in the simulation. The exponential law sets in very fast, after a length corresponding to a few bounces.

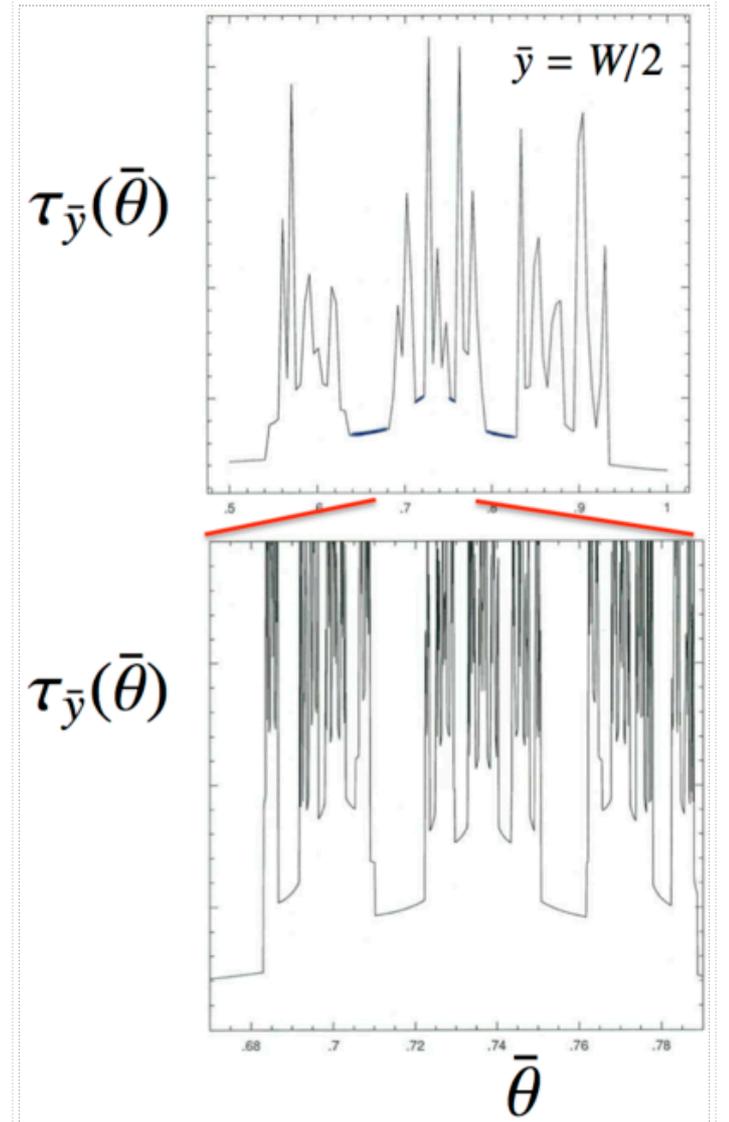

Figure 4: Time-delay function (in arbitrary units) as a function of the injection angle $\bar{\theta}$ (in radians) for a stadium billiard (with $R/W = 1$ as shown in the inset of Figure 9). The trajectories are launched at the center of the incoming lead ($\bar{y} = W/2$). The self-similar character of $\tau_{\bar{y}}(\bar{\theta})$ is illustrated by the blow-up in the lower panel (with better $\bar{\theta}$-resolution) of the central part of the upper panel.

The appearance of a single scale characterizing the length-distribution is a consequence of ergodic motion over the whole energy surface while in the scattering region (Bauer, 1990-a). The value of the escape rate can be estimated from general arguments of ergodicity in the case of chaotic cavities with small openings, where the typical trajectory bounces around many times before it escapes (Jensen, 1991-a). Assuming that the instantaneous distribution of trajectories is uniform on the energy surface, the escape rate is simply given by $\gamma = F/\mathcal{A}$, where $F$ is the flux through the holes (equal to the size of the holes times $v/\pi$, the factor of $\pi$ arising from the integration over the departing angles), and $\mathcal{A}$ is the area of the two-dimensional scattering domain. In the case of small holes this simple estimate reproduces remarkably well the escape rates obtained from the numerical determination of the survival probability sampling over classical trajectories.

In the integrable case, the particle moves over only that part of the energy surface allowed by the conserved quantities, and there is not a single scale for the length-distribution. In situations with multiple scales power-law distributions are observed (Bauer, 1990-a; Oakeshott, 1992-a; Lai, 1992-a). For the case of a rectangular cavity, an approximate $L^{-3}$ dependence for the length-distribution is obtained (dashed lines in Figure 5.a and .c). For integrable cavities, the length-distribution depends on the chosen set of initial conditions for sampling the trajectories. In the case of Figure 5 a uniform distribution of $\bar{y}$ along the entrance lead and a $\cos\bar{\theta}$ weighted angular distribution as initial conditions (consistently with the classical limit of the quantum problem) have been used. Not all integrable systems are alike concerning the length distribution, as circular billiards exhibit an exponential decay over same range of lengths (Legrand, 1991-a; Lin, 1993-a). It should be kept in mind that the long tails of the length-distributions are in general irrelevant from the point of view of mesoscopic transport, in view of the physical cutoffs encountered in ballistic microstructures (see Sec. 1.1.6).

# Effective area distribution

The effect of a magnetic field perpendicular to the plane of electrons on quantum transport depends on the area accumulated by the classical scattering trajectories. Scattering trajectories are open, and therefore do not have a well defined enclosed area. Instead, the *effective area* of a trajectory *s* can be defined from the circulation of the vector potential

$$\Theta_s = \frac{2\pi}{B} \int_{C_s} \mathbf{A} \cdot d\mathbf{r} .  \qquad (35)$$

- $\mathbf{A}$ is the vector potential defining the magnetic field $\mathbf{B} = B\hat{\mathbf{z}}$ .
- $C_s$ is the path followed by the trajectory inside the cavity.

If *s* were a closed trajectory, $\Theta_s$ would be equal to $2\pi$ times the enclosed area. Unlike the scattering time, $\Theta_s$ can be positive or negative. For a chaotic dynamics the distribution of effective areas depends on a single scale, similarly to the case of the length-distribution. Numerical calculations and analytic arguments (Berry, 1986-a; Jalabert, 1990-a; Oakeshott, 1992-a; Lecheminant, 1993-a) yield a distribution

$$N(\Theta) \propto \exp(-\alpha_{cl}|\Theta|) ,  \qquad (36)$$

where the parameter $\alpha_{cl}$ can be interpreted as the inverse of the typical effective area enclosed by a scattering trajectory. Figure 5.b presents the distribution $N(\Theta)$ (only for positive $\Theta$, solid line) obtained from the simulation of classical trajectories in a stadium cavity, in good agreement with the proposed distribution. Exploiting the ergodicity of the chaotic dynamics in the scattering domain, and assuming that the area is accumulated in a random-walk fashion, the parameter $\alpha_{cl}$ can be related to the escape rate and the typical length scale of the cavity (Jensen, 1991-a, Doron, 1991-a).

Scattering trajectories yield an effective area which is not gauge-invariant. However, the large (in absolute value) effective areas are associated with long trajectories bouncing many times, which are then constituted by many loops and two extreme "legs" in and out of the cavity. The dominant contribution comes from the loops, which is gauge-invariant. Changing the gauge in the numerical simulations modifies the distribution for small $\Theta$, but not the exponent $\alpha_{cl}$ governing the distribution of large $\Theta$.

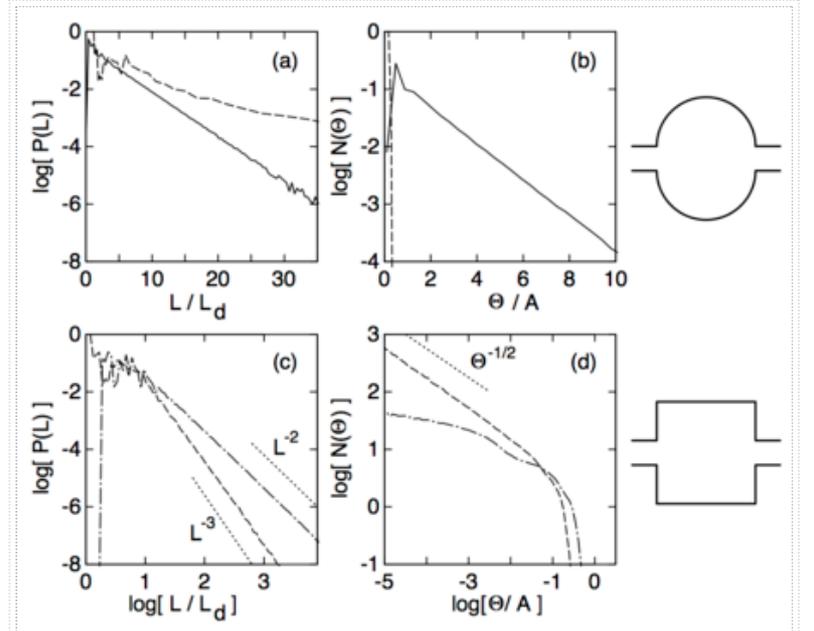

Figure 5: Classical distributions of length [(a),(c)] and effective area [(b),(d)] for the stadium (solid lines) and rectangular (dashed lines) billiards shown at the right. In the stadium [(a),(b)] both distributions are close to exponential, after a short transient region, and they are very different from the distributions for the rectangle, which show power-law behavior. A double-logarithmic scale is used in [(c),(d)] to compare the distributions for the square with different power-laws (indicated by dotted lines). The area distribution for the square in panel (b) is cutoff already for very small values of the effective area, and the corresponding power-law is visible in panel (d) by blowing up the interval of small values of $\Theta/A$. In panels (c) and (d) the dash-dotted lines are, respectively, the two-particle distributions of length and area-differences for pairs of trajectories starting with the same angle. Unlike the chaotic case, the two-particle distributions, that determine the conductance fluctuations through Eq. (50), do not factorize as the product of two one-particle distributions. $L_d$ denotes the direct length between the leads and $A$ is the area of the cavity. (From Ref. (Baranger, 1993-r), copyright 1993, American Institute of Physics.)

In the integrable case the effective area distributions are typically power-laws. For a rectangular cavity, the area distribution (dashed in Figure 5.d) exhibits (before a sharp cutoff) an approximate $\Theta^{-1/2}$ dependence.

# Semiclassical description of ballistic transport

**Refs. (Baranger, 1993-r; Stone, 1995-r; Nakamura, 1997-b; Baranger, 1999-r; Jalabert, 2000-r; Richter, 2000-b)**

## Quantum interference in clean cavities

The conductance fluctuations are the trademark of the mesoscopic regime, and the magnetic field appears as the main tuning parameter in experiments. The theoretical study of conductance fluctuations and other interference phenomena, like weak-localization and shot noise, is usually done by a combination of quantum numerical calculations and semiclassical expansions. Quantum mechanical calculations based on the recursive Green function method (Lee, 1981-a) allow to extract the transmission coefficient for a confinement potential that mimics the electrostatic potential defining the quantum dot. In the experimentally realizable microstructures such a potential results from the imposed gate voltages, the additional short or long-range disorder, and the self-consistent screening. This detailed information is however rather difficult to extract for the quantum dots used in quantum transport measurements (Nixon, 1990-a; Stopa, 1996-a).

Therefore, the crude approximation of a clean billiard with hard walls and a flat potential inside is usually made in theoretical studies concerning quantum chaos. This choice is the simplest for numerical and analytic calculations, and allows to treat the case of a classical dynamics that exhibits hard chaos. The applicability of this approximation to the experimentally achievable micro-cavities depends on the fabrication details (Marcus, 1993b-r; Taylor, 1997; Sachrajda, 1998-a; Marlow, 2006-a). Self-consistent electrostatic calculations of clean quantum dots show that both, hard and soft electrostatic confinement, could be encountered (see Figure 22) depending on the potentials used for the heterostructures and the particular geometry.

Figure 6 shows the transmission coefficient of an asymmetric *clean* cavity as a function of the incoming flux $kW/\pi$ ($\mu_1 = \hbar^2 k^2/2m$, the integer part of $kW/\pi$ is the number of propagating channels $N$). The overall behavior of the transmitted flux is a linear increase with $k$ due to the two-dimensional character of the problem. The classical limit of the semiclassical approximation (presented in Sec. 2.3.3), corresponding to the neglect of quantum interference, reproduces the slope of this secular behavior, which is noted as "classical".

Superimposed to the secular behavior, there are fine-structure fluctuations characteristic of the cavity under study. These *conductance fluctuations*, analogous to those of disordered metals, also appear when the Fermi energy is fixed and the magnetic field is used as a control parameter. The conductance fluctuations are characterized by their magnitude, $\langle(\delta T)^2\rangle$, and the correlation scale as a function of wave-vector $k_c$ (or magnetic field $B_c$). The numerical results indicate that these characteristic scales do not change when going into the semiclassical limit of large $kW$.

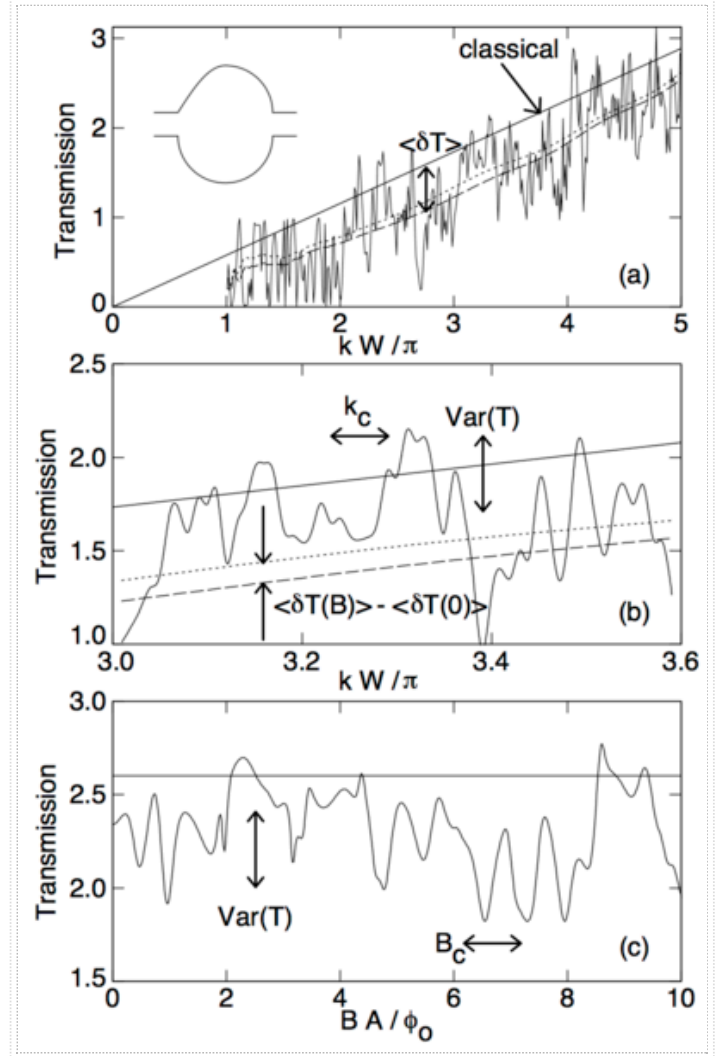

Figure 6: Transmission through the cavity shown in the inset of panel (a) as a function of the wave-vector $k$ [(a) and (b)] or magnetic field $B$ [(c)]. The straight solid line is the classical transmission coefficient (calculated in Sec. 2.3.3). The fluctuating solid line is the numerically obtained quantum transmission coefficient, and the dashed (dotted) line is its smoothed counterpart at zero magnetic field (a magnetic field such that $BA/\Phi_0 = 0.25$). $\langle\delta T\rangle$ stands for the offset between the classical and quantum transmission coefficients. The difference of $\langle\delta T\rangle$ with and without magnetic field is a measure of the weak-localization in the cavity. $\mathrm{Var}(T)$ quantifies the fluctuations of the transmission coefficient around its secular value. The scales $k_c$ and $B_c$ are, respectively, the momentum and magnetic-field correlation lengths. $W$ is the width of the leads, $A$ is the area of the cavity and $\Phi_0 = hc/e$ is the flux quantum. (From Ref. (Baranger, 1993-r), copyright 1993, American Institute of Physics.)

The secular behavior of the transmission coefficient (dashed line) lies below its classical value (smooth solid line). Such departure arises from quantum effects in the regime $kW/\pi \lesssim 1$ (i.e. the transmission coefficient is negligible before the opening of the first mode). The above defined classical limit only reproduces the slope of the large-$kW$ smoothed transmission coefficient, but the shift $\langle \delta T \rangle$ does not disappear in the large-$kW$ limit. The presence of a weak magnetic field tends to decrease such an offset, yielding a secular behavior (dotted line) that runs higher than in the $B=0$ case. This is the *weak-localization effect* for ballistic cavities (Baranger, 1993-a). The reason for choosing an asymmetric cavity is that the ballistic weak-localization effect is strongly dependent on the spatial symmetries of the cavity (Baranger, 1996-a).

The numerical results of Figure 6 show that the conductance fluctuations and the weak-localization effect, first discussed in the context of disordered mesoscopic conductors, are also present in ballistic mesoscopic cavities. The differences among these two types of mesoscopic systems call for a rethinking of the appropriate definition of averages, as well as the concept of universality, in the ballistic regime.

# Semiclassical scattering amplitudes

## Semiclassical Green function
**Refs. (Ozorio, 1988-b; Gutzwiller, 1989-r; Gutzwiller, 1990-b; Brack, 1997-b; Stöckmann, 1999-b; Haake, 2001-b)**

The Green function is the Fourier transform of the propagator. The Van Vleck expression for the latter, together with a stationary-phase integration on the time variable, leads to the semiclassical approximation (SCA) for the Green function

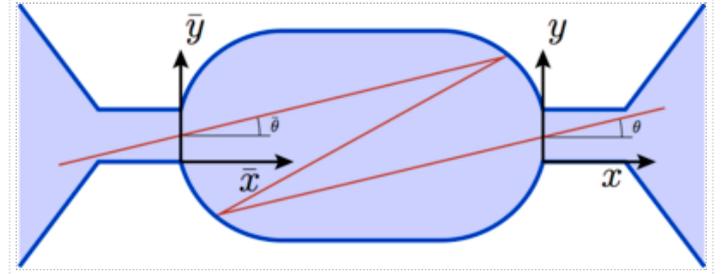

Figure 7: Typical classical trajectory entering the cavity at point $(\bar{x}, \bar{y})$ with and angle $\bar{\theta}$ and leaving it at point $(x, y)$ with and angle $\theta$.

$$\mathcal{G}(\mathbf{r}, \bar{\mathbf{r}}, \varepsilon) = \frac{2\pi}{(2\pi i\hbar)^{(f+1)/2}} \sum_{s(\bar{\mathbf{r}},\mathbf{r})} \sqrt{D_s} \, \exp\left[\frac{i}{\hbar} S_s(\mathbf{r}, \bar{\mathbf{r}}; \varepsilon) - i\frac{\pi}{2}\nu_s\right]. \tag{37}$$

- $f$ is the number of degrees of freedom (dimensions of the configuration space). The choice of $f=2$ is adopted throughout since it is the relevant case for low-dimensional transport.
- $s(\bar{\mathbf{r}}, \mathbf{r})$ stands for the classical trajectories going from $\bar{\mathbf{r}}$ to $\mathbf{r}$ with energy $\varepsilon$. (The argument $\varepsilon$ is not explicitly written.)
- $S_s = \int_{C_s} \mathbf{p} \cdot d\mathbf{q}$ is the action integral along the path $C_s$. In the case of billiards without magnetic field $S_s = \hbar k L_s$, where $L_s$ is the trajectory length.
- $D_s$ describes the evolution of the classical probability and can be expressed as a determinant of second derivatives of the action. For two-dimensional cavities connected to leads, as in Figure 7, we have $D_s = \dfrac{M_e}{v \cos\theta} \left|\left(\dfrac{\partial \bar{\theta}}{\partial y}\right)_{\bar{y}}\right|$.
- $\bar{\theta}$ and $\theta$ are, respectively, the incoming and outgoing angles of the trajectory $s$ with the $x$-axis.
- $v$ is the velocity of the scattering particles.
- $\nu_s$ is the Maslov index counting the number of constant-energy conjugate points along the path $C_s$, as well as the phase acquired at the bounces with hard walls.

## Semiclassical transmission amplitudes

The transmission amplitudes determining the conductance of a cavity (like the ones in Figure 1 and Figure 6) through Eq. (16) admit a semiclassical form, which can be obtained by inserting the semiclassical Green function (37) into Eq. (24), and then performing, in the case of a large number of modes $N$, the $\bar{y}$ and $y$ integrals through two successive stationary-phase approximations. For hard-wall leads, the stationary points $\bar{y}_0$ and $y_0$ are, respectively, given by

$$\left(\frac{\partial S}{\partial \bar{y}}\right)_y = -\frac{\hat{\bar{a}}\hbar\pi}{W}, \quad \hat{\bar{a}} = \pm a; \qquad \left(\frac{\partial S}{\partial y}\right)_{\bar{y}} = -\frac{\hat{b}\hbar\pi}{W}, \quad \hat{b} = \pm b. \tag{38}$$

The dominant trajectories are those where the initial and final transverse momentum equal, respectively, the momentum of the corresponding transverse (channel) wave-function. The semiclassical expression for the transmission amplitude can then be cast as (Jalabert, 1990-a)

$$t_{ba} = -\frac{\sqrt{2\pi i\hbar}}{2W} \sum_{\hat{\bar{a}}=\pm a} \sum_{\hat{b}=\pm b} \sum_{s(\hat{b},\hat{\bar{a}})} \mathrm{sgn}(\hat{\bar{a}}\hat{b}) \sqrt{\widetilde{D}_s} \exp\left[\frac{i}{\hbar}\widetilde{S}_s(\hat{b},\hat{\bar{a}};\varepsilon) - i\frac{\pi}{2}\widetilde{\nu}_s\right]. \tag{39}$$

- $s(\hat{b},\hat{\bar{a}})$ stands for the classical trajectories, of energy $\varepsilon$, with entering and exiting angles $\theta_{\hat{\bar{a}}}$ and $\theta_{\hat{b}}$ such that $\sin\theta_{\hat{\bar{a}}} = \frac{\hat{\bar{a}}\pi}{kW}$ and $\sin\theta_{\hat{b}} = \frac{\hat{b}\pi}{kW}$, respectively.

- $\widetilde{S}(\hat{b},\hat{\bar{a}};\varepsilon) = S(y_0,\bar{y}_0;\varepsilon) + \left(\frac{\hbar\pi\hat{\bar{a}}}{W}\right)\bar{y}_0 - \left(\frac{\hbar\pi\hat{b}}{W}\right)y_0$ is the reduced action and $\bar{y}_0(y_0)$ is the stationary point for the $\bar{y}(y)$ integration. In the case of billiards without magnetic field $\widetilde{S} = \hbar k\widetilde{L}$, with $\widetilde{L} = L + k\bar{y}_0\sin\theta_{\hat{\bar{a}}} - ky_0\sin\theta_{\hat{b}}$.

- For two-dimensional billiards $\widetilde{D}_s = \frac{1}{M_e v \cos\theta}\left|\left(\frac{\partial \bar{y}}{\partial \theta}\right)_{\bar{\theta}}\right|$.

- $\widetilde{\nu} = \nu + H\left(\left(\frac{\partial\bar{\theta}}{\partial\bar{y}}\right)_y\right) + H\left(\left(\frac{\partial\theta}{\partial y}\right)_{\bar{\theta}}\right)$, where $H$ is the Heaviside step function.

## Semiclassical reflection amplitudes

For the semiclassical reflection amplitude (24) there are two kinds of trajectories contributing to $\mathcal{G}(0,y,0,\bar{y};\varepsilon)$; those that penetrate into the cavity and those which go directly from $\bar{y}$ to $y$ staying on the cross section of the lead. It is only trajectories of the first kind which contribute to the semiclassical reflection amplitude given in terms of trajectories leaving and returning to the cross section at the left lead, with appropriate quantized angles. The trajectories of the second kind merely cancel the term $\delta_{ba}$.

## Convergence and generalizations of the semiclassical expansions

The semiclassical transmission amplitude (39) is, for an open system, the analogous of the Gutzwiller trace formula for the density of states of a closed system (Gutzwiller, 1971-a). In the chaotic case, both formulas are expressed as a sum over isolated classical trajectories, allowing to establish the connection between classical and quantum properties. The main difference between the scattering and energy-level problems, at the semiclassical level, is that the trace formula involves the sum over periodic orbits while the transmission amplitude is given by open trajectories that go across the scattering region. In chaotic systems the number of trajectories connecting two given points grows exponentially with the trajectory length. In open systems the trajectories can escape the scattering region, therefore their proliferation is much weaker than in the close case (although still exponential). Therefore, the convergence of semiclassical propagators in chaotic scattering will not encounter the difficulties of the trace formula. From the quantum point of view, since the Gutzwiller trace formula aims to reproduce a delta-function spectrum, it can be conditionally convergent at most. On the contrary, the quantum transmission amplitude is a smooth function of the Fermi energy (away from the thresholds at the opening of new modes), and the semiclassical sum can be absolutely convergent (depending on the value of the fractal dimension $d$ of the strange repeller governing the chaotic scattering (Jensen, 1994-a)). Moreover, when applied to mesoscopic systems, the semiclassical expansions should be truncated by physical cutoffs (Sec. 1.1.6).

Chaotic scattering problems have been studied by Miller (Miller, 1974-a), in the context of molecular collisions, in terms of the semiclassical propagator in the momentum representation. In such case the relevant sum is over classical trajectories with fixed incident and outgoing momenta. Eq. (39) is a mixed position-momentum representation of the Green function, and it can be adapted to handle finite magnetic fields, soft walls in the leads (Baranger, 1991-a), and tunneling in the cavities (Schreier, 1998-a).

For direct trajectories that traverse the cavity without collisions with the walls, or in cases where the dynamics of the cavity is integrable, the classical trajectories are not necessarily isolated, and only one of the two stationary-phase integrations leading to (39) can be performed (see Sec. 2.7). The corresponding semiclassical expressions of the transmission amplitudes are sums over *families of trajectories* (Pichaureau, 1999-a), in analogy with the Berry-Tabor formula for the density of states of integrable systems (Berry, 1976-a).

The numerical evaluation of the semiclassical transmission amplitudes can be addressed once the classical trajectories are characterized. Such a difficult task has been carried out for simple geometries like that of a circular scattering domain (Lin, 1996-a), where short trajectory (Ishio, 1995-a; Schreier, 1998-a) and diffraction (Schwieters, 1996-a) effects have been highlighted. The semiclassical approach can be generalized to include diffraction effects in the transmission and reflection amplitudes (Vattay, 1997-r; Wirtz, 2003-a; Březinova, 2010-a), through "ghost paths", or diffractive trajectories (like reflections off the mouth of an exiting lead). These effects are particularly important in the extreme quantum limit of $N = 1$.

The semiclassical transmission amplitude (39) has been generalized to the case where the spin-degeneracy is broken by a spin-orbit coupling acting in the scattering region (Zaitsev, 2005-a).

# Transmission coefficients, average values, and fluctuations

## Transmission coefficients

The transmission coefficients between two modes are obtained from the magnitude squared of the corresponding transmission amplitudes. Thus, in a semiclassical approach, they are given by sums over *pairs* of trajectories. Focusing in the case of billiards, it is convenient to scale out the energy (or wave-vector) dependence and write the transmission coefficient between modes $a$ and $b$ as

$$T_{ba}(k) = \frac{1}{2}\left(\frac{\pi}{kW}\right) \sum_{\hat{a},\tilde{a}=\pm a} \sum_{\hat{b},\tilde{b}=\pm b} \sum_{s(\hat{a},\hat{b})} \sum_{u(\tilde{a},\tilde{b})} F_{s,u}(k) \ . \tag{40}$$

- $F_{s,u}(k) = \sqrt{\tilde{A}_s \tilde{A}_u} \exp[ik(\tilde{L}_s - \tilde{L}_u) + i\pi\phi_{s,u}]$.
- $s(\hat{a}, \hat{b})$ labels the paths with extreme angles $\theta_{\hat{a}}$ and $\theta_{\hat{b}}$.
- $u(\tilde{a}, \tilde{b})$ labels the paths with extreme angles $\theta_{\tilde{a}}$ and $\theta_{\tilde{b}}$.
- $\tilde{A}_s = (\hbar k/W)\tilde{D}_s$ is a geometrical factor, independent on energy.
- $\phi_{s,u} = (\tilde{\nu}_u - \tilde{\nu}_s)/2 + \hat{a} + \hat{b} + \tilde{a} + \tilde{b}$.

## Wave-vector and energy averaged values

The expression of the transmission coefficients (40) is valid in the semiclassical (large-$k$) limit and therefore some kind of average has to be defined in order to relate it with the highly structured curve of the transmission in Figure 6. In a ballistic system, the ensemble average is not relevant since only a single cavity is at stake. The appropriate average is over wave-vector (or energy), defined for an arbitrarily observable $O(k)$ as

$$\langle O \rangle = \lim_{q \to \infty} \frac{1}{q} \int_{q_c}^{q_c+q} dk \ O(k) \ , \qquad \frac{q_c W}{\pi} \gg 1 \ . \tag{41}$$

This average is particularly suited for analytical calculations, since it provides a rigorous treatment of the $k$-dependent transmission coefficients, but it is not appropriate for dealing with experimental or numerical results, where only a finite $k$-range is accessible. In practice, an average over many quasi-periods of the function $T(k)$, yields results consistent with (41). This approach is usually adopted in quantum chaos studies by performing a local energy average (Blümel, 1988-a). At the experimental level, the average of the transmission coefficient over a finite energy-range yields the finite-temperature conductance.

## Transmission probability

The secular behavior of the transmission across two-dimensional cavities is linearly increasing with $k$ (outgoing flux proportional to the incoming flux). The *transmission probability* is then defined by the average

$$\mathcal{T} = \left\langle \frac{\pi}{kW} T(k) \right\rangle \ . \tag{42}$$

In the large $kW$-limit the modes are closely spaced in angle, and the sums over modes can be converted into integrals over angles

$$\sum_{a}^{N} \sum_{\hat{a}=\pm a} \to \left(\frac{kW}{\pi}\right) \int_{-1}^{1} \mathrm{d}(\sin\hat{\theta}) \, .$$

Thus, the $k$-dependence only remains in the phase factors. Exchanging the angle-integrals with the $k$-average,

$$\mathcal{T} = \frac{1}{2} \int_{-1}^{1} \mathrm{d}(\sin\bar{\theta}) \int_{-1}^{1} \mathrm{d}(\sin\theta) \sum_{\bar{\theta}'=\pm\bar{\theta}} \sum_{\theta'=\pm\theta} \sum_{s(\bar{\theta},\theta)} \sum_{u(\bar{\theta}',\theta')} \sqrt{\widetilde{A}_s \widetilde{A}_u} \, \langle \exp[ik(\widetilde{L}_s - \widetilde{L}_u) + i\pi\phi_{s,u}] \rangle \, . \tag{43}$$

The evaluation of the average leads to $k(\widetilde{L}_s - \widetilde{L}_u) + \pi\phi_{s,u} = 0$. In the absence of symmetries, such a relation is only possible if $s = u$. Quantum interference is therefore absent in the resulting *diagonal* term. Changing variables, from the outgoing angle $\theta$ to the initial position $\bar{y}$, results in

$$\mathcal{T} = \frac{1}{2} \int_{-1}^{1} \mathrm{d}(\sin\bar{\theta}) \int_{0}^{W} \frac{\mathrm{d}\bar{y}}{W} f(\bar{y},\bar{\theta}) \, . \tag{44}$$

- $f(\bar{y},\bar{\theta}) = 1$ if the trajectory with initial conditions $(\bar{y},\bar{\theta})$ is transmitted.
- $f(\bar{y},\bar{\theta}) = 0$ if the trajectory with initial conditions $(\bar{y},\bar{\theta})$ is reflected.

The expression (44) is a purely classical one, with the intuitive interpretation of a transmission probability. $\mathcal{T}$ can also be obtained from a Boltzmann equation approach (Baranger, 1991-a). The transmission probability is experimentally relevant when the temperature is high enough to kill the interference effects, and thus it has been used to understand the early experiments on transport in ballistic junctions at Helium temperatures (Roukes, 1987-a; Ford, 1989-a; Beenakker, 1988-a).

The numerical implementation of Eq. (44) is easily done by sampling the classical trajectories with random choices of the initial position and initial angles (with a weight of $\cos\bar{\theta}$). This purely classical procedure yields values of $\mathcal{T}$ which are consistent with the slope of the quantum numerical results (see Figure 6).

### Transmission shift and fluctuations

The secular (smoothed) behavior of the $T(k)$ (dashed line in Figure 6) is characterized by the slope $\mathcal{T}$ of its asymptote and the transmission shift, defined by the average

$$\langle \delta T \rangle = \left\langle \left( T(k) - \left(\frac{kW}{\pi}\right)\mathcal{T} \right) \right\rangle \, . \tag{45}$$

The fluctuations with respect to the secular behavior are

$$\delta T(k) = T(k) - \left(\frac{kW}{\pi}\mathcal{T} + \langle \delta T \rangle\right) \, . \tag{46}$$

The data in Figure 6 presents the striking feature that the typical size of the fluctuations $\delta T(k)$ is of order 1, independently of the $k$-interval. These conductance fluctuations are considered in Sec. 2.4 from a semiclassical approach (Jalabert, 1990-a) based on the semiclassical treatment of the $S$-matrix fluctuations as a function of energy, introduced by Gutzwiller (Gutzwiller, 1983-a), and later developed (Blümel, 1988-a; Gaspard, 1989-a; Doron, 1991-a). The dependence of $\langle \delta T \rangle$ on magnetic field amounts to the ballistic weak-localization effect (treated in 2.5).

## Conductance fluctuations in classically chaotic cavities

### Wave-vector dependent conductance fluctuations and power spectrum

The conductance fluctuations can be studied through the *wave-vector dependent correlation function* of the transmission

$$C_K(\Delta k) = \sum_{a,b}^{N} \sum_{a',b'}^{N} C_{K,bab'a'}(\Delta k) \, . \tag{47}$$

- $C_{K,bab'a'}(\Delta k) = \langle \delta T_{ba}(k + \Delta k) \delta T^*_{b'a'}(k) \rangle$.

The correlation function is characterized by the typical size $C_K(0)$ of the conductance fluctuations and the correlation length $k_c$ giving the scale for the decay in the variable $\Delta k$. The Fourier power spectrum

$$\widehat{C}_K(x) = \int d(\Delta k)\, C_K(\Delta k)\, e^{ix\Delta k} \tag{48}$$

is particularly useful in order to separate the different length scales appearing in the conductance fluctuations.

## Semiclassical approach to the correlation length of the conductance fluctuations

The semiclassical expression for $C_{K,bab'a'}(\Delta k)$ involves sums over terms depending on four trajectories, say $s(\hat{a},\hat{b})$, $u(\tilde{a},\tilde{b})$, $s'(\hat{a}',\hat{b}')$, $u'(\tilde{a}',\tilde{b}')$. The terms with $s=u$ and $s'=u'$ are excluded due to the subtraction of the average values. $C_K(\Delta k)$ results from the sum of a large number of terms that, in general, have very different phases, making the semiclassical calculation very difficult. A special contribution is that of the diagonal terms, defined by $\hat{a}=\hat{a}'$, $\tilde{a}=\tilde{a}'$, $\hat{b}=\hat{b}'$, $\tilde{b}=\tilde{b}'$, $s=s'$, and $u=u'$. Keeping only these terms leads to the so-called *diagonal approximation* $C_K^D(\Delta k)$. Such an approximation is not justified for the calculation of $C_K(0)$ since only represents a small fraction of the total number of terms contributes. In Sec. 2.6 it is discussed the appropriate way of incorporating the contributions of the off-diagonal terms.

Under the assumption that the correlation function $C_K(\Delta k)$ and its diagonal contribution $C_K^D(\Delta k)$ have both a functional dependence on $\Delta k$ governed by the same parameter, the correlation length $k_c$, it is useful to address the semiclassical calculation of the latter. Converting the sums over modes of Eq. (47) into integrals leads to

$$C_K^D(\Delta k) = \frac{1}{4} \int_{-1}^{1} d(\sin\bar{\theta}) \int_{-1}^{1} d(\sin\theta) \sum_{\bar{\theta}'=\pm\bar{\theta}} \sum_{\theta'=\pm\theta} \sum_{s(\bar{\theta},\theta)} \sum_{u(\bar{\theta}',\theta')} {}' \tilde{A}_s \tilde{A}_u \exp\left[i\Delta k(L_s - L_u)\right]. \tag{49}$$

- The "prime" in the summation over trajectories indicates that the terms $s=u$ are excluded.

The Fourier power spectrum of the diagonal component verifies

$$\widehat{C}_K^D(x) \propto \int_0^{\infty} dL\, P(L+x)\, P(L). \tag{50}$$

- $P(L) = \frac{1}{4}\int_{-1}^{1} d(\sin\theta) \int_{-1}^{1} d(\sin\theta') \sum_{u(\theta,\theta')} \tilde{A}_u \delta(L - L_u)$ is the classical distribution of lengths $L$.
- The trajectories are taken to be uniformly distributed in the sine of the angle in the definition of the classical distribution of lengths.
- The angular constraints linking trajectories $u$ and $s$ are neglected.
- The constraint $u \neq s$ is ignored due to the proliferation of long paths.
- $x$ is taken positive.

As discussed in 1.4.4, for chaotic billiards the distribution of lengths is exponential for large $L$ (and independent on the injection conditions), while there may be deviations at small $L$. Using the exponential form for all lengths,

$$\widehat{C}_K^D(x) \propto e^{-\gamma_{cl}x} \quad , \quad C_K^D(\Delta k) = \frac{C_k^D(0)}{1+(\Delta k/\gamma_{cl})^2}. \tag{51}$$

For billiards the correlation length $k_c = \gamma_{cl}$ is $k$-independent, implying that the conductance fluctuations persist (and remain invariant) in the large-$kW$ limit. In generic chaotic systems, the energy-correlation functions can be obtained from a semiclassical analysis and an energy-average over intervals small in the classical scale (such that the trajectories are unchanged) but large in the quantum scale (containing many oscillations of the transmission coefficient) leading to (Blümel, 1988-a; Doron, 1991-a)

$$C_E^D(\Delta\varepsilon) = \frac{C_E^D(0)}{1+\left(\Delta\varepsilon/(\hbar\gamma)\right)^2}. \tag{52}$$

The conductance fluctuations are thus on a scale $\varepsilon_c = \hbar\gamma$ that is much larger than the level spacing $\Delta$. Due to the wide openings of the cavity, transport occurs in the regime of *overlapping resonances*. This regime has been extensively studied in Nuclear Physics, in the context of compound nuclei, characterized by the Ericson fluctuations (Ericson, 1960-a). The photoexcitation cross sections of rubidium Rydberg states in crossed, electric and magnetic fields, provide another physical example characterized by the Ericson regime (Madroñero, 2005-a). One-dimensional models of chaotic scattering obtained as an open variant of the kicked rotator exhibit transmission fluctuations, with a quantum correlation length which is well described by the numerically computed classical escape rates (Borgonovi, 1992-a).

## Semiclassical approach to the magnetic filed correlation length for chaotic cavities

The conductance fluctuations as a function of the magnetic field are relevant from the experimental point of view, since an external field is a very useful control variable. The *magnetic filed dependent correlation function* is defined as an average over $k$

$$C_B(\Delta B) = \langle \delta T(k, B + \Delta B)\, \delta T^*(k,B) \rangle \,. \tag{53}$$

In analyzing experimental or numerical data, averages over finite $k$ or $B$-intervals (small enough not to appreciably modify the classical dynamics) are generally used.

The derivation of the magnetic field correlation length (Jalabert, 1990-a) follows similar lines as in the case of the $k$-correlation length of Sec. 2.4.2. The diagonal approximation consists in the identification of trajectories $s$ and $s'$, as well as $u$ and $u'$. Such pair of trajectories are not identical, since they correspond to different Hamiltonians. However, the shadowing theorem ensures the existence of pairs of nearby trajectories with the same boundary conditions and an action difference $[S_s(B + \Delta B) - S_s(B)]/\hbar = \Theta_s \Delta B/\Phi_0$. The diagonal term $C_B^D(\Delta B)$ is then selected from the general expression (53) yielding

$$C_B^D(\Delta B) = \frac{1}{4} \int_{-1}^{1} d(\sin\bar\theta) \int_{-1}^{1} d(\sin\theta) \sum_{\bar\theta'=\pm\bar\theta} \sum_{\theta'=\pm\theta} \sum_{s(\bar\theta,\theta)} \sum_{u(\bar\theta',\theta')}{}' \tilde{A}_s \tilde{A}_u \exp\left[i\frac{\Delta B}{\Phi_0}(\Theta_s - \Theta_u)\right] . \tag{54}$$

- $\Phi_0 = hc/e$ is the flux quantum.

The Fourier power spectrum of the diagonal component verifies

$$\widehat{C}_B^D(\eta) \propto \int_{-\infty}^{\infty} d\Theta\, N(\Theta + \eta)\, N(\Theta) \,. \tag{55}$$

- $N(\Theta)$ is the effective area distribution.

Using the exponential form (36) of the distribution of effective areas $N(\Theta)$, for all values of $\Theta$ results in (Jalabert, 1990-a)

$$\widehat{C}_B^D(\eta) \propto e^{-\alpha_{cl}|\eta|}\left(1 + \alpha_{cl}|\eta|\right) \quad , \quad C_B^D(\Delta B) = \frac{C_B^D(0)}{[1 + (\Delta B/\alpha_{cl}\Phi_0)^2]^2} \,. \tag{56}$$

Under the expectation that for a chaotic cavity only one characteristic scale appears in the correlation function $C_B(\Delta B)$ and its diagonal component $C_B^D(\Delta B)$, the magnetic field correlation length is $B_c = \alpha_{cl}\Phi_0$. The independence of this field scale on $k$ implies that the $B$-dependent conductance fluctuations persist (and remain invariant) in the large-$kW$ limit.

The expressions (51), (52) and (56) are typical quantum chaos connections, as they relate mesurable properties of a *quantum system* ($k_c$, $\varepsilon_c$, $B_c$) with *classical quantities* ($\gamma_{cl}$, $\gamma$, $\alpha_{cl}$) determined by the underlying classically chaotic dynamics.

# Quantum numerical calculations of conductance fluctuations in classically chaotic cavities

The regime of validity of the semiclassical predictions can be tested by quantum numerical calculations. The conductance through a cavity within a one-particle description can be obtained, as a function of the Fermi energy $\varepsilon_F$ of the reservoirs or a perpendicular magnetic field $B$. The study of fluctuations requires a definition of the average values, inducing some degree of arbitrariness when dealing with finite data. The use of the *power-spectra* $\widehat{C}_K(x)$ and $\widehat{C}_B(\eta)$, directly obtainable from the Fourier power of the raw data was proposed to circumvent this problem (Jalabert, 1990-a).

The analysis in terms of the power-spectrum is useful because it allows to distinguish the fluctuations according to their length scales. This distinction is not only of technical nature when comparing with numerical simulations, but it is also important on physical grounds. In experimentally relevant mesoscopic systems there are various cut-off lengths beyond which the simple disorder-free, one-particle models are not applicable, and the connection with Quantum Chaos looses its meaning (see 1.1.6).

Figure 8 shows the smoothed power spectra $\widehat{C}_K(x)$ for an asymmetric cavity for the cases where the number of propagating modes in the leads is $N = 1$ (triangles) and $N = 21$ (squares). The later case is well represented by Eq. (51) over a wide range of $x$. There are, nevertheless, deviations for small lengths ($x \simeq L_d$) and for large $x$. The deviations for small lengths are understandable since the chaotic nature of the dynamics (and the statistical treatment of the trajectories) cannot give an appropriate description for short trajectory lengths. The deviations for large lengths arise from the limitations of semiclassics and the diagonal approximations, and they become more important upon reducing $k$. Even if the data for $N = 1$ has important departures from the semiclassical prediction (51), the slope obtained by fitting data up to $x/L_d \approx 20$ is remarkably accurate. The quantum correlation lengths $\gamma_{qm}$ obtained by the linear fitting of $\widehat{C}_K(x)$ are in very good agreement with the classical escape rate $\gamma_{cl}$ for a different geometries encompassing a large span of values of $\gamma_{cl}$ (see Figure 9.a).

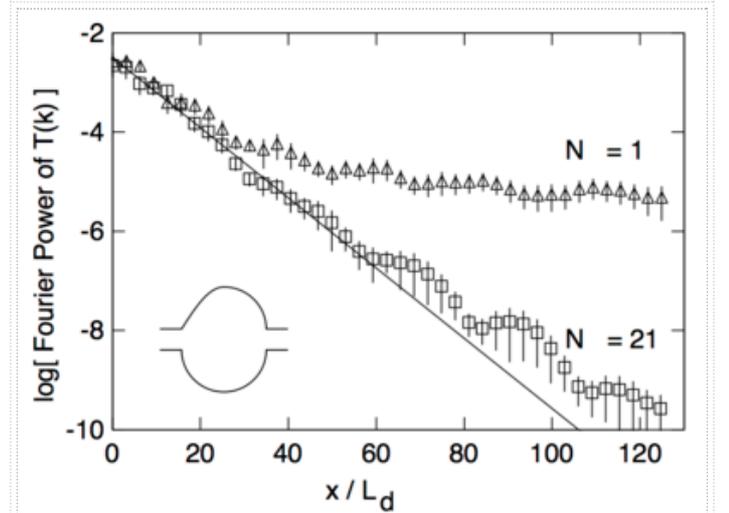

Figure 8: Power spectrum of the transmission $T(k)$ for the chaotic structure shown in the inset with two electron densities corresponding to $N = 21$ (squares) and $N = 1$ (triangles) open channels. In the former case, there is a very good agreement with Eq. (51), while in the extreme quantum limit when only one mode is propagating in the leads, the agreement with the semiclassical theory is restricted to a small interval of $x$. (From Ref. (Baranger, 1993-r), copyright 1993, American Institute of Physics.)

For the conductance fluctuations as a function of magnetic field, the numerically obtained $\widehat{C}_B(\eta)$ shows good agreement with the semiclassical prediction (56) for an intermediate range of values of $\eta$ (Figure 10). The analysis in terms of the power-spectrum is necessary since the correlation function $C_B(\Delta B)$ follows the semiclassical prediction for a restricted interval of $\Delta B$ leaving aside the contributions from very short and very long trajectories (Jalabert, 1990-a). The good agreement between $\alpha_{qm}$ (obtained by the fitting to the quantum calculations) and $\alpha_{cl}$ (obtained from the simulation of the classical dynamics) is presented in Figure 9.b, where $\alpha_{cl}$ is varied over roughly two orders of magnitude by changing the size of the structures considered. For the four-probe structure, the fluctuations studied are those of the Hall resistance (which can be expressed as a function of the transmission coefficient between leads) (Büttiker, 1986-a; Beenakker, 1988-a; Baranger, 1991-a).

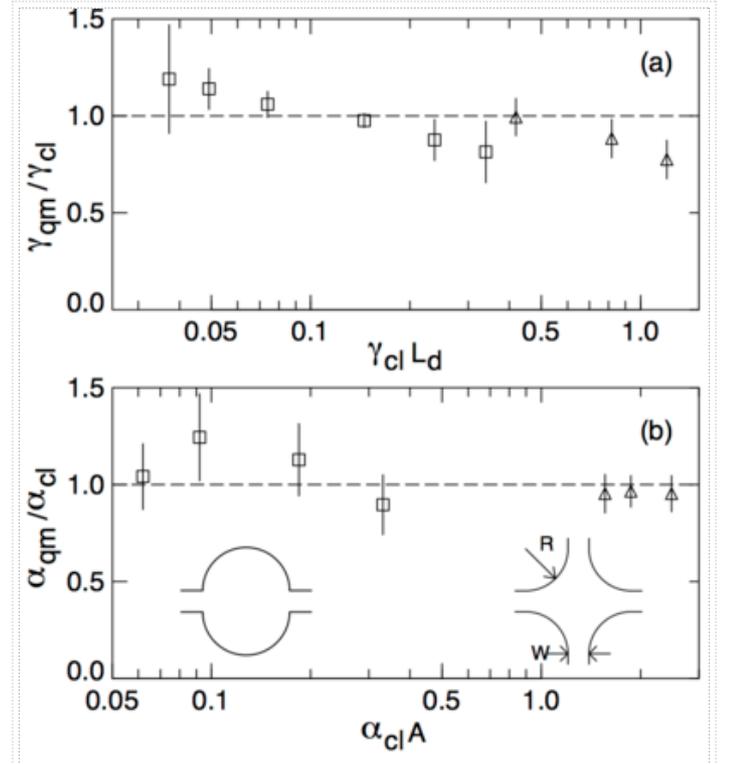

Figure 9: (a) Ratio of the numerically obtained wave-vector correlation length $k_c$ to the classical escape rate $\gamma_{cl}$ as a function of $\gamma_{cl}$ for both types of structures shown as insets; four-disc structure (triangles) with $R/W = 1, 2, 4$, and stadium (squares) with $R/W = 0.5, 1, 2, 4, 6, 8$. (b) Ratio of magnetic field correlation length to the exponent of the distribution of effective areas $\alpha_{cl}$ as a function of $\alpha_{cl}$ for the four-disc structure with $R/W = 1, 2, 4$ (triangles), and the open stadium with $R/W = 1, 2, 4, 6$ (squares). $R$ is the radius of the circles composing the stadium, or that of the disks of the four-disc structure. $W$ is the width of the leads. (Adapted from Ref. (Jalabert, 1990-a), copyright 1990, American Physical Society.)

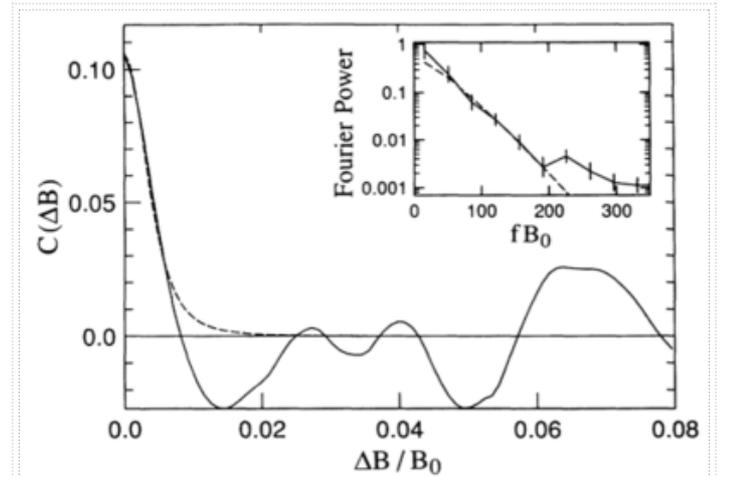

Figure 10: Magnetic-field correlation function obtained from a magnetoconductance curve of a stadium (see panel (c) in Figure 6). The dashed line is the semiclassical prediction for the correlation function. Inset: the smoothed power spectrum (solid) and best fit to the form (56) of $\widehat{C}_B(\eta)$ (dashed). (From Ref. (Jalabert, 1990-a), copyright 1990, American Physical Society.)

The $k$-independence of $\gamma_{qm}$, $C_K(0)$, $\alpha_{qm}$ and $C_B(0)$ is approximately respected in the numerical simulations away of the quantum limit of small $N$. The numerical quantum calculation validate the conjecture that the correlation functions in a chaotic cavity have *a unique characteristic length*, determined by the underlying classical dynamics, and therefore extracting these lengths from the diagonal part of the correlations is appropriate. The $k$-independence obtained for the size of the conductance oscillations arising from quantum calculations, as well as for the magnitude of the diagonal part of the correlation functions, points towards the universal character of the conductance fluctuations in chaotic ballistic systems.

Multiple-connected structures, like an open Sinai billiard, have been studied numerically and with the semiclassical approximations of Secs. 2.2.2 and 2.2.3, resulting in an Aharonov-Bohm periodicity induced by the central scatterer (Kawabata, 1997-a; Ree, 1999-a).

# Weak-localization in classically chaotic cavities

As indicated in Sec. 2.3.4 and in the discussion of Figure 6, the shift $\langle \delta T \rangle$ is sensitive to a perpendicular magnetic field. The presence of a small field increases the average conductance. This effect is called the ballistic weak-localization (Baranger, 1993-a), by analogy with the disordered case. It is important to realize that it is an average effect. Only after removing the (large) conductance fluctuations by the $k$-average, the (small) difference between the secular behaviors with and without magnetic field emerges. The two-probe conductance is an even function of the magnetic field, therefore in a given sample, $g(B)$ may have a *maximum or a minimum* at $B=0$. The two possible cases are observed on individual samples, experimentally (Keller, 1994-a) and in the numerical simulations.

## Semiclassical approach to coherent backscattering

The expression (44) of the transmission probability $\mathcal{T}$ is obtained in the diagonal approximation of pairing each trajectory with itself, thus neglecting the possibility that different paths may have the same effective action. However, time-reversal (Baranger, 1993-a) or spatial (Whitney, 2009-a) symmetries induce degeneracies among actions of symmetry-related pairs of trajectories, resulting in the non-vanishing off-diagonal terms. Achieving exact geometrical symmetries of the confining potential in actual microstructures is quite difficult, due to limitations in the fabrication procedure. But the time-reversal symmetry is exactly fulfilled in the absence of an external magnetic field.

At zero magnetic field, the interference of time-revered trajectories does not appear in the transmission coefficients $T_{ba}$, nor in the non-diagonal (in modes) reflection coefficients $R_{ba}$ having $b \neq a$, but only in the diagonal reflection coefficients $R_{aa}$. The reflection coefficients $R_{ba}$ and the *reflection probability* $\mathcal{R}$ are given, respectively, by semiclassical expressions analogous to (40) and (44), where the contributing trajectories start and exit at the entrance lead. Three kinds of pairs contribute to $R_{aa}$: those of identical trajectories (giving rise to $\mathcal{R}$), those of time-revered trajectories (contributing to $\langle \delta R \rangle = \langle (R - (kW/\pi)\mathcal{R}) \rangle$), and those with different actions. As in the case of conductance fluctuations, the off-diagonal terms are difficult to evaluate. In Sec. 2.6 it is discussed an appropriate way of incorporating their contribution.

The assumption of a single magnetic field scale for $\delta R(B)$ and for its diagonal component $\delta R^D(B)$ justifies attempting a semiclassical calculation of the latter. For magnetic fields $B$ weak enough not to modify appreciably the classical trajectories, the action difference between two time-reversed paths $s$ and $u$ is $S_s - S_u = 2\hbar\Theta_s B/\Phi_0$, and after changing the sum over modes by an integral over initial angles, the diagonal correction to the total reflection coefficient writes

$$\langle \delta R^D(B) \rangle = \frac{1}{2} \int_{-1}^{1} d(\sin\bar{\theta}) \sum_{s(\bar{\theta},\pm\theta)} \widetilde{A}_s \exp\left[i \frac{2B}{\Phi_0} \Theta_s\right], \tag{57}$$

which yields an order unity ($k$-independent) contribution containing only classical parameters (and $\Phi_0$). Assuming that in a chaotic system there is a uniform distribution of exiting angles, and that the distribution (34) of effective areas is valid even when the initial and final angles of the trajectories are constrained (Baranger, 1993-a),

$$\langle \delta R^D(B) \rangle = \frac{\mathcal{R}}{1 + (2B/\alpha_{cl}\Phi_0)^2}. \tag{58}$$

The Lorentzian line-shape is governed by the same parameter $\alpha_{cl}$ of the conductance fluctuations (up to a factor of 2). The diagonal reflection coefficients $R_{aa}$ are on average twice as large as the typical off-diagonal terms (of the order of $\mathcal{R}/N$). This factor of 2 enhancement, is known as *elastic enhancement* in the contexts of in disordered systems Akkermans, 1995-b and Nuclear Physics (Iida, 1990-a; Lewenkop, 1991-a).

The *coherent backscattering* $\langle \delta R^D(0) \rangle - \langle \delta R^D(B) \rangle$ is one of the contributions to the weak-localization $\langle \delta R(0) \rangle - \langle \delta R(B) \rangle$. Due to the unitarity condition, the weak-localization can also be expressed as $\langle \delta T(B) \rangle - \langle \delta T(0) \rangle$. Since there are no time-symmetry related pairs of trajectories contributing to $\langle \delta T(0) \rangle$, we see that the non-diagonal terms (in mode and trajectory indices) are relevant to evaluate the magnitude of the weak-localization effect. Similarly as in the case of conductance fluctuations, a chaotic cavity is expected to be characterized by a single field scale. Therefore, the weak-localization is conjectured to have the same Lorentzian line-shape (58), with the width given by the classical parameter $\alpha_{cl}$.

## Quantum numerical calculations of weak-localization and coherent backscattering

Figure 11 shows the numerically obtained quantum transmission through of a chaotic cavity (lower-right) at $B = 0$ (solid) and its smoothed trace (dashed), as a function of $k$. The application of a small field results in another rugged transmission (not shown), which when smoothed (dotted), lies above the corresponding $B = 0$ result, illustrating the ballistic weak-localization effect. The behavior of the smoothed traces close to $B = 0$ is well represented by a Lorentzian (inset) line-shape, with the width given by the classical parameter $\alpha_{cl}$, in agreement with the conjecture that the weak-localization and coherent backscattering have the same line-shape. The non-chaotic cavity of the upper right presents an approximately linear line-shape close to $B = 0$ (dashed in the inset).

The numerical quantum calculations allow to separately evaluate the weak-localization and the coherent backscattering effects. For the two structures of Figure 12 the field-dependent part of the (smoothed) total reflection coefficient $\langle R(0) \rangle - \langle R(B) \rangle$ (solid) is split in its diagonal (dashed) and off-diagonal (dotted) parts. For the structure with the stopper $\langle \delta R^D \rangle$ is approximately independent of $k$, and its magnitude is within 30% of $\mathcal{R}$. The elastic enhancement factor goes approximately from 2 to 1 when the field is turned on, in good agreement with the semiclassical prediction. An important off-diagonal contribution of opposite sign, not accessible by the diagonal approximation used in 2.3.4, reduces the weak-localization effect with respect to the coherent backscattering. The structure without stoppers exhibits similar features, but has a reduced weak-localization effect. Also, the magnitude of the coherent backscattering differs considerably from $\mathcal{R}$ and there is an important net variation as a function of $k$. These discrepancies with the semiclassical diagonal approximation are due to the presence of short paths and the approximate nature of the uniformity assumption used to obtain Eq. (58). Eliminating the effect of the short paths is thus necessary in order to approach the universal regime (see Sec. 3.4).

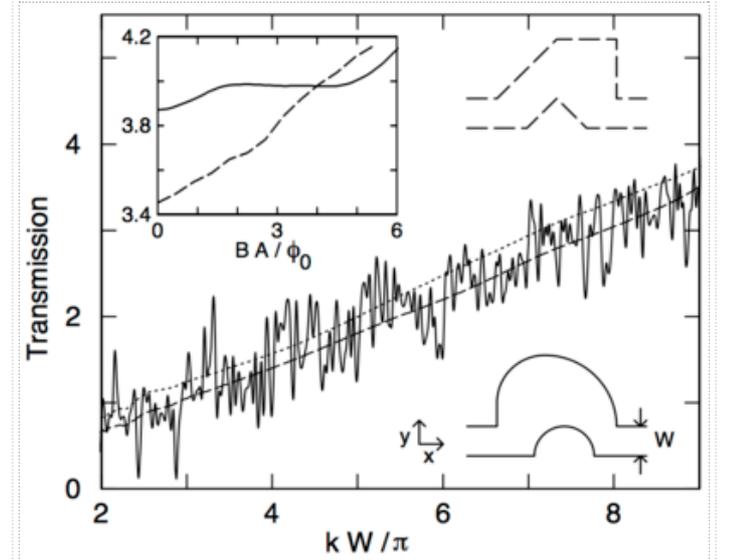

Figure 11: Transmission as a function of wave-vector for the half-stadium structure shown in the bottom right. The zero-temperature fluctuations (solid) are eliminated through the smoothing induced a temperature corresponding to 20 correlation lengths. The offset resulting from changing the magnetic field from $B = 0$ (dashed) to $B = 2\Phi_0/A$ (dotted) corresponds to the average magnetoconductance effect. $A$ is the area of the structures. Inset: smoothed transmission coefficient as a function of the flux through the cavity ($kW/\pi = 9.5$) showing the difference between the chaotic structure in the bottom right (solid) and the regular structure in the top right (dashed). The geometries chosen for the numerical calculations follow the shapes of the microstructures of Figure 18 used in Refs. (Keller, 1994-a; Keller, 1996-a). (From Ref. (Baranger, 1993-a), copyright 1993, American Institute of Physics.)

## Beyond the diagonal approximation: loop and Ehrenfest-time corrections

**Refs. (Waltner, 2010; Waltner, 2011)**

### Off-diagonal terms

The diagonal semiclassical approximations used in 2.4.2, 2.4.3, and 2.5.1 for the study of the conductance fluctuations and weak-localization only considered the contribution of terms pairing equal or time-reversed trajectories. While this approach is expected to yield the correct line-widths of the conductance fluctuations and the weak-localization, it does not respect unitarity, and it cannot provide the magnitude of these effects.

The simplistic view that terms with pairs of trajectories with actions that are not exactly equal cancel upon energy averaging has two shortcomings. On the one side, due to the exponential proliferation of trajectories with length, there are pairs with a very small action difference, as compared to $\hbar$ (Argaman, 1995-a; Argaman, 1996-a). On the other hand, the $N$ diagonal reflection coefficients $R_{aa}$ are less numerous than the $N(N-1)/2$ off-diagonal coefficients. Since the diagonal approximation yields the former to order $(1/N)^0$, the latter should be obtained to order (1/*N*) in order to keep the consistency of a (1/*N*) expansion.

## Loop contributions

Based on analogous cases of disordered (Akkermans, 2007-b) and closed chaotic systems (Sieber, 2001-a; Sieber, 2002-a), Richter and Sieber (Richter, 2002-a) proposed that the off-diagonal terms surviving the energy averages are given by pairs of trajectories which remain close (in configuration space), and only differ in whether they undergo or avoid a self-intersection with a small crossing angle $\varphi$. Two of these trajectories are sketched, respectively, by the solid and dashed traces in Figure 13 (the specularity of the reflections has not been respected in order to have the encounter region restricted to a few bounces). In a ballistic microstructure the very long trajectories are sensitive to the smooth disorder, and therefore they experience a small lateral deflection from the straight lines classically obtained in the clean case. However, the global stability of chaotic dynamics makes this distinction unimportant for the present analysis.

The action difference of the so-called Richter-Sieber pairs can be obtained by linearizing the dynamics in the vicinity of the encounter (Sieber, 2002-a)

$$\Delta S(\varphi) = \frac{p^2 \varphi^2}{2 M_\mathrm{e} \lambda} \,. \tag{59}$$

- $\varphi$ is the crossing angle.
- $p$ is the momentum.
- $\lambda$ is the Lyapunov exponent.

In addition to the action difference (59), the calculation of the off-diagonal contributions necessitates the knowledge of the number of self-crossings $P(\varphi, \tau)\mathrm{d}\varphi$ in the range between $\varphi$ and $\varphi + \mathrm{d}\varphi$ for orbits with time $\tau$. This is a complicated problem depending on the nature of the classical dynamics. The assumption of ergodicity leads to (Richter, 2002-a)

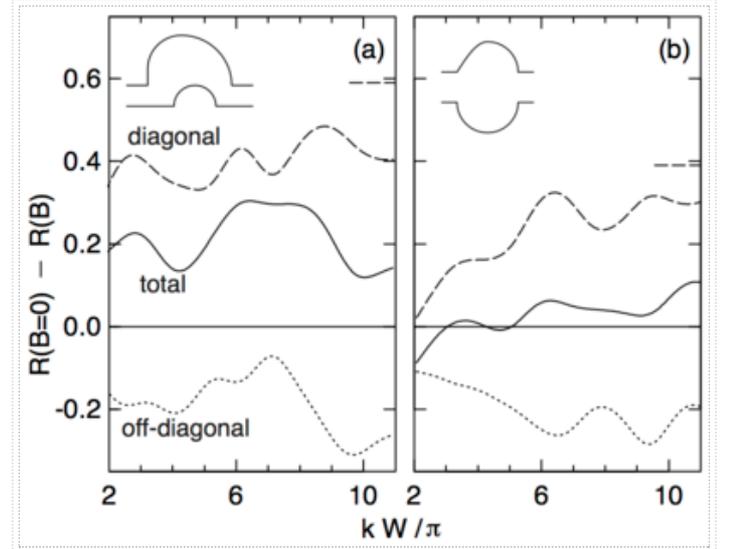

Figure 12: Change in the total (smoothed) reflection coefficient (solid), as well as the diagonal (dashed) and off-diagonal (dotted) parts, upon changing $B$ from 0 to $2\alpha_{cl}\Phi_0$. The dashed ticks on the right mark the classical value of the reflection probability $\mathcal{R}$. The weak-localization (total) correction has a positive coherent backscattering component (diagonal) and another contribution with opposite sign (off-diagonal). Blocking the short trajectories allows to approach the universal results of the semiclassical theory. (From Ref. (Baranger, 1993-a), copyright 1993, American Institute of Physics.)

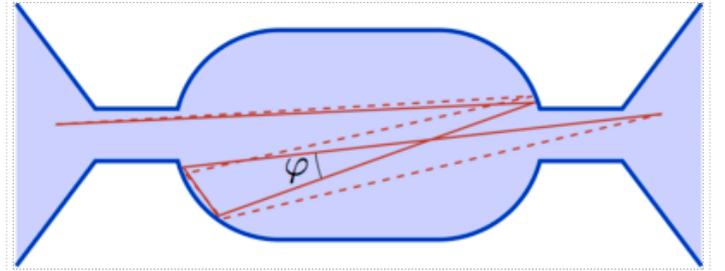

Figure 13: Schematic picture of a pair of interfering trajectories characterized by a crossing and an avoided crossing with a small angle $\varphi$. Such a pair gives rise to the loop contribution of the weak-localization correction of the represented cavity.

$$P(\varphi, \tau) = \frac{M_\mathrm{e} C^2}{\Sigma(\varepsilon)} \left(\tau - \tau_{\min}(\varphi)\right)^2 \, \sin\varphi \,. \tag{60}$$

- $\Sigma(\varepsilon)$ denotes the phase-space volume of the system at energy $\varepsilon$.
- $\tau_{\min}$ is the minimal time to form a closed loop, $\tau_{\min} = (2/\lambda)\ln(c/\varphi)$ for a uniform hyperbolic dynamics (with $C$ of order unity).

From (59) and (60), the *loop contributions* to the scattering coefficients are (Richter, 2002-a)

$$T_{ba}^{\mathrm{loop}}(k) = R_{ba}^{\mathrm{loop}}(k) = \frac{4\pi\hbar}{\Sigma(\varepsilon)} \int_0^\pi \mathrm{d}\varphi \int_{2\tau_{\min}}^\infty \mathrm{d}\tau \, e^{-\gamma(\tau - \tau_{\min})} P(\varphi, \tau) \, \cos\left(\frac{p^2 \varphi^2}{2 M_\mathrm{e} \lambda}\right) = -\frac{1}{4N^2} \,. \tag{61}$$

These contributions restore the unitarity condition that the diagonal approximation of 2.5.1 failed to fulfill, and agree with the random-matrix theory results of Sec. 3.4. The agreement with random-matrix theory results for all the observables considered is not surprising, since the ergodic hypothesis used in the semiclassical loop calculations is equivalent to the assumptions upon which the use of random-matrix theory is justified. In both cases there is no system parameter other that the number of modes *N*, and therefore all the system specific information of the original system has been washed out by the ergodic hypothesis.

The loops formed by off-diagonal trajectory-pairs are traversed in opposite directions, and therefore these orbits acquire an additional phase difference in the presence of a weak magnetic field. Since the flux enclosed in the loops is governed by the area distribution (36), the magnetic field dependence of the off-diagonal terms is the same that of the diagonal ones, confirming the conjecture that coherent backscattering and weak-localization have the same line-shape in chaotic cavities.

More refined arguments take into account the interplay between encounters and the proximity of the leads, as well as the possibility of pairs of trajectories with multiple encounters (Heusler, 2006-a). Diagrammatic rules have been developed (Müller, 2007-a) towards a systematic expansion in orders of $1/N$ for weak-localization, conductance fluctuations and shot noise. While in the first case only pairs of trajectories are relevant, the calculations for the other two observables require considering quadruples of trajectories. The correct handling of multiple encounters becomes then crucial, and lead to full agreement with random-matrix theory.

## Ehrenfest-time corrections

The semiclassical approach of the previous sections, based on the diagonal and loop approximations, makes extensive use of the ergodic hypothesis for predicting the magnitude of the weak-localization and the conductance fluctuations. In an open system such idealization amounts to neglect the contribution from terms related with direct, lead connecting trajectories (see 2.5.2), as well as that arising from trajectories that stay short times in the cavity. The relative importance of the latter is quantified by the average staying time in the cavity, or by its inverse, the escape rate $\gamma$ presented in 1.4.3. In order to determine the universal features of quantum transport, this last time scale should be compared with the Ehrenfest time, defined by the time it takes for a minimal wave-packet to spread and cover the entire cavity. In a chaotic cavity the stretching in phase-space is exponential and

$$\tau_E = \frac{1}{\lambda} \ln\left[\frac{pa}{\hbar}\right]. \tag{62}$$

- $\tau_E$ is the Ehrenfest time.
- $a$ is typical size of the structure.
- $\lambda$ is the Lyapunov exponent.
- $p/\hbar$ is the initial spread of a minimal wave-packet.

The Ehrenfest time separates times where the evolution of a particle follows essentially the classical dynamics from times when it is dominated by wave interference. The interference phenomena are then restricted to trajectories larger than $\tau_E$. Imposing this cutoff in the semiclassical sums over pairs of trajectories, the result (61) takes the form (Adagideli, 2003-a)

$$T_{ba}^{\text{loop}}(k, \tau_E) = -\frac{1}{4N^2} \exp\left[-\gamma \tau_E\right]. \tag{63}$$

This result has been originally derived from field theoretical methods (Aleiner, 1996-a) relying on a small amount of disorder in an otherwise clean, extended, two-dimensional structure (i.e. a Lorentz gas of hard disks with a superimposed smooth disorder). Taking as $a$ the typical size of the disks, the separation of the dynamics on scales shorter and longer than the Ehrenfest time allows to treat correlations of the disorder potential for short times and the use the diffusion equation for long times.

The Ehrenfest-time corrections are difficult to detect in ballistic cavities like the one simulated in Figure 6, since their observation requires achieving very large wave-vectors $k$. This is why most of the numerical checks, like that of the weak-localization correction (63), have been done through simulations in the open quantum kicked rotator (a-Rahav05).

Coherent backscattering is not affected by the Ehrenfest-time correction (Rahav, 2006-a). Contrary to the weak-localization, the variance of the conductance is found to be independent of the Ehrenfest time (Tworzydlo, 2004-a; Jacquod, 2004-a; Brouwer, 2006-a). In the semiclassical limit, the non-diagonal quantum corrections are shown to have a universal parametric dependence which is not described by random-matrix theory (Brouwer, 2007-a). The shot noise (Brouwer, 2007p-a) and the distribution of waiting times for electrons traversing a quantum dot (Waltner, 2011-a) have been shown to be affected by Ehrenfest-time corrections.

# Integrable and mixed dynamics

## Direct trajectories

The semiclassical form (39) of the transmission amplitude is not valid when the contributing trajectories are not isolated, but belong to *families* (see 2.2.4). For geometries with the two leads facing each other, the contribution to the diagonal transmission amplitude from direct trajectories is given by (Baranger, 1993-r)

$$t_{aa}^d = -\exp\left[\frac{ikL_d}{\cos\theta}\right]\left\{(1-\rho\tan\theta)\exp[-i\pi a\rho] + \frac{1}{\pi a}\sin[\pi a\rho]\right\}. \tag{64}$$

- $\sin\theta = \frac{a\pi}{kW}$.
- $\rho = \frac{L_d}{W}\tan\theta$.

The off-diagonal terms ($a \neq b$) vanish if $a$ and $b$ have different parity. When families of trajectories are relevant, only the $\bar{y}$-integration in (24) can be done by stationary-phase approximation, while the $y$-integration leading to Eq. (64) is done exactly (allowing to incorporate diffractive effects). For modes with the same parity the off-diagonal terms are not zero, but they are significant only for $b \simeq a$. Ref. (Lin, 1996-a) generalized these results to the case in which the leads are not collinear.

The contribution from the family of direct trajectories has a different dependence on $\hbar$ (or $k$) than that of the isolated trajectories. The number of modes that support direct trajectories is $N(W/L_d)$, and therefore the effect of direct trajectories does not disappear in the semiclassical limit. Direct trajectories are relevant in certain geometries, and their presence hampers the straightforward comparison between the semiclassical theory with numerical calculations or experimental data. Thus, many of the numerical simulations incorporate "stoppers" in the billiards which eliminate this effect. At the experimental level various approaches have been used: displacing the leads (Keller, 1994-a), having an angle smaller than $\pi$ between the two leads (Marcus, 1992-a; Huibers, 1998), or using stoppers inside the cavity (Keller, 1996-a; Lee, 1997-a).

## Scattering through a rectangular cavity

The case of the square is rather special among integrable systems since the conserved quantities of the cavity are the same as in the leads. The families of trajectories can be found by going to an extended space spanned by copies of the original cavity, and treat them like direct trajectories. A continuous-fraction approach allows to identify the families of trajectories and calculate the semiclassical transmission amplitudes for the scattering through a rectangular billiard (Pichaureau, 1999-a). The resulting conductance fluctuations are not universal, but increase with $k$.

Quantum mechanical calculations for a square cavity allowed to identify the peaks of the Fourier transform of the transmission amplitude with the families (or bundles) of classical trajectories contributing in the semiclassical expansion (Wirtz, 1997-a). The inclusion of diffractive paths leads to a good quantitative agreement with the quantum calculations (Wirtz, 2003-a).

## Circular billiards

The circular billiard is particularly interesting because it has been experimentally studied (Marcus, 1992-a; Berry, 1994b-a; Chang, 1994-a; Persson, 1995-a; Lee, 1997-a), and it is an integrable geometry where the semiclassical transmission amplitude (39) is applicable since the contributing trajectories are isolated. Also, the proliferation of trajectories with the number of bounces is much weaker than for the chaotic case, allowing for the explicit summation of Eq. (39).

Lin and Jensen (Lin, 1996-a) undertook such a calculation considering trajectories up to 100 bounces. Going into the semiclassical limit, by increasing the number of modes $N$ or the width of the leads, resulted in a better fulfillment of the unitarity condition $T+R=N$ (only a 1% deviation is obtained for $N=20$). The direct semiclassical sum yielded a coherent backscattering that is significantly reduced by off-diagonal contributions to the total reflection.

The signature of classical trajectories in the numerically obtained quantum transmission amplitudes has been established for circular billiards (Ishio, 1995-a; Schwieters, 1996-a; Schreier, 1998-a). In particular, the Fourier transform of the transmission amplitudes shows strong peaks for lengths corresponding to the classical trajectories contributing in the semiclassical expansion (39). Since the injection angle depends on $k$, a given trajectory contributes to (39) only over a limited energy range. This is why in geometries with stable trajectories, like the circle, the Fourier peaks are more pronounced than for the stadium billiard. Numerical calculations for circular billiards show that $\langle(\delta T)^2\rangle$ increases with $k$ (Ishio, 1995-a), consistently with the behavior found for another integrable case (the square cavity of 2.7.2). Diffraction corrections were shown to be important in the small-$N$ case (Březinova, 2010-a).

## Mixed dynamics

Cavities with hyperbolic and regular classical dynamics are the most commonly studied cases of ballistic transport. However, the behavior with a mixed phase-space, containing both chaotic and regular regions, is the most generic situation for a dynamical system. It is also experimentally relevant since the microstructures do not have perfect hard-wall confining potentials and are not disorder free.

Ketzmerick considered the problem of a dynamical system with mixed phase-space (Ketzmerick, 1996-a), where the trapping generated by the infinite hierarchy of cantori leads to a power-law for the escape rate of the cavity $P(\tau) \propto \tau^{-\Upsilon}$, with the exponent $\Upsilon > 1$. From a semiclassical diagonal approximation to the conductance, Ketzmerick proposed that the graph of $g$ versus $\varepsilon$ has (in the case $\Upsilon < 2$) the statistical properties of fractional Brownian motion with fractal dimension $d = 2 - \Upsilon/2$.

Numerical simulations by Huckestein and collaborators (Huckestein, 2000-a) in cavities with mixed dynamics connected to leads yielded the power-law distribution of the classical escape rate, but the quantum curve $g(\varepsilon)$ failed to exhibit fractal behavior. The effect of isolated resonances (Hufnagel, 2001) has been invoked to explain these numerical results, adding subtle issues to the description of quantum transport through cavities with generic mixed classical dynamics (Takagaki, 2000-a; Louis, 2000-a).

# Random-matrix theory for ballistic cavities

**Refs. (Brouwer, 1997-t; Beenakker, 1997-r; Alhassid, 2000-r; Mello, 1999-r; Mello, 2004-b)**

## Random-matrix ensembles

**Refs. (Bohigas, 1989-r; Guhr, 1998-r; Haake, 2001-b; Fyodorov, 2011-i)**

Random-matrix theory (RMT) has been applied to study the statistical properties in a variety of physical problems, ranging from Nuclear Physics to spectral distribution in small quantum systems and conductance fluctuations in disordered mesoscopic conductors. The basic assumption of these approaches is that the matrix describing the problem at hand is the most random one among those verifying the required symmetries and constraints of the system under study. The Hamiltonian matrix is the relevant one for analyzing spectral statistics of complex systems (disordered or classically chaotic). The mean-level spacing appears as the sole constraint in this case, and different ensembles are obtained according to the symmetries that may exist in addition to the Hermitian character of the Hamiltonian. The transfer matrix, incorporating the constraint of the elastic mean-free-path, is the appropriate tool to study the conductance fluctuations in quasi-one dimensional disordered systems.

Systems where all scattering processes are equally probable can be characterized by the Dyson's circular scattering ensembles of unitary matrices $S$ (that in the case of quantum transport take the form (9)). There exist three main symmetry classes of scattering matrices according to the possible additional symmetries beyond the condition $SS^\dagger = I$, which are usually characterized by the value of a parameter $\beta$:

- $\beta = 1$ - *Circular Orthogonal Ensemble* (COE) - absence of magnetic field and spin-orbit scattering ($S^\mathrm{T} = S$).
- $\beta = 2$ - *Circular Unitary Ensemble* (CUE) - non-zero magnetic field.
- $\beta = 4$ - *Circular Symplectic Ensemble* (CSE) - spin-orbit scattering and no magnetic field ($S$ is a self-dual quaternion matrix).

In the same way that the Bohigas-Giannoni-Schmit conjecture (Bohigas, 1984-a) identifies the statistical properties of the spectrum of Hermitian matrices with those of classically chaotic systems, Bümel and Smilansky proposed that chaotic scattering is represented by COE and CUE scattering matrices, and furthermore, they derived the statistical properties of the eigenphase distribution from a semiclassical analysis (Blümel, 1990-a).

## Invariant measures in the circular ensembles

The basic assumption of equal probability of all possible scattering process translates into a uniform distribution over the matrix ensemble. The probability $\mathcal{P}_\beta(S_0, dS)$ of obtaining a matrix $S$ in a neighborhood $dS$ of some given $S_0$ is independent on $S_0$,

$$\mathcal{P}_\beta(S_0, dS) = \frac{1}{V_\beta} \, \mu_\beta(dS) \, . \tag{65}$$

- $\mu_\beta(dS)$ is the $\beta$-dependent measure in the neighborhood $dS$ of $S_0$ obtained from the condition of invariance under unitarity transformations that preserve the symmetries of $S_0$.
- $V_\beta = \int \mu_\beta(dS)$ is the total volume of the matrix space.

In Dyson's original approach, $\mu_\beta(dS)$ was expressed in eigenvalue-eigenvector coordinates (Dyson, 1962-a). This is a suitable representation to obtain the distribution of the eigenphases, but it is not appropriate for the study of transport through the quantum dot.

A transport property $A$ can generally be expressed as a *linear statistic*, that is, a sum $A = \sum_{n=1}^{N} a(\lambda_n)$ over the $\lambda_n$ parameters of the polar decomposition (introduced in Sec. 1.3.7). For instance, according to Eqs. (16) and (21), the dimensionless conductance is characterized by the function $a(\lambda) = (1 + \lambda)^{-1}$.

The $\lambda$-parameters are not simply related to the eigenvalues of the scattering matrix. Expressed in the coordinates of the polar decomposition (20), the invariant measure of the COE ensemble can be written as (Jalabert, 1995-a)

$$\mu_1(dS) = \prod_{n=1}^{N} \frac{1}{(1+\lambda_n)^{3/2}} \prod_{nm} \left| \frac{1}{1+\lambda_n} - \frac{1}{1+\lambda_m} \right| \prod_{m=1}^{N} d\lambda_n \prod_{l=1}^{2} \mu(du_l) . \tag{66}$$

- $\mu(du_l)$ is the invariant (Haar) measure of the unitary matrices $u_l$.

## Joint-distribution and density of the $\lambda$-parameters

The measure (66), together with its generalization to the other circular ensembles (Jalabert, 1995-a; Frahm, 1995-a) allow to write the join distribution of the $\lambda$-parameters as a Gibbs distribution,

$$\mathcal{P}(\{\lambda_n\}) = \frac{1}{Z} \exp[-\beta \mathcal{H}(\{\lambda_n\})] . \tag{67}$$

- $Z$ is a normalization constant.
- $\beta \in \{1, 2, 4\}$ determines the ensemble and plays the role of an inverse temperature.
- $\mathcal{H}(\{\lambda_n\}) = -\sum_{i<j} \ln|\lambda_i - \lambda_j| + \sum_i V_\beta(\lambda_i)$ is an effective Hamiltonian characterized by a logarithmic pairwise interaction and a one-body confining potential.
- $V_\beta(\lambda) = \left(N + \frac{2-\beta}{2\beta}\right) \ln(1+\lambda)$.

The confining potential is $\beta$-independent to order $N$, but not to order $N^0$, leading to a density $\rho(\lambda) = \rho_N(\lambda) + \delta\rho(\lambda)$. The first contribution, of order $N$ yields the "Boltzmann conductance", while the $\beta$-dependent correction $\delta\rho$, of order $N^0$, is responsible for the weak-localization effect. Working Eq. (67) order by order in $N$ in the limit $N \gg 1$ results in

$$\rho_N(\lambda) = \frac{N}{\pi(1+\lambda)\sqrt{\lambda}} \quad \delta\rho(\lambda) = \left(\frac{\beta-2}{4\beta}\right) \delta_+(\lambda) . \tag{68}$$

- $\delta_+$ is the one-sided delta-function satisfying $\int_0^\infty d\lambda \, \delta_+(\lambda) = 1$.

The transmission eigenvalue density $\rho_N(T) = \rho_N(\lambda) |d\lambda/dT|$ has a *bimodal* distribution with peaks near unit and near zero transmission.

## Average values, weak-localization, and conductance fluctuations for $N \gg 1$

Taking the transmission as the linear statistics leads, in the case $N \gg 1$, to (Baranger, 1994-a; Jalabert, 1994-a)

$$\langle T \rangle = \frac{1}{2} N + \langle \delta T \rangle . \tag{69}$$

- $\langle \delta T \rangle = \frac{\beta - 2}{4\beta}$.
- $\langle (\delta T)^2 \rangle = \frac{1}{8\beta}$.

In the presence of a magnetic field $\beta = 2$, and then $\langle T \rangle = \langle R \rangle = N/2$. This equality between transmission and reflection coefficients is the quantum analog of what is expected from the "ergodic" exploration of the dot boundaries by the classical trajectories, and it is broken by quantum interference once the magnetic field is eliminated. Taking the difference between the values of $\langle \delta T \rangle$ corresponding to $\beta = 1$ and $\beta = 2$ results in the *universal ballistic the weak-localization correction* of $-1/4$. Analogously, the difference between the values of $\langle \delta T \rangle$ corresponding to $\beta = 4$ and $\beta = 2$ results in an anti-localization correction of $1/8$. Thus, according to RMT, the *magnitude of the* *conductance fluctuations is universal*, and only depending of the parameter $\beta$. A reduction factor of 2 is obtained for the variance of the conductance when a magnetic field is applied inducing the transition form the COE to the CUE.

Figure 14 shows the weak-localization and the conductance fluctuations obtained from quantum numerical calculations for an asymmetric structure (the "stomach") where direct paths and whispering gallery trajectories have been blocked. After performing extensive averaging over the Fermi energy, the magnetic field, and the position of the stopper, the universal values predicted by random-matrix theory are obtained for $N \geq 4$.

The distribution (68) has been extended to other situations. For instance, the joint probability density of reflection eigenvalues has been determined for the case of chaotic cavities with non ideal leads (with ballistic or tunnel contacts and having an arbitrary number of propagating modes in the leads) (Jarosz, 2015-a).

## Conductance distributions for small $N$

The case $N = 1$ corresponds to a quantum dot which is coupled to the reservoirs by two quantum point contacts with a quantized conductance $G_0 = 2e^2/h$. The probability distribution (68) reduces in this case to $P(\lambda) = (1/2)\beta(1+\lambda)^{-1-\beta/2}$, yielding a transmission distribution (Baranger, 1994-a, Jalabert, 1994-a)

$$w(T) = \frac{1}{2}\beta T^{-1+\beta/2}, \quad 0 \leq T \leq 1. \tag{70}$$

This is a remarkable result. In the presence of magnetic field ($\beta = 2$), any value of the transmission between 0 and 1 is equally likely. In non-zero field it is more probable to find a small than a large transmission, provided that the boundary scattering preserves spin-rotation symmetry ($\beta = 1$). In the presence of spin-orbit scattering at the boundary ($\beta = 4$), however, a large conductance is more probable than a small one.

Figure 15 shows the numerical quantum simulations of the transmission distribution for the structure shown in the inset of Figure 14 for small values of $N$. For $N = 1$, a very good agreement with Eq. (70) is obtained in both cases (without and with magnetic field). This highly non-Gaussian behavior disappears when increasing $N$, and for $N = 3$ the transmission distribution is well approximated by a Gaussian in the cases without and with magnetic field.

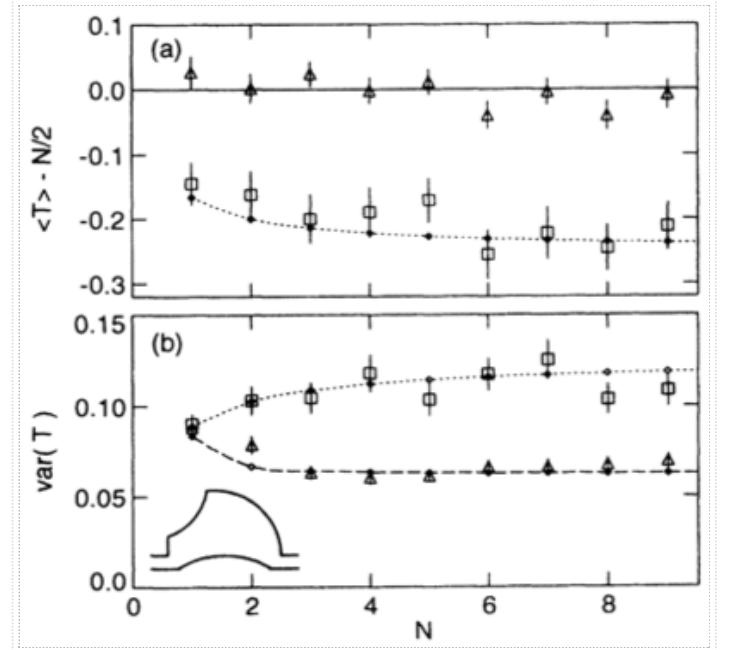

Figure 14: The magnitude of the (a) weak-localization correction and (b) conductance fluctuations as a function of the number of modes in the leads $N$. The numerical results for $B = 0$ (squares) agree with the prediction of the COE (dotted line), while those for $B \neq 0$ (triangles) agree with the CUE (dashed line). The inset shows a typical cavity used for the calculations, where the effect of direct and short trajectories is eliminated. The numerical results involve averaging over energy (50 points at fixed $N$), 6 different cavities (obtained by changing the placement of the stoppers), and 2 magnetic fields ($BA/\phi_0 = 2, 4$ in the case with $B \neq 0$). $A$ ls the area of the cavity. (From Ref. (Baranger, 1994-a), copyright 1994, American Physical Society.)

## Shot-noise

The shot-noise power $P$, described by Eq. (19), is associated with the linear statistics $a(\lambda) = P_0\lambda(1+\lambda)^{-2}$, leading to (Jalabert, 1994-a)

$$\langle P \rangle = \frac{1}{8}N P_0 = \frac{1}{4} P_{\text{Poisson}}. \tag{71}$$

The 1/4 reduction factor with respect to the uncorrelated case in a chaotic dot is to be compared with the 1/3 reduction of shot noise in a diffusive conductor (Beenakker, 1992-a). Since $\langle P \rangle$ is $\beta$-independent, there is *no* weak-localization correction in the shot noise of a chaotic dot, in contrast to the case of a diffusive conductor (De Jong, 1992-a).

# Universality in random-matrix theory and semiclassics

By imposing $S^\dagger = S$, the random-matrix theory of circular ensembles is free from the problems with the unitarity condition that the diagonal semiclassical approximation faces in the description of quantum transport. Random-matrix theory yields universal results, equivalent to those of the loop-corrected semiclassical approximation, but in a considerable simpler fashion. It is important to remark that the universal values for the weak-localization correction and the conductance fluctuations obtained within RMT, or the loop-corrected SCA, rely on the ergodicity of the underlying classical dynamics, and therefore apply to a very restricted set of structures.

Obviously, random-matrix theory is not of any help when dealing with cavities with integrable or mixed dynamics, and semiclassics remains the preferred tool in these cases. Even when the underlying dynamics of the dot is chaotic, the observation of universal behavior is only possible if the role of short trajectories is negligable ($\gamma \tau_E \ll 1$) ($\gamma$ is the escape rate and $\tau_E$ the Ehrenfest time given by Eq. (62)) and the direct trajectories are blocked. Therefore, the numerical simulations of Figure 14 and Figure 15 were performed in cavities where the geometry were chosen to minimize the effect of direct and short trajectories.

The effect of the short-time dynamics (*direct process*) can be incorporated in an information-theory approach through the Poisson's kernel (Doron, 1992-a; Baranger, 1996p-a). However, the simplicity and the usefulness of the random-matrix approach is diminished in this case.

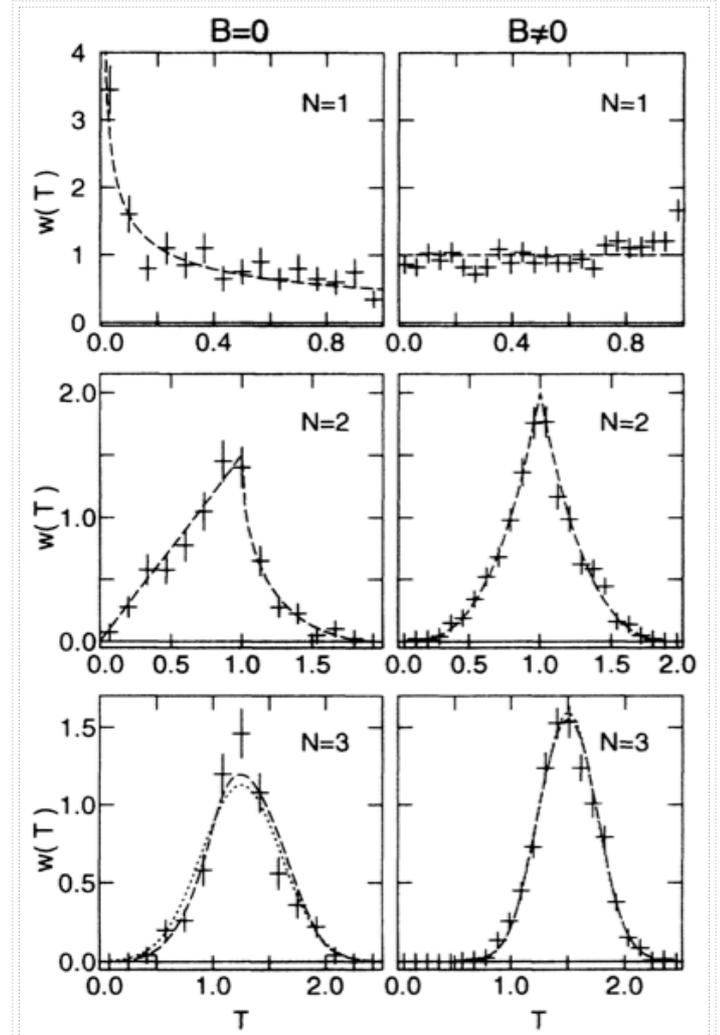

Figure 15: The transmission distribution at fixed number of modes $N = 1, 2, 3$ in both, the absence (first column) and presence (second column) of a magnetic field for the geometry of Figure 14. The numerical results (plusses with statistical error bars) are in good agreement with the predictions of the circular ensembles (dashed lines). For $N = 3$ the distribution approaches a Gaussian (dotted lines). (From Ref. (Baranger, 1994-a), copyright 1994, American Physical Society.)

The connection (28) between the Hamiltonian and scattering matrices allows to establish the relationships between the statistical distribution of both matrices. The random-matrix hypothesis for the Hamiltonian of a chaotic dot (or the zero-dimensional non-linear sigma model) coupled to leads yields equivalent results to those of Eq. (69) (Iida, 1990-a; Lewenkopf, 1991-a), and allows to calculate the crossover $\beta = 1$ to $\beta = 2$ as a function of magnetic field (Pluhǎr, 1994-a).

## Spatial symmetries and effective channels

Geometrical symmetries of the ballistic cavities translate into a block structure of $S$ (Baranger, 1996-a). The weak-localization correction and the conductance fluctuations are different from those of Sec. 3.4, as they depend on the additional symmetries of $S$. The SCA for the conductance of nearly symmetric quantum dots yields results equivalent to those of the RMT and allows to study the transition between the different symmetry classes (Whitney, 2009-a).

Random-matrix theory can be helpful for the study of decoherence process. A phenomenological way of introducing decoherence is by attaching a virtual lead that draws no current but provides a channel for phase breaking (Büttiker, 1986p-a). It amounts to consider that the term $\hat{H}_{s-env}$ of the Hamiltonian (2) simply represents the coupling to a lead that does not take a net current. This additional lead can be incorporated in a random-matrix theory approach (Baranger, 1995-a; Brouwer, 1995-a; Brouwer, 1997-a). Confronting the experimental results with the random-matrix theory predictions allows to determine the number of effective channels in the virtual lead, providing a useful estimation of $L_\Phi$.

# Experiments in open ballistic quantum dots

**Refs. (Westervelt, 1999-r; Bird, 1999-r; Bird, 2003-b)**

## Conductance fluctuations in ballistic microstructures

**Refs. (Marcus, 1993-r; Marcus, 1997-r; Bird, 1997-r; Huibers, 1999-t; Hackens, 2005-t)**

### Early experiments on conductance fluctuations

The statistical analysis of the low-temperature magnetoconductance of ballistic QDs defined in *GaAs/AlGaAs* heterostructures was first performed by Marcus and collaborators (Marcus, 1992-a). Two geometrical shapes (stadium and circle) were lithographically patterned (each in two different samples) in order to achieve a steep-walled electrostatic confinement. The leads were oriented at right angles of each other aiming to reduce transmission via direct trajectories (see insets in Figure 16). The transport mean-free-path was estimated to be 2.6 $\mu$m, a few times larger than the size of the structures ($a \simeq 0.6\ \mu$m), indicating that the ballistic regime was atteined. The number of conducting channels in the contacts was between $N=1$ and $N=3$, and the change could be achieved without substantially affecting the size of the device itself. These low numbers of modes place the experiment somehow at the limit of applicability of the semiclassical theory.

The magnetoconductance was reported to be reproducible under thermal cycling, demonstrating that the fluctuations were a fingerprint of these samples. The traces corresponding to the stadium and the circle (Figure 16) presented some likeness, but differed in the detail of the fluctuations. Such a difference could be quantified by following the analysis of the power-spectrum of Sec. 2.4. The power-spectrum of the stadium cavities showed a good agreement with Eq. (56) over three orders of magnitude in power. Deviations for large and small areas were observed (see Figure 17 and the inset in Figure 21.b). The conductance fluctuations in the circular billiard appeared to be more structured (more weight in the high harmonics of the power spectrum) when compared with the case of the stadium. The measurable difference in the transport through the two structures then appeared in the larger weight of the high harmonics of the magneto conductance for the case of the circle.

The magnetic field scale of the fluctuations was found to be consistent with the semiclassical prediction and it was increasing with the mean conductance through the dot in samples having up to $N=5$ incoming channels (Marcus, 1994-a). Such a behavior was in line with the expectation that a larger mean conductance is related with wider openings, and thus with larger escape rates and $\alpha$-parameters.

# Ballistic conductance fluctuations: different shapes and different control parameters

The systematic study of conductance fluctuations requires a considerable amount of averaging. A given magnetoconductance curve offers only a limited interval for averaging, since once the cyclotron radius becomes comparable to the size of the structure, the nature of the classical dynamics may change. In order to cope with this problem, alternative types of averages have been developed by tuning the Fermi energy (Keller, 1994-a, Keller, 1996-a; Zozoulenko, 1997-a), and/or thermal cycling the sample (Berry, 1994-a; Berry, 1994p-a), as well as small distortions in the shape of the cavity (Chan, 1995-a).

The fact that thermal cycling could effectively produce different samples, by re-accommodation of impurities, hinted the importance of the smooth disorder in these microstructures (that were ballistic, but not clean). The small amount of small-angle scattering affected the very long trajectories, without altering the statistical signature of chaotic trajectories (Berry, 1994-a). The power-spectrum of the conductance fluctuations in a circular cavity with a central bar ("pacman billiard") was found to be well fitted by the theoretical prediction (56), valid for a classically chaotic structure (Berry, 1994p-a). In the unpatterned circular cavity, the previous fit was considerably poorer, and the characteristic field was found to be smaller (by a factor of 3) than in the case of the packman billiard. Such difference follows from the fact that a barrier inside the cavity drives the spectrum of effective areas towards smaller values (Schreier, 1998-a).

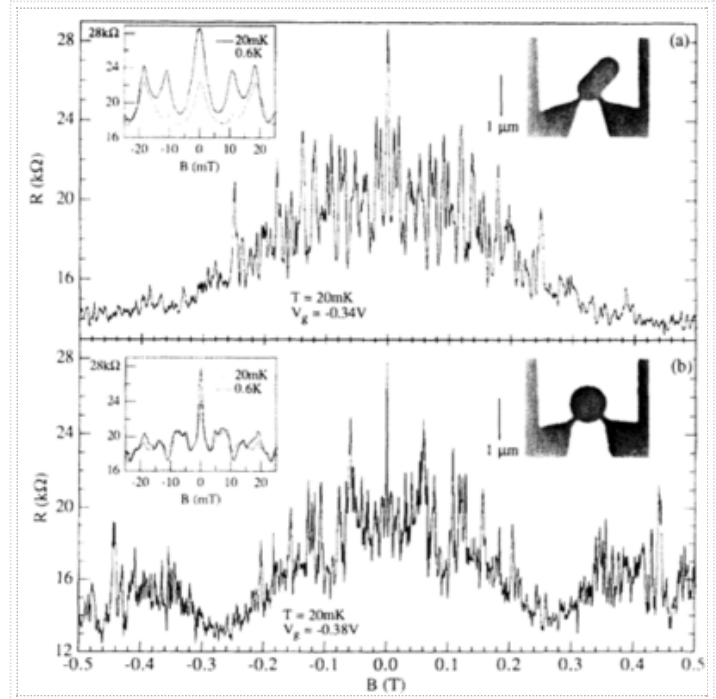

Figure 16: Resistance as a function of perpendicular magnetic field B for (a) stadium (b) circle, both with $N = 1$ fully transmitted modes in leads. Insets: Zero-field peaks at 20 mK (solid) and 0.6 K (dashed), and electron micrographs of devices. (From Ref. (Marcus, 1992-a), copyright 1992, American Physical Society.)

The exponential decay of the power spectrum found in circular cavities was signed by steps at characteristic frequencies corresponding to integral fractions of flux quanta through the dot (Persson, 1995-a).

Keller *et al.* (Keller, 1994-a; Keller, 1996-a) fabricated microstructures where the electron density (and hence $k_F$) was tunable while maintaining the geometry approximately fixed. Different shapes were considered: a stadium where the leads were not aligned, an asymmetric half-stadium with a stopper of direct trajectories ("stomach billiard"), and a polygonal shape with a stopper (see Figure 18). The $k$ and $B$ dependence of the conductance (Figure 19) allowed to show that the correlation length $B_c = \alpha \Phi_0$ of the $B$-dependent conductance fluctuations for the chaotic cavities is approximately $k$-independent (inset), in agreement with Eq. (56). However, the conductance fluctuations of the polygonal geometry did not show qualitative differences with those of the chaotic case. This departure from the theoretical prediction for clean integrable cavities was attributed to the effect of residual disorder. The signature of short paths was established by analyzing the peaks in the power spectrum of the $k$-dependent conductance (Keller, 1994-a).

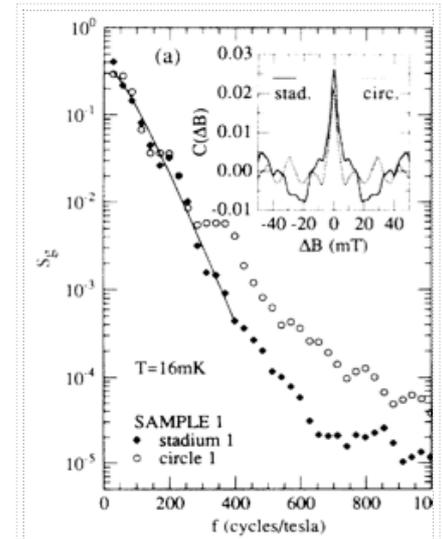

Figure 17: Averaged power spectra of conductance fluctuations for stadium (solid diamonds) and circle (open circles) with $N = 3$ transverse modes in leads. Solid curves are fits of semiclassical theory, Eq. (56), to the stadium data. Insets: Autocorrelation of stadium (solid) and circle (dashed) for 0.01 T $< B <$ 0.29 T. (From Ref. (Marcus, 1992-a), copyright 1992, American Physical Society.)

## Conductance fluctuations and inelastic scattering

In the completely coherent picture of Sec. 2 the parameter $\alpha_{cl}$ governing the area distribution in a chaotic cavity is given by the geometry and the escape rate $\gamma$. At the phenomenological level, phase breaking can be modeled as an extra lead of the dot, which draws no net current but hinders electrons for participating in interference effects (Büttiker86p-a) (see Sec. 3.8). Within this picture, the measured $\alpha$-parameter depends on both, the classical escape time $\gamma^{-1}$ and the phase-breaking time $\tau_\Phi = L_\Phi/v$. Therefore, measuring the conductance fluctuations of a given sample at different openings allows to extract $\tau_\Phi$ (Marcus, 1993-a).

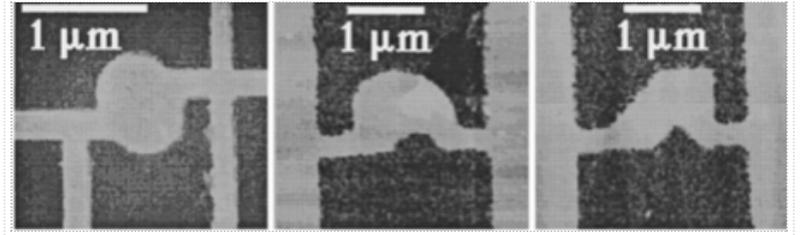

Figure 18: Cavity images of the three shapes (sadium, stomach, and polygon billiards) studied in Ref. (Keller, 1996-a). The light areas are metal and the dark areas are the GaAs surface. (From , copyright 1996, American Physical Society.)

The temperature-dependence of $\tau_\Phi$ in *GaAs* microstructures (Clarke, 1995-a), and its saturation below 100 mK (Huibers, 1999-a), shed light into the influence of electron-electron interactions for the dephasing process occurring in a quantum dot. Similar conclusions were extracted from measurements in *InGaAs* microstructures, while the quantitative differences (larger values of $\tau_\Phi$) were explained by the different material properties (Hackens, 2002-a).

The role of the effective channels describing incoherent process was further put in evidence by the failure of obtaining the RMT prediction (70) for the non-Gaussian conductance distribution in the case of only one ($N = 1$) open channel in the leads (Huibers, 1998-a). Only when a virtual lead with a temperature-dependent number of effective channels was taken into account, a good fitting to the random-matrix theory results was achieved, resulting in the Gaussian distribution relevant for the case $N \geq 3$.

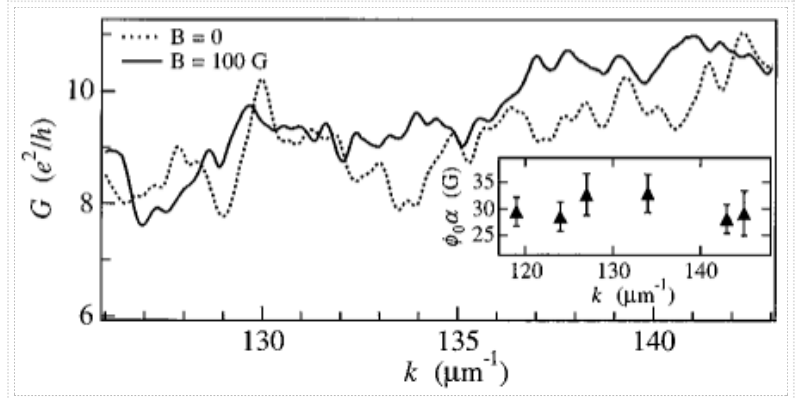

Figure 19: Conductance fluctuations $G(k)$ for the stomach billiard for magnetic fields $B = 0$ and $B = 100$ G ($\approx 3.5\ \alpha\Phi_0$). Inset: $\alpha\Phi_0$ versus $k$ showing that the change of $k$ does not affect the typical area enclosed. (From Ref. (Keller, 1996-a), copyright 1996, American Physical Society.)

The determination of $L_\Phi$ through measurements of the conductance fluctuations provides an example where the theoretical ideas of Quantum Chaos are useful to probe fundamental properties of condensed matter systems.

## Weak-localization in ballistic microstructures

**Refs. (Keller, 1995-t; Bird, 1997-r; Chang, 1997-r)**

### Early experiments on weak-localization

The magnetoresistance traces of Figure 16 (Marcus, 1992-a) exhibit prominent peaks around $B = 0$. Separating these peaks from the background of conductance fluctuations requires some kind of average. The energy-averaged magnetoconductance traces performed by Keller and collaborators (Keller, 1996-a) yielded a conductance minimum at $B = 0$, i.e. the so called weak-localization peak. At a fixed value of $k$, the conductance of a given sample presented, as a function of $B$, an extreme at $B=0$, but not necessarily a minimum (i.e. in Figure 19 the $k$-dependent conductance at finite field is not always above that of $B=0$). This observation is consistent with the lack of self-averaging of ballistic cavities. The weak-localization of the nominally chaotic billiards showed a good quantitative agreement with the numerical results of Sec. 2.5.2. However, the polygon billiard failed to exhibit the weak-localization features expected for the integrable classical dynamics.

The use of sub-micron stadium-shaped quantum dots (with up to $N = 7$ modes) cycled at room temperature allowed Berry and collaborators (Berry, 1994-a) to obtain average values and separate the weak-localization peak from the conductance fluctuations, making it possible the comparison against the theoretical results obtained for chaotic cavities in the semiclassical limit. The line-shape of the peak was found to be Lorentzian, in agreement with the semiclassical prediction (58). Moreover, the field scales of the weak-localization and conductance fluctuations were found to be related by the factor of 2 that discussed in Sec. 2.5.

# Weak-localization in arrays of microstructures

Chang and collaborators (Chang, 1994-a) fabricated arrays of microstructures on a high-quality *GaAs/AlGaAs* buffer. Three different shapes were considered: stadium, circle, and rectangle. In each case, 48 cavities, nominally identical but actually slightly different due to uncontrollable shape distortions and residual disorder, were connected as 6 rows in a series of 8 in parallel. Thus, the conductance fluctuations were suppressed by the ensemble average (at $T = 50$ mK). The resulting weak-localization peak was found to be Lorentzian for the stadium cavities and triangular for the circular ones (see Figure 20), in agreement with the semiclassical prediction and detailed numerical calculations (Baranger, 1999-r). Rectangular cavities, however, failed to yield a cusp of the magnetoresistance at $B = 0$ expected for this integrable geometry.

# Weak-localization and conductance fluctuations resulting from ensemble averages

Microstructures admitting small shape distortions (less than 5 % in the area) by tuning the voltage of lateral gates (inset of Figure 21.a) were developed by Chan, Marcus, and collaborators (Chan, 1995-a), allowing to study weak-localization and conductance fluctuations in the same sample. The lithographic shape of the cavity did not correspond to a chaotic geometry. But the possible integrability of the dynamics was expected to be broken by the shape distortions employed for averaging. Moreover, in these relatively large structures, smooth disorder affected the long trajectories. The number $N$ of open channels was not in the semiclassical regime, as it was tuned to a value of 2. Conductance was studied as a function of magnetic field and electrostatic shape distortion, allowing to gather very good statistics. The fluctuations as a function of magnetic field showed very good agreement with Eq. (56) (inset of Figure 21.b), despite the above mentioned factors that could limit the applicability of the semiclassical theory of transport through a chaotic cavity.

The shape-distortion fluctuations yielded an exponential power spectrum, in agreement with the calculations of Bruus and Stone (Bruus, 1994-a) showing that the semiclassical formalism of Sec. 2 could be extended to this case. A Lorentzian shape for the weak-localization peak was obtained, with a width related to the characteristic field of the conductance fluctuations, as predicted by semiclassical theory. The magnitude of the shape-dependent conductance fluctuations at non-zero field had a factor of 2 reduction with respect to the zero-field value, and the line shape of $\langle (\delta T)^2 \rangle$ was found to be a squared Lorentzian, in agreement with theory (Efetov, 1995-a). The rich statistics that this type of structures allowed to gather was used to extract the moments of the conductance, as well as the whole conductance distribution in order to compare with the random-matrix theory predictions of Sec. 3.

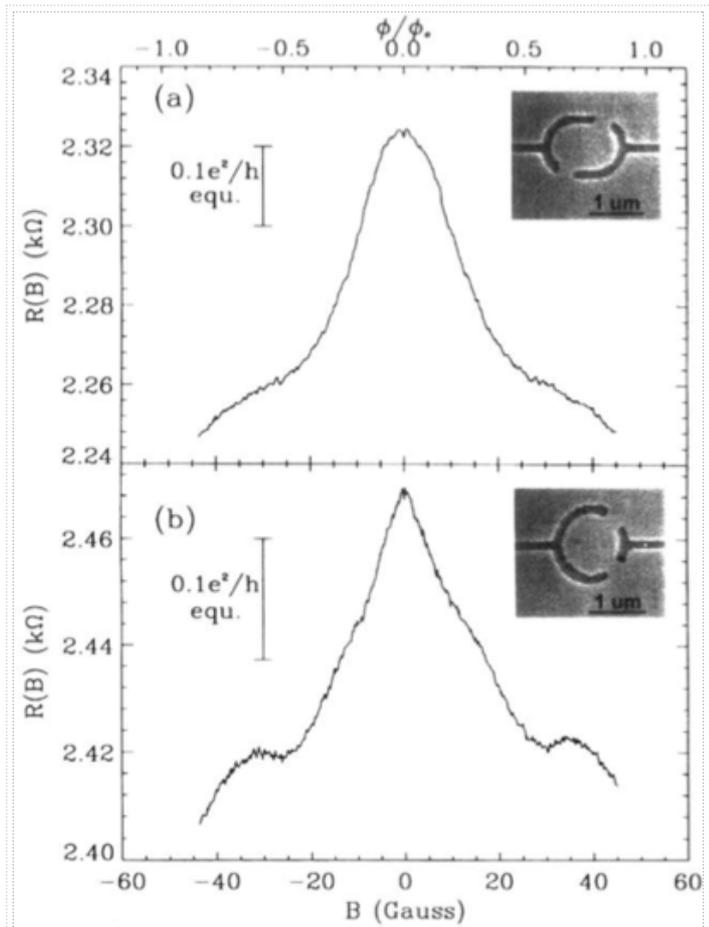

Figure 20: The magnetoresistance for (a) 48 stadium cavities and (b) 48 circle cavities, normalized to a single cavity. The weak-localization peak line shape shows a Lorentzian behavior for the chaotic, stadium cavities. In contrast, the line shape for the non-chaotic, circle cavities exhibits a triangular shape (linearly decreasing). The vertical bars indicate the corresponding change in conductance, showing that the magnitude of the measured weak-localization effect is considerably smaller than the universal values predicted by Eq. (69). Insets: electron micrographs of the cavities. (From Ref. (Chang, 1994-a), copyright 1994, American Physical Society.)

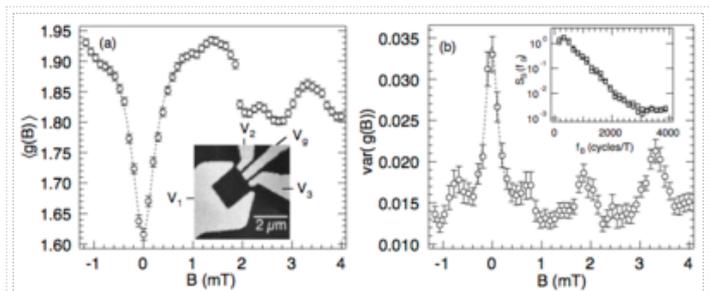

Figure 21: (a) Shape-averaged conductance showing a weak-localization peak fitted by a Lorentzian (dashed). Inset: electron micrograph of device (the gate voltage $V_g$ is used to produce shape distortions. (b) Variance of shape-distortion conductance fluctuations (in units of $(e^2/h)^2$) Inset: power spectral density in magnetic field fitted by Eq. (56). (Adapted from Ref. (Chan, 1995-a), copyright 1995, American Physical Society.)

# Weak-localization in different geometries

Bird and collaborators (Bird, 1995-a) used the thermal smearing of the conductance fluctuations to measure the weak-localization correction in rectangular cavities. The peak line-shape changed its profile from Lorentzian to triangular as the quantum point contacts at the entrance of the cavity were closed. The transition occurred for $N \simeq 2$, demonstrating the non-trivial role played by the contacts when the cavity is fed from leads with s small number of modes. Resistance measurements and numerical analysis (Zozoulenko, 1997-a; Zozoulenko, 1998-a) on square cavities suggested that, depending on the geometry of the contacts, transport through the cavity could be mediated by just a few resonant levels or specific families of trajectories, illustrating the importance of the geometry and the injection conditions in the integrable case (Ouchterlony, 1999-a). Different weak-localization profiles have been obtained in quantum numerical calculations within a fixed geometry by varying the election energy and the softness of the confining potential (Akis, 1999-a).

Lee, Faini and Mailly used shape and energy averages to extract the weak-localization peak of chaotic (stadium and stomach) and integrable (circular and rectangular) cavities (Lee, 1997-a). The former exhibited a Lorentzian line-shape, consistently with the theoretical prediction (58). However, among the integrable cavities, only the rectangle showed the expected triangular shape, while the circle yielded a Lorentzian. Chang has proposed (Chang, 1997-r) that the discrepancy between the results of Refs. (Chang, 1994-a) and (Lee, 1997-a) was due to the shorter physical cut-offs that were present in the latter experiment, hindering the long trajectories to exhibit the signatures of the integrable dynamics.

Square-shaped ballistic cavities filled with antidot arrays resulted in a cusp-like weak-localization peak, while the empty cavities of equal geometry showed a Lorentzian peak (Lütjering, 1996-a). These experimental results, that seemed to contradict the theoretical predictions for classically chaotic and integrable systems, were explained by invoking a mixed-dynamics in the case of the filled cavity and by the imperfections (boundary roughness and small-angle scattering) for the nominally regular system.

Experiments and numerical calculations have primarily focused on the differentiation between Lorentzian and linear line-shapes of the weak-localization peak according to the underlying classical dynamics. In this context it is important to recall that a complete semiclassical theory of weak-localization only exits for the case of a chaotic dynamics.

# Fractal conductance fluctuations

**Refs. (Micolich, 2000-t; Micolich, 2013-r; Pilgrim, 2014-t)**

## Soft-wall effects

A stadium and a Sinai billiard, which are paradigms of chaotic dynamics, become mixed systems when fabricated by lithographic methods that result in a soft-wall confinement. These two geometries were patterned with purposely soft confining potentials and very wide leads (0.7 $\mu$m) allowing most of the trajectories to rapidly exit the structures (Taylor, 1997-a; Sachrajda, 1998-a). The resulting conductance fluctuations were claimed to have a fractal nature over two orders of magnitude in magnetic field (Micolich, 1998-a; Sachrajda, 1998-a).

In square cavities, the soft confinement was found not to be effective in the smearing of the conductance fluctuations in comparison with the hard-wall scenario (Ouchterlony, 1999-a).

## Small-angle scattering effects

Marlow and collaborators (Marlow, 2006-a) performed a comprehensive comparison of magnetoconductance fluctuations in 30 devices spanning the ballistic, quasi-ballistic, and diffusive regimes, concluding that all of them exhibit identical fractal behavior. Billiards made on *GaAs/AlGaAs* heterojunctions, associated with relatively "soft" confinement potentials were compared with others made on *GaInAs/InP* heterostructures exhibiting "hard" confinement (according to the simulations reproduced in Figure 22). The similar behavior, in terms of conductance fluctuations, of billiards with both kinds of confinement contradicts the claim (Sachrajda, 1998) that the fractal conductance fluctuations are associated with the mixed dynamics resulting from a soft confinement .

This comparative study concluded that the origin of the fractal conductance fluctuations found in all devices is the small-angle scattering that deflects electron trajectories away from straight paths (Marlow, 2006-a). Such a picture was associated with the interpretation of scanning gate microscopy (SGM) studies in 2DEG (Topinka, 2001-a) and quantum dots (Crook, 2003-a) in terms of the drifting of electron trajectories due to small-angle scattering.

Concerning the link claimed with SGM studies, it is important to remark that the interpretation of SGM measurements as traces of the classical electron paths has been questioned for nanostructures surrounded by a 2DEG (Jalabert, 2010-a, Gorini, 2013-a) and QDs (Kozikov, 2013b-a). Moreover, the scale over which the bending of the SGM traces is appreciable corresponds to several $\mu$m in high-mobility samples.

## Disorder effects

In order to further study the effects of disorder on conductance fluctuations, a comparison was established between two nominally identical geometries, where one of them was patterned on a standard modulation-doped heterojunction, and the other on an undoped (gated) heterostructure (See, 2012-a). Consistently with previous findings (Berry, 1994-a), the magnetoconductance fingerprints of the modulation-doped sample changed under thermal cycling. While in the undoped case the magnetoconductance was reproducible. This different behavior demonstrated that the role of disorder was negligible in the undoped case. However, the statistical analysis of the magnetoconductance presented similar features in both cases, with quantitative differences that could be attributed to the different electron density achieved with and without doping.

It was then concluded the the presence of small-angle scattering in a modulation-doped structure is not capable of amplifying the fractal behavior of the magneto conductance fluctuations induced by a soft-confinement.

## Fractal analysis versus power-spectrum analysis

While various experiments have observed a fractal nature in the conductance fluctuations of quantum dots, alternative interpretations have been put forward invoking the effect of soft-wall confinement and/or small-angle scattering. It is important to remark the difficulty in evaluating the fractal character of $g(B)$ (which is like a "time series") , given the restricted $B$-interval usually available and the limited spectral content of the curve (Takagaki, 2000-a; Louis, 2000-a; Hufnagel, 2001-a) . The power-law $C_B(\eta) \propto \eta^{-(\Upsilon+1)}$ expected for the fractal conductance fluctuations (the exponent $\Upsilon$ is introduced in Sec. 2.7.4) might be difficult to distinguish from the semiclassical prediction (56) if only a reduced $\eta$-range is available.

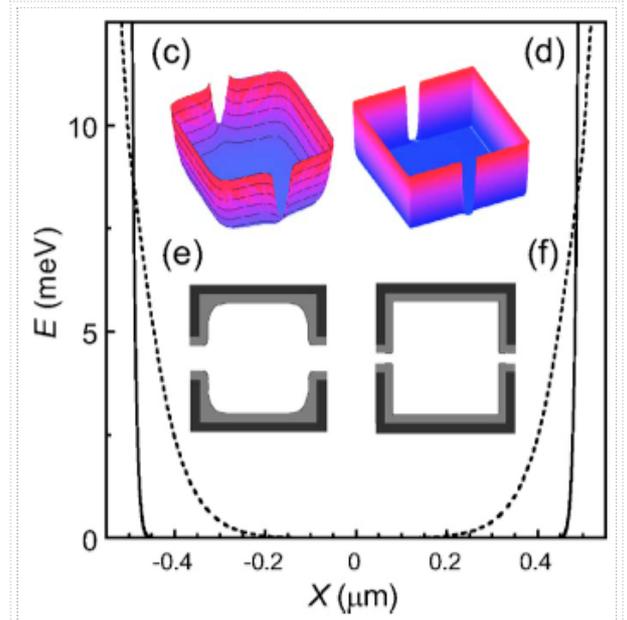

Figure 22: Cross section of the calculated potential energy $E$ versus the spatial location $X$ across the billiard's central region created by square-shaped top gates. The dashed and solid lines are for structures made in *GaAs/AlGaAs* and *GaInAs/InP* heterostructures, respectively. The potential energy profile in the plane of the 2DEG and the top views of the resulting geometries are, respectively, shown in (c) and (e) for *GaAs/AlGaAs* and in (d) and (f) for *GaInAs/InP* . Gray indicates the depletion region in the 2DEG. (Adapted from Ref. (Marlow, 2006-a), copyright 2006, American Physical Society.)

The difficulties arising from the limited set of data, as well as the relevance of the characteristic lengths of the problem, were recognized in the early studies of quantum chaos in ballistic transport. Thus, the analysis in terms of the power-spectrum of the conductance was proposed (Jalabert, 1990-a), and later used to interpret the statistical properties of the measured magnetoconductance (Marcus, 1992-a). As emphasized in Sec. 2.4.4, the applicability of Eq. (56) is restricted to $\eta$-intervals that leave aside the short non-universal trajectories, as well as the very long ones that are sensitive to cutoffs from physical effects like disorder and decoherence, while the averaging of the power-spectrum aids to wash out other non-universal features.

In a closed cavity, the soft-wall confinement and the small-angle scattering generically drive nominally chaotic or integrable geometries into a mixed classical dynamics. The effect of a weak smooth disorder in quantum dots has been analyzed in the context of orbital magnetism (Richter, 1996-a; Richter, 1996p-a), and the small extra phase that an electron trajectory picks up was found not to be important in high-mobility samples. The global stability of a chaotic system is not altered by the small perturbation of a smooth disorder. Integrable geometries are more sensitive to weak disorder, but when the physical quantity under study is signed by contributions coming from short trajectories, the effect of disorder can be very small. In an open system it is difficult to evaluate how much softness in the confinement and small-angle scattering is needed to drive the classical dynamics into a mixed one. This is due to the fact that the very long trajectories, which are the ones most affected by the perturbations, might contribute very little to the transport properties.

Few studies exist to quantify the transition to mixed dynamics in an open system due to smooth confinement and weak small-anlge scattering. Among them, quantum calculations performed in order to describe the weak-localization experiments, showed that weak disorder does not necessarily mask the different behavior predicted between clean chaotic and integrable cavities (Chang, 1994-a; Baranger, 1999-r).

# Shot noise

The Random Matrix Theory prediction (71) for the shot noise of a ballistic cavity was tested (Oberholzer, 2001-a) on cavities defined by two point contacts lithographically designed on a narrow (8 $\mu$m) Hall bar. The shot-noise power exhibited a reduction factor of approximately 1/4 with respect to the uncorrelated value $P_\text{Poisson}$ of Eq. (18) (see Figure 23). The best agreement with the theoretical prediction was achieved for the symmetric case of $\eta = G_\text{L}/G_\text{R} = 1$, and assuming a small mode mixing in the quantum point contacts. The left (right) quantum-point contact operated with $N_\text{L}$ ($N_\text{R}$) open modes resulting in a conductance $G_\text{L}$ ($G_\text{R}$) slightly away from the conductance plateaus. The non-perfect transmission of the quantum point contacts resulted in an additional source of noise that needed to be corrected for the comparison against the RMT prediction.

Later measurements (Oberholzer, 2002-a) varying $N_\text{L}$ and $N_\text{R}$ in a systematic way confirmed that (71) is fulfilled for the case of fairly closed contacts $N_\text{L}, N_\text{R} \leq 5$. In more open cavities the shot-noise power was suppressed by a factor $e^{-\gamma \tau_E}$, similarly to the loop corrections (63) of the transmission coefficients, and in agreement with the prediction of Ref. (Agam, 2000-a). The universal result is thus obtained only in the quantum regime when the dwell time $\gamma^{-1}$ is much larger than the Ehrenfest time $\tau_E$. In the opposite case, when transport is dominated by short trajectories, the deterministic character of the classical dynamics hiders the observation of shot noise.

In small cavities with large openings the suppression of the shot noise has been related with the existence of broad resonances that support direct process well described by deterministic classical dynamics (Nazmitdinov, 2002-a).

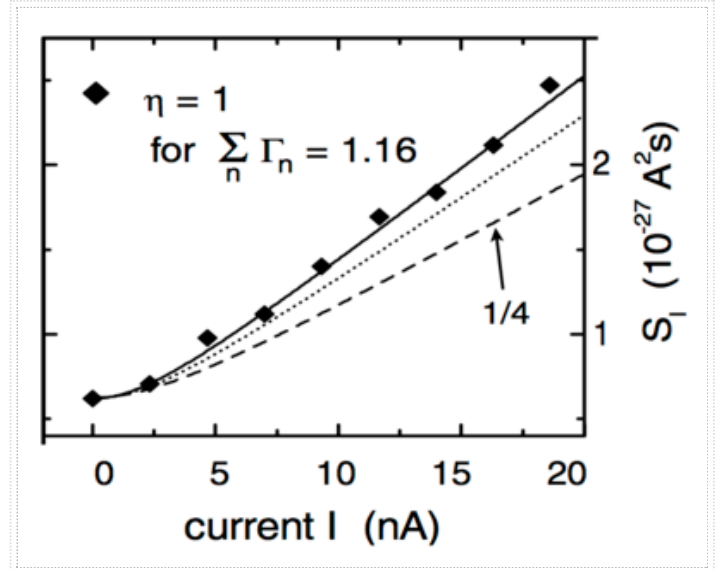

Figure 23: Shot noise $S$ (noted $P$ in the text) as a function of the current $I$ for a cavity defined by two point contacts in the symmetric case of both contacts equally opened $\eta = 1$. The ratio of the number of modes $\eta = G_\text{L}/G_\text{R} = N_\text{L}/N_\text{R}$ is adjusted by independently varying the openings of the left and right contacts. The experimental points are represented by filled squares fitted to the solid line. The dashed line corresponds to the prediction without mode mixing (the modes of the QPCs are completely open or closed) and the dotted line accounts of a slight mode mixing (the transmission coefficients of the two open modes are, respectively, $\Gamma_1 = 0.9$ and $\Gamma_2 = 0.26$). (Adapted from Ref. Oberholzer, 2001-a, copyright 2001, American Physical Society.)

# Coulomb blockade and quantum chaos in ballistic quantum dots

## Coulomb blockade in weakly coupled quatum dots

**Refs. (Kastner, 1992-r; Kouwenhoven, 1997-r; Kouwenhoven, 1999-r; Aleiner, 2002-r)**

### Conditions for observing Coulomb blockade

In quantum dots that are sufficiently small and weakly connected to the leads, at low temperatures, the Coulomb repulsion of electrons cannot be ignored. The conditions for entering into the interaction-dominated regime are

$$G \ll \frac{e^2}{h} \tag{72}$$

$$k_\text{B} T \ll \frac{e^2}{C} \tag{73}$$

- $G$ is the conductance of the dot.

- $e^2/C$ is the charging energy of a single electron in the dot.
- $C$ is the capacitance of the dot.
- $T$ is the temperature.

The condition (72) is obtained when the coupling to the leads is so weak that the typical broadening of the levels in the dot ($\Gamma$) becomes smaller than the typical level spacing ($\Delta$), since the dimensionless conductance can be estimated as the ratio $\Gamma/\Delta$. Such a situation generically appears when tunnel barriers are imposed at the entrance and exit of the dot (as sketched in Figure 24), and thus the number of electrons within the dot is almost a good quantum number.

When the condition (73) holds, the incoming electrons from the reservoirs do not have enough energy to overcome the gap in the tunneling density of states and cannot participate in transport (Figure 24). This suppression of the conductance is known as *Coulomb blockade* (CB).

The CB can be overcome by varying the Fermi energy of the reservoirs or by changing the electrostatic energy of the dot with a voltage $V_g$ applied to a nearby gate. The linear conductance is then an oscillating function of $\varepsilon_F$ or $V_g$, and each period corresponds to a *single electron tunneling* through the dot.

## Coulomb blockade oscillations

The low-temperature ($T = 50$ mK) conductance of a quantum dot like that of Figure 25 displays, as a function of the gate voltage $V_g$, an alternation of peaks and valleys. These are the so-called *Coulomb blockade oscillations*, presented in Figure 26 (Meriav, 1990-a). The quasi-period depends on the geometrical dimensions of the dot (which determine its capacitance).

The change of $V_g$ between two conductance peaks corresponding to $\nu$ and $\nu + 1$ electrons in the dot can be estimated as

$$\Delta V_{g,\nu} = \frac{C}{eC_g}\left(\Delta\varepsilon_\nu + \frac{e^2}{C}\right) \quad (74)$$

- $C = C_1 + C_2 + C_g$ is the total capacitance of the dot.
- $C_g$ is the dot-gate capacitance.
- $C_{1(2)}$ is the capacitance across the barrier with lead 1(2).
- The charging energy $e^2/C$ is supposed to fully account the interaction effects of the incoming electron with the dot.
- $\Delta\varepsilon_\nu$ is the energy spacing between the $\nu$ and $\nu + 1$ single-particle (non-interacting) levels in the dot, assumed to be independent on the charging of the dot.

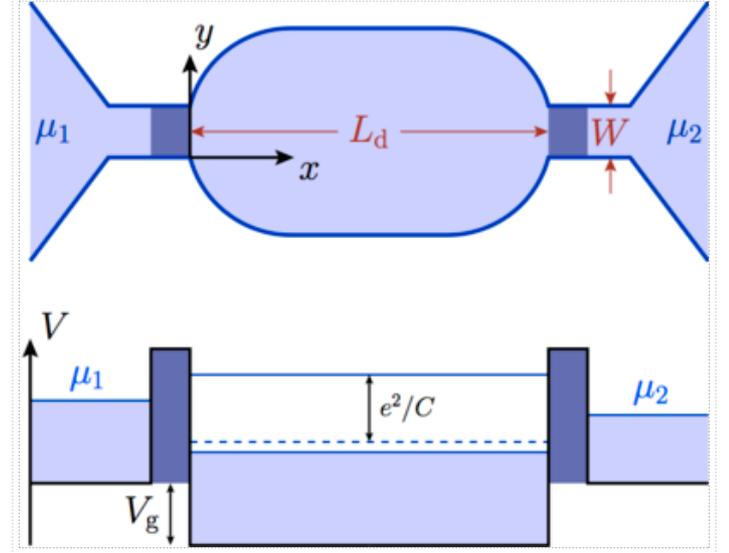

Figure 24: Upper panel: sketch of a ballistic cavity coupled to reservoirs (with electrochemical potentials $\mu_1$ and $\mu_2$) through tunnel barriers (dark gray-blue) placed at connections with the leads. The electrons are confined to the light-blue regions. $W$ is the width of the leads, $L_d$ denotes the distance between the entrance and exit points in the cavity. Lower panel: cut of the potential landscape along the longitudinal direction. The gate potential $V_g$ allows to change the electrostatic potential within the dot. In the numerical calculations the change is supposed to be a rigid shift without deformations in the shape of the cavity. The occupied singe-particle states are indicated by the colored sector of the energy diagram. The dashed-blue horizontal line stands for the first (lower energy) non-interacting single-particle state. The solid-blue horizontal line represents the first available state taking into account the charging energy $e^2/C$.

Eq. (74) reposes over the drastic assumptions that the electrostatic potential defining the dot changes under its successive fillings only through rigid shifts given by the charging energy. Such a simplification is appropriate to describe the Coulomb blockade features of dots with, roughly, more than hundred electrons. This so-called *constant-interaction model* (CIM) has the advantage of reducing a genuine many-body problem into an effective single-particle one. Such a connection allows to apply the ideas of Quantum Chaos developed in the one-particle case.

In semiconductor-based quantum dots $\Delta \ll e^2/C$, but the first energy scale is in general not completely negligible with respect to the second one (as is the case of metallic grains), and therefore the period of the conductance oscillations presents a small modulation in $V_g$.

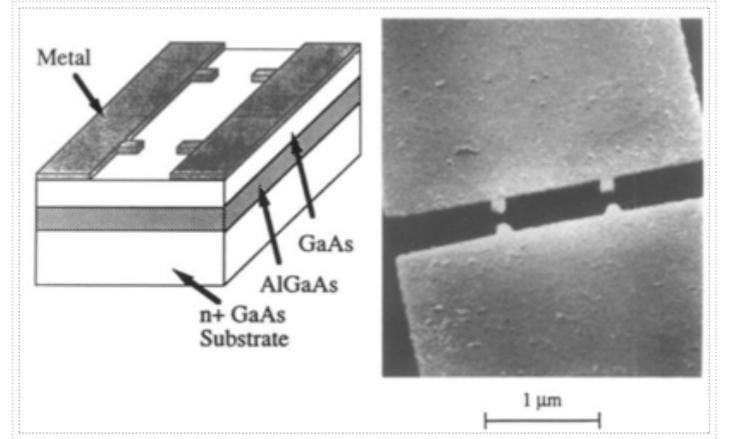

Figure 25: Schematic drawing of device structure (left) and scanning electron micrograph of one of the samples (right) used to measure Coulomb blockade oscillations. The electron gas forms at the top *GaAs/AlGaAs* interface, with a density controlled by the substrate voltage $V_g$. The patterned metallic electrodes on top determine a narrow channel with two potential barriers, defining an electron island. (Adapted from Ref. (Meriav, 1990-a), copyright 1990, American Physical Society.)

## Coulomb blockade regimes

The thermal energy $k_B T$ plays an important role in defining the Coulomb blockade physics. Leaving aside the case of very small dots, where Kondo physics emerges, and within the condition (73) for observing the discreteness of the charge, different regimes can be achieved:

- $\Gamma \ll \Delta \ll k_B T \ll e^2/C$ - many dot levels are excited by thermal fluctuations ( *classical Coulomb blockade*).
- $\Gamma \ll k_B T \ll \Delta \ll e^2/C$ - only one dot level participates to transport and the scattering of electrons with different incoming energies is relevant ( *incoherent quantum Coulomb blockade* with sequential tunneling).
- $k_B T \ll \Gamma \ll \Delta \ll e^2/C$ - only one dot level participates to transport and only the scattering of electrons at a single energy is relevant ( *coherent quantum Coulomb blockade* with resonant tunneling).

## Conductance close to a resonance

From the experimental point of view, and in view of Quantum Chaos studies, the case $\Gamma \ll k_B T \ll \Delta \ll e^2/C$ is the most relevant one. The incoherent character of transport allows the use of rate-equations to obtain the line-shape of the conductance versus $V_g$ around the $\nu$-th resonance (Beenakker, 1991-a)

$$\frac{G}{G_{\max}} = \cosh^2\left(\frac{\delta}{2k_B T}\right) \qquad (75)$$

- $\delta = \dfrac{eC_g}{C}|V_g - V_{g,\nu}|$.
- $V_{g,\nu}$ is the value of the gate voltage at which the $\nu$-th resonance is observed.
- $G_{\max} = \dfrac{e^2}{h}\dfrac{\pi}{2k_B T}\dfrac{\Gamma_\nu^{(1)}\Gamma_\nu^{(2)}}{\Gamma_\nu^{(1)}+\Gamma_\nu^{(2)}} = \dfrac{e^2}{h}\dfrac{\pi \Gamma}{2k_B T}\alpha_\nu$ is the peak height of the $\nu$-th resonance.
- $\Gamma_\nu^{(l)} = \left|\gamma_\nu^{(l)}\right|^2$ is the *partial width* of the $\nu$-th level into the left or right lead, $l = 1$ or 2 respectively. The partial-width amplitude $\gamma_\nu^{(l)}$ of Eq. (26) is taken for $c = (l,1)$ since in the CB problem only the lowest transverse mode (with the largest penetration factor) contributes to electronic transport. The index $a = 1$ is then not explicitly written.
- $\Gamma_\nu = \Gamma_\nu^{(1)} + \Gamma_\nu^{(2)}$ is the *resonance width* of the $\nu$-th level.
- $\alpha_\nu = \dfrac{\Gamma_\nu^{(1)}\Gamma_\nu^{(2)}}{\Gamma_\nu \Gamma}$ is the peak amplitude normalized by the mean resonance width $\Gamma$.

# Statistical theory of Coulomb blockade peak-height distribution in chaotic quantum dots

**Refs. (Beenakker, 1997-r; Alhassid, 2000-r; Gökçedağ, 2002-t)**

The large fluctuations in the amplitude of adjacent peaks experimentally observed at low magnetic field (see Figure 26) were proposed to arise from the chaotic nature of the eigenstates of irregular quantum dots (Jalabert, 1992-a). The penetration factors $P_c$, being smooth monotonous functions of the energy of the incoming electrons, could not be at the origin of the peak-height fluctuations. The statistical approach can be established from Eq. (75) by using the RMT predictions for the wave-functions of classically chaotic systems, and relating the latter to the transport properties through the quasi-one dimensional version of the R-matrix theory (see Sec. 1.3.10).

## Wave-function statistics
**Refs. (Mirlin, 2000-r; Urbina, 2013-r)**

The maximum-entropy principle, expected to be applicable in classically chaotic systems, results in a Gaussian probability distribution of the wave-functions. In this approach, the eigenfunction $\psi_\nu$ associated with the eigenvalue $\varepsilon_\nu = \hbar^2 k_\nu^2 / 2M_e$ has a probability density

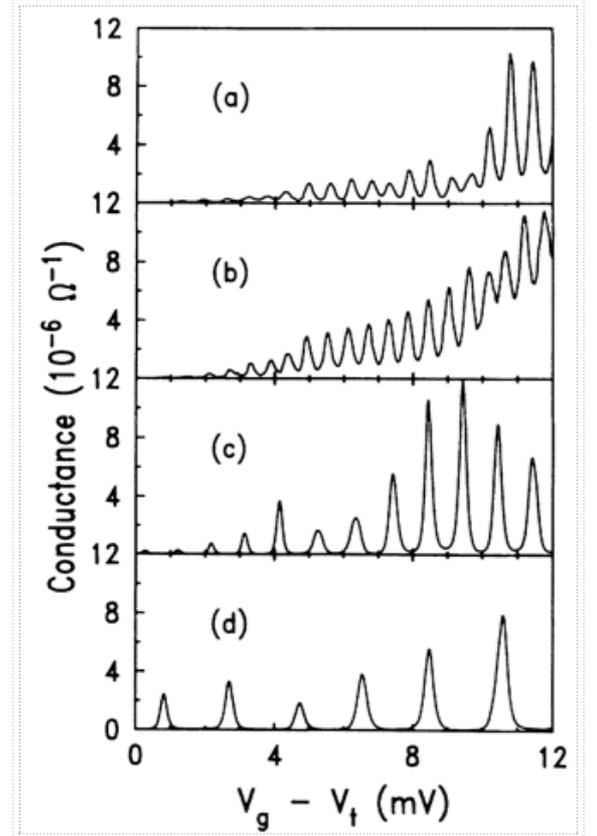

Figure 26: Quasi-periodic oscillations of conductance versus gate voltage $V_g$, measured from a sample-dependent threshold $V_t$. (a) and (b) are for two samples with the same electrode geometry and hence show the same period. (c) and (d) show data for progressively shorter distances between the two constrictions, with a corresponding increase in period. Each oscillation corresponds to the addition of a single electron between the barriers. The peak-height exhibits an overall increase as a function of $V_g$, with superimposed peak-to-peak fluctuations. (Adapted from Ref. (Meriav, 1990-a), copyright 1990, American Physical Society.)

$$\mathcal{P}(\psi_\nu, \psi_\nu^*) = \mathcal{N} \exp\left[-\frac{\beta}{2} \int_\mathcal{A} d\mathbf{r} \int_\mathcal{A} d\mathbf{r}' \ \psi_\nu(\mathbf{r})^* \ \mathcal{K}(\mathbf{r},\mathbf{r}';k_\nu) \ \psi_\nu(\mathbf{r}')\right]. \tag{76}$$

- $\mathcal{A}$ is the area of the two-dimensional billiard.
- $\mathcal{N}$ is the normalization constant.
- $\beta = 2$ in the absence of time-reversal symmetry ($B \neq 0$). In this case $\psi_\nu$ and $\psi_\nu^*$ are independent random variables.
- $\beta = 1$ when time-reversal symmetry is present ($B = 0$). The wave-functions can be taken to be real and thus the number of independent degrees of freedom is reduced respect to that of the $\beta = 2$ case.
- $\langle \psi_\nu \rangle = 0$, where the angular brackets denote the average over $\mathcal{P}(\psi_\nu, \psi_\nu^*)$.
- The kernel $\mathcal{K}$ is the inverse of the two-point correlation function $R$. That is, $\int_\mathcal{A} d\mathbf{r} \ \mathcal{K}(\mathbf{r},\mathbf{r}';k) \ R(\mathbf{r},\mathbf{r}';k) = \delta(\mathbf{r}-\mathbf{r}')$, with $R(\mathbf{r},\mathbf{r}';k_\nu) = \langle \psi_\nu(\mathbf{r})^* \ \psi_\nu(\mathbf{r}') \rangle$.
- The eigenfunctions belonging to different resonances are supposed to be uncorrelated, so that $\mathcal{P}(\psi_1, \psi_2, \ldots, \psi_1^*, \psi_2^*, \ldots) = \prod_\nu \mathcal{P}(\psi_\nu, \psi_\nu^*)$.

According to the Voros-Berry conjecture, the Wigner function for a classically chaotic system is ergodically distributed on the energy manifold of phase-space (Voros, 1976-a; Berry, 1977-a), and the two-point correlation function is given by

$$R(\mathbf{r},\mathbf{r}';k_\nu) = \frac{1}{\mathcal{A}} \ J_0(k_\nu|\mathbf{r}-\mathbf{r}'|) . \tag{77}$$

- $J_0$ is the is zeroth Bessel function of the first kind.

An important shortcoming of the Gaussian distribution (76) is that the wave-function normalization is only satisfied on average, and not by the individual realizations. In addition, a crucial limitation of the two-point correlation function (77) is the fact that it ignores boundary effects. That is, **r** and **r'** are supposed to be in "the bulk" of the chaotic system. These two limitations are particularly relevant in the context of ballistic nanotructures when calculating properties that are sensitive to the $n > 1$ moments of the distribution (Narimanov, 2001-a).

## Partial-width amplitude distribution

The partial-width amplitudes $\gamma_\nu^{(l)}$ are linearly related with the internal wave-functions $\psi_\nu$ (see Eq. (26)). Thus, according to (76), their distribution should also be Gaussian. In the case $B = 0$, the partial-width amplitudes are real and their distribution can be expressed as

$$\mathcal{P}(\gamma_\nu^{(1)}, \gamma_\nu^{(2)}) = \frac{1}{2\pi\sigma_\nu^2 \sqrt{1-\rho_\nu^2}} \, \exp\left[-\frac{\left(\gamma_\nu^{(1)}\right)^2 + \left(\gamma_\nu^{(2)}\right)^2 - 2\rho_\nu \gamma_\nu^{(1)} \gamma_\nu^{(2)}}{2\sigma_\nu^2(1-\rho_\nu^2)}\right]. \tag{78}$$

- $\sigma_\nu^2 = \langle \gamma_\nu^{(1)} \gamma_\nu^{(1)} \rangle = \langle \gamma_\nu^{(2)} \gamma_\nu^{(2)} \rangle = \langle \Gamma_\nu^{(1)} \rangle = \langle \Gamma_\nu^{(2)} \rangle$ is the average partial width. For simplicity, the couplings to the two leads are taken to be symmetric.
- $\rho_\nu = \frac{1}{\sigma_\nu^2} \langle \gamma_\nu^{(1)} \gamma_\nu^{(2)} \rangle$ is the channel correlation.

The parameters $\sigma_\nu^2$ and $\rho_\nu$ are obtained through integrals of the two-point correlation function $R(\mathbf{r}, \mathbf{r}'; k_\nu)$. According to (78), the probability of obtaining a given partial width $\Gamma_\nu^{(l)}$ follows the Porter-Thomas (chi-squared of order 1) distribution

$$\mathcal{P}\left(\Gamma_\nu^{(l)}\right) = \frac{1}{\sqrt{\pi \Gamma \Gamma_\nu^{(l)}}} \, \exp\left[-\frac{\Gamma_\nu^{(l)}}{\Gamma}\right]. \tag{79}$$

- $\Gamma = 2\langle \Gamma_\nu^{(l)} \rangle$ is the average width of the resonances.

## Peak-height distributions

In the case of a dot with left-right symmetry there is only one independent partial width since $\Gamma_\nu^{(1)} = \Gamma_\nu^{(2)}$. The normalized peak amplitude $\alpha = \Gamma_\nu^{(1)}/\Gamma$ is therefore described by the Porter-Thomas distribution (79). A generic quantum dot is expected not to exhibit spatial symmetries, therefore the joint probability distribution of $\Gamma_\nu^{(1)}$ and $\Gamma_\nu^{(2)}$ is relevant. Neglecting the wave-function and channel correlations ($\rho_\nu = 0$), the $B = 0$ distribution of the normalized peak amplitude $\alpha$ is given by (Jalabert, 1992-a)

$$\mathcal{P}_{(\beta=1)} = \sqrt{\frac{2}{\pi\alpha}} \, e^{-2\alpha}. \tag{80}$$

In presence of a magnetic field large enough to break the time-reversal symmetry in the dot, the peak-height distribution is (Jalabert, 1992-a; Prigodin, 1993-a)

$$\mathcal{P}_{(\beta=2)} = 2e^{-4\alpha} \int_0^\infty dz \, \sqrt{\frac{z+4\alpha}{z}} \, e^{-z} = 4\alpha[K_0(2\alpha) + K_1(2\alpha)]e^{-2\alpha}. \tag{81}$$

- $K_n$ are the modified Bessel functions.

Both distributions are *highly non-Gaussian*, and in particular in the $\beta = 1$ case the small values of $\alpha$ are dominant. Eqs. (80) and (81) have been numerically verified for non-interacting dots lacking spatial symmetries and with a classically chaotic dynamics (Jalabert, 1992-a; Bruus, 1994-a). Moreover, the departure from the random-matrix distributions obtained for integrable and nearly integrable geometries demonstrated that the distribution of level widths provides a tool for the Quantum Chaos task of differentiating quantum properties according to the underlying classical dynamics.

Ignoring wave-function correlations corresponds to the basic random-matrix theory assumption that the components of the Hamiltonian eigenvectors are uniquely linked by the normalization condition (that itself becomes irrelevant in the limit of a large dimension of the Hamiltonian matrix). The previous RMT predictions have been generalized to the case of an arbitrary number of possibly correlated channels (Alhassid, 1997-a).

The relevance of single-particle models for the description of Coulomb blockade oscillations relays on the applicability of the constant-interaction model. A theoretical support for this approach was provided by density-functional calculations of realistic structures (Stopa, 1996-a), yielding good agreement with the peak amplitude distributions (80) and (81). Moreover, exact diagonalizations in small dots found that the conductance peak-height statistics is independent of the interaction strength (Berkovits, 1998p-a). That is, identical to the statistics predicted by the CIM using single-electron random matrix theory.

# Experiments on peak-height statistics in ballistic quantum dots

**Refs. (Chang, 1997-r; Marcus, 1997-r; Patel, 2002-t)**

The peak amplitude distributions (80) and (81) were experimentally verified by Chang *et al.* (Chang, 1996-a) and Folk *et al.* (Folk, 1996-a). The first set of data was collected in relatively small quantum dots ($a \sim 0.25$ $\mu$m, $l_T \sim 0.4$ $\mu$m), each containing approximately 100 electrons. A temperature of 75 mK allowed to work in the incoherent single-level regime. The peak-height distributions without and with a magnetic field $B$ were found to be strongly non-Gaussian, and a clear difference could be established between the two cases. The mean-resonanace width $\Gamma$ was used as a single-parameter fit (independent of the magnetic field), leading to a good agreement with the theoretical predictions.

Figure 27 presents the experimentally measured peak-height distributions (filled dots with error bars) for the second set of data (Folk, 1996-a), showing that they are well fitted by Eqs. (80) and (81) for $B = 0$ and $B \neq 0$, respectively (solid lines). The quantum dots (micrograph shown in the panel of the lower inset) were relatively large ($a \sim 0.6$ $\mu$m, $l_T \sim 9$ $\mu$m), containing approximately 1000 electrons, and their shape could be distorted by the effect of nearby gates in order to collect statistically significant data. The inset in the upper panel presents a typical sequence of peaks from which the data was taken. The correlations observable between neighboring peaks is at odds with the basic random-matrix theory assumption of uncorrelated eigenstates.

The channel correlations, neglected in obtaining Eqs. (80) and (81), were calculated from the spatial wave-function correlations (77), and shown to account for the local correlations of the peak amplitudes, without affecting the full peak-height distribution (Narimanov, 2001-a). Short trajectories connecting the exiting points were found to affect the tails of the conductance distribution (Kaplan, 2000-a). The robustness of the peak-height distribution with respect to wave-function correlations (Vallejos, 1999-a), as well as with respect to interaction effects (Stopa, 1996-a; Berkovits, 1998b-a) is at the origin of the good agreement of the experimental results with the random-matrix theory predictions (80) and (81).

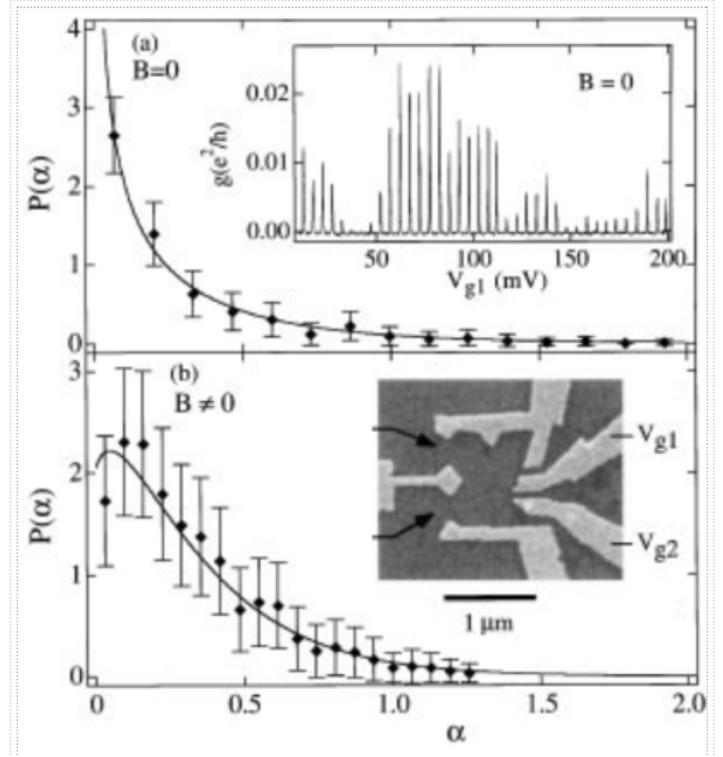

Figure 27: Distribution of the dimensionless peak amplitude for (a) the orthogonal ensemble ($B = 0$) and (b) the unitary ensemble ($B \neq 0$). The filled dots with error bars represent the experimental measurements and the solid lines the random-matrix predictions (80) and (81). Lower inset: micrograph of a quantum dot that can be deformed by the gate voltages $V_{g1}$ and $V_{g2}$. The arrows indicate the quantum-point contacts defining the entrance and exit of the dot. Upper inset: example of a set of peaks from which the distribution was obtained. (Adapted from Ref. (Folk, 1996-a), copyright 1996, American Physical Society.)

The lack of peak-height correlations in Chang *et al.* measurements (Chang, 1996-a) could be explained by the greater importance of disorder (smaller $l_T/a$), in comparison with the other experiments, since the correlator (77) is exponentially suppressed for distances larger than the elastic-mean free-path (Mirlin, 2000-r).

# Peak-spacing statistics

**Refs. (Alhassid, 2000-r; Aleiner, 2002-r; Ullmo, 2008-r)**

According to (74) the spacing between successive Coulomb blockade peaks is given by the charging energy and the single-particle eigenvalue separation. Within the constant-interaction model the peak spacing distribution of an irregular dot should be bimodal; with a maximum around zero spacing (when the highest orbital level gets its two spin-degenerate states occupied) and another branch reproducing the level spacing of a classically chaotic system (Bohigas, 1984-a).

The first measurements of the peak-spacing statistics (Sivan, 1996-a; Simmel, 1997-a) yielded distributions far away from bimodal, and exhibiting a width of the order of the charging energy. However, these results were shown to be affected by charge switching events. Later work (Patel, 1998-a; Lüscherl, 2001-a) avoiding these spurious effects obtained a Gaussian peak-spacing distribution with a width of the order of the single-particle level spacing. Exact diagonalization of small dots have also yielded a Gaussian spacing distribution (Berkovits, 1998-a).

The disagreement of the measured peak-spacing distribution with the simple random-matrix theory prediction is to be contrasted with the good agreement found for the peak-height distribution, and points to the limits of applicability of the CIM to account for fluctuations of the typical energies of the problem. Various theoretical proposals have been developed to improve the constant-interaction model. On one hand, it was put forward the fact that under the addition of new electrons and the variation of the gate voltage, the quantum dot is deformed (the "scrambling" effect). A lateral gate does not act as the idealized case of Figure 24, but produces shape distortions as the gate voltage changes (Vallejos, 1998-a; Vallejos, 1999-a). Other proposals considered the effect of the residual interactions (beyond the mean field) and the fluctuations of the charging energy have been calculated within the Random Phase Approximation (Blanter, 1997-a). The electronic spin appeared as an essential ingredient to explain the main features of the experimentally observed peak-spacing distribution. In particular exchange effects were incorporated within the so called *universal Hamiltonian* (Kurland, 2000-a),

The need to go beyond the constant-charging model for the description of the peak-spacing fluctuations of the Coulomb blockade oscillations takes this problem outside the realm of Quantum Chaos applied to single-particle classically chaotic systems.

# Phase-locking between Coulomb blockade peaks: experiments in Aharonov-Bohm interferometers

**Refs. (Hackenbroich, 2001-r)**

## Two-terminal conductance experiments

A quantum dot operating in the Coulomb blockade regime with only a single transverse channel of the leads participating in the electronic transport is characterized by the complex $V_g$-dependent transmission amplitude $t = |t| e^{i\alpha}$. The squared modulus $|t|^2$ is experimentally accessible since it is proportional to the conductance through the dot, while the *transmission phase $\alpha$* cannot be directly measured.

In an attempt to access the transmission phase, Yacoby *et al.* (Yacoby, 1995-a) embedded, in one of the arms of a phase-coherent Aharonov-Bohm ring, a quantum dot operating in the CB regime (see the sketch in Figure 28 and the micrograph of the actual device in Figure 29).

The conductance through the Aharonov-Bohm ring is a periodic function of the enclosed magnetic flux, and therefore it can be written as

$$g_{\text{AB}}(V_g, \Phi) = g_{\text{AB}}^{(0)}(V_g) + \sum_p g_{\text{AB}}^{(p)}(V_g) \cos\left(2\pi p \frac{\Phi}{\Phi_0} + \beta_p(V_g)\right), \tag{82}$$

- $V_g$ is the gate voltage that changes the electrostatics of the dot.
- $\Phi$ is the magnetic flux through the ring.
- $\Phi_0 = hc/e$ is the flux quantum.
- $\beta_p(V_g)$ are the $V_g$-dependent phases associated with the different harmonics $p$.

In a two-terminal device as the one of Ref. (Yacoby, 1995-a) the Onsager reciprocity relations dictate that $g_{\text{AB}}$ is an even function of $\Phi$ (Büttiker, 1986-a). Thus, $\beta_p = 0, \pi$ are the only two possible values (Levy Yeyati, 1995-a). Consequently, when $\beta_1$ was monitored through the conductance measurements, the only two values obtained were 0 and $\pi$, rather than the targeted transmission phase $\alpha$.

Of particular interest were the abrupt changes of $\beta_1$ between the two allowed values obtained, at values of $V_g$ corresponding to Coulomb blockade resonances (i.e. when an electron is added to the QD), as well as in the conductance valleys in-between each and every two consecutive Coulomb blockade resonances. Equivalent points in successive Coulomb blockade peaks exhibited Aharonov-Bohm oscillations which were in phase, since going from one peak to the next one always encompassed two phase lapses of $\pi$ (see Figure 30). The *phase-locking* between different peaks was unexpected, and thus posed a serious theoretical challenge for its understanding.

## Open-ring experiments

In a second generation of phase-sensitive experiments, Schuster *et al.* (Schuster, 1997-a) lifted the reciprocity constraints by adopting a multi-terminal device through the opening of the arms of the ring to additional grounded terminals. The dashed lines on the arms of the ring in the sketch of Figure 28 stand for the additional terminals. Working with such a "leaky" interferometer suppresses processes with multiple windings around the ring, and an appropriate tuning of the opening of the ring arms in this multi-terminal setup allows for the identification of $\beta_1$ with $\alpha$ (Aharony, 2002-a).

Without the reciprocity constraint, Ref. (Schuster, 1997-a) obtained a Breit-Wigner behavior of the phase, with a smooth increase of $\pi$ every time a CB resonance is crossed (as expected from Friedel's sum rule), as well as *systematic phase lapses* of $\pi$ in-between each and every two consecutive Coulomb blockade resonances. These phase lapses between resonances in the multi-terminal case demonstrated that the phase-locking of consecutive peaks was a generic effect of the transmission phase in the Coulomb blockade regime.

## Experiments in very small quantum-dots

The quantum dots in the devices of Refs. (Yacoby, 1995-a) and (Schuster, 1997-a) had linear dimensions of the order of 0.4-0.5 $\mu$m and operated with hundreds of electrons. The working temperatures were $T \approx 80$ mK and the transport mean-free-path was estimated to be $l_T = 10 - 15$ $\mu$m, placing these experiments in the ballistic regime. A third generation of phase-sensitive experiments (Avinum-Kalish, 2005-a) targeted smaller dots (of the order of 0.1 $\mu$m) with zero to few tens of electrons, and even lower temperatures ($T \approx 30$ mK). The estimations of the energy scales involved in the experiments were: the thermal energy $k_B T \approx 0.003$ meV, the level spacing $\Delta \approx 0.5$ meV, the level width $\Gamma \approx 0.03 - 0.3$ meV, and the charging energy $e^2/C = 1 - 3$ meV.

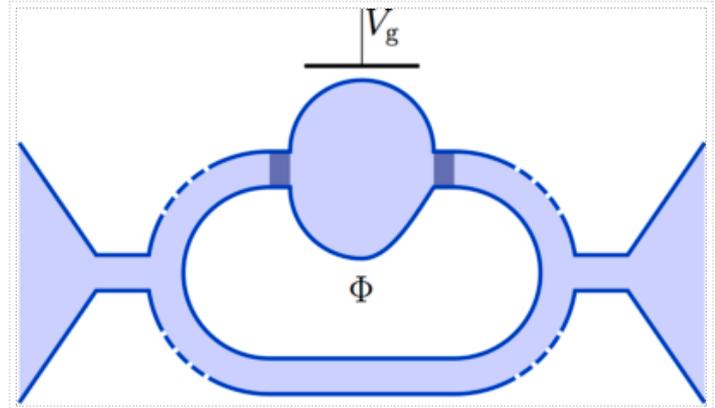

Figure 28: Schematic description of an Aharanov-Bohm interferometer, threaded by a magnetic flux $\Phi$, with a quantum dot embedded in its upper arm. The dot is capacitively coupled to a plunger gate potential $V_g$, which tunes its electrochemical potential, and thus changes its electronic occupancy. The unsymmetrical shape of the dot reflects the presumed lack of geometric symmetry in experimentally realizable microstructures. The dark segments at the entrances of the dot represent tunnel barriers. The dashed lines on the arms of the interferometer stand for a number of possible additional leads.

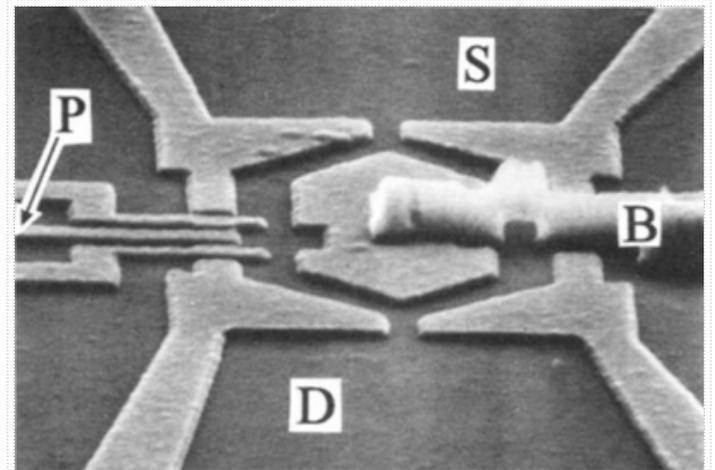

Figure 29: Scanning electron micrograph of the Aharanov-Bohm interferometer with a quantum dot defined on its left arm. The clearer regions are the metal gates. The electron paths interfere when going from the source (S) to the drain (D) along the two arms of the ring. The central metallic island is biased via an air bridge (B) extending to the right. The electrostatic potential in the dot is controlled by the voltage $V_g$ applied in the plunger (P). (Adapted from Ref. (Yacoby, 1995-a), copyright 1995, American Physical Society.)

The key observation in these experiments was that, as the number of electrons on the quantum dot was reduced from 20 down to 0, the phase $\beta_1$ underwent a crossover from the regime with alternating $\pi$ jumps at and in-between CB resonances, to a regime where phase lapses in-between resonances occurred in a random fashion. Thus the complete phase-locking of Coulomb peaks was absent for very small dots.

# Phase-locking between Coulomb blockade peaks: tendency arising from wave-function correlations

The complete phase-locking observed in relatively large quantum dots (lost in the case of very small dots) appeared in contradiction with the expected results from the random-matrix theory applied to the constant-interaction model (which predicts random phase lapses for each pair of consecutive peaks). Consequently, an important theoretical effort was devoted to the understanding of this phenomenon.

Some theoretical works attempted to explain the phase-locking of the Coulomb blockade resonances within the constant-interaction model (Oreg, 2007-a), while other approaches (Silvestrov, 2007-a) proposed going beyond this model and considered genuine many-body effects. In particular, numerical calculations were performed in lattice models representing interacting fermions, and an interpretation was proposed in which the electronic correlations could induce a mode switching mechanism between a broad level well coupled to the leads and nearly narrow levels (Karrasch, 2007-a). However, detailed many-body numerical calculations disputed such a view, showing that electron-electron interactions do not generically change the tendency towards phase-locking (Molina, 2013-a).

## Sign-rule for the transmission zeros

The phase-slip of $\pi$ in the transmission phase between resonances is associated with the vanishing of $t$, and the switching between the $\eta = 0$ and $\eta = \pi$ branches characterizing the time-reversal symmetric case (see Eq. (32)). The evolution of $t$ in the complex plane as a function of $\varepsilon$ (or $kL_\mathrm{d}$) is presented in Figure 31.b, obtained from numerical calculations for a non-interacting quantum dot connected to leads through tunnel barriers.

Within the CIM, the transmission amplitude takes the Breit-Wigner form (29) and the behavior of $t(\varepsilon)$ between resonances is generically dictated by the partial-width amplitudes $\gamma_\nu^{(1)}$ and $\gamma_\nu^{(2)}$ (for decaying, respectively, into the first mode of the left (1) and right (2) leads) of the levels $\nu$ that are nearby (in energy).

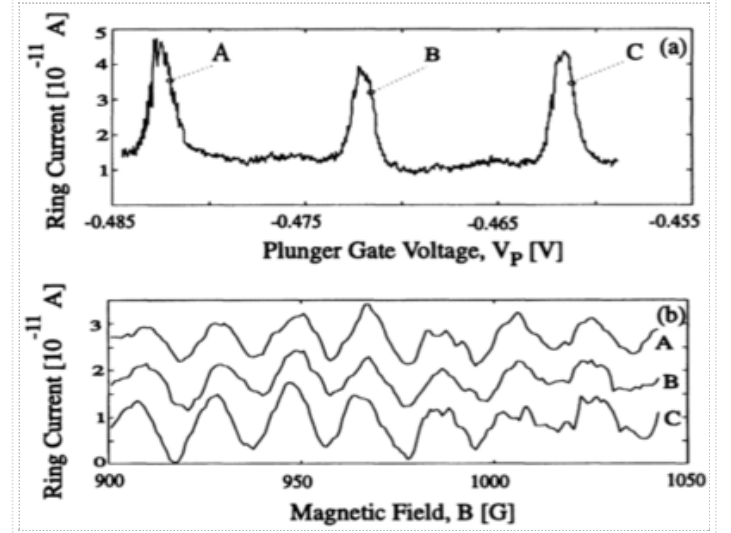

Figure 30: (a) A series of Coulomb blockade peaks as a function of the plunger gate voltage $V_\mathrm{p}$ ($V_\mathrm{g}$ in the text). (b) The corresponding current oscillations taken at the marked points A, B, and C in successive peaks of the ring's current. All oscillations are seen to be in phase. (Adapted from Ref. (Yacoby, 1995-a), copyright 1995, American Physical Society.)

The appearance of transmission zeros is governed by a *sign rule* (Levy Yeyati, 2000-a) according to the sign of $D_\nu = \gamma_\nu^{(1)} \gamma_\nu^{(2)} \gamma_{\nu+1}^{(1)} \gamma_{\nu+1}^{(2)}$

- If $D_\nu < 0$ there is no transmission zero or there is an even number of zeros between resonances $\nu$ and $\nu + 1$. Two consecutive resonances have a relative phase of $\pi$.
- If $D_\nu > 0$ there is an odd number of transmission zeros between resonances $\nu$ and $\nu + 1$. Two consecutive resonances are in phase.

On one hand, in cases where the partial-width amplitudes are not strongly fluctuating among the eigenstates, the $\varepsilon$-dependence of $t$ between two resonances $\nu$ and $\nu + 1$ is dictated by the values of $\gamma_\nu^{(1)}, \gamma_\nu^{(2)}, \gamma_{\nu+1}^{(1)}$, and $\gamma_{\nu+1}^{(2)}$ (*restricted off-resonance* behavior). As a consequence of (29), there can be either none or one transmission zero in the interval (according to the sign rule).

On the other hand, in cases where the partial-width amplitudes are strongly fluctuating among the eigenstates, the $\varepsilon$-dependence of $t$ between two resonances $\nu$ and $\nu + 1$ can be dictated by the value of partial-width amplitudes corresponding to far-away states (*unrestricted off-resonance* behavior), and any number of transmission zeros could in principle appear in the interval.

The sign-rule is given only in terms of the values of the partial-width amplitudes at the resonances. The details along the energy interval between the two resonances are not relevant. This is particularly important in the context of Coulomb blockade, since the single-particle description of the constant-interaction model is properly applied only close to the resonances. Moreover, the sign-rule can be expressed only in terms of the wave-functions of the isolated dot, and thus the details of the dot-lead coupling appear as unimportant.

A random and uncorrelated distribution of the partial-width amplitudes has equal probabilities for obtaining a negative or positive $D_\nu$, that is, $\mathcal{P}(D_\nu < 0) = \mathcal{P}(D_\nu > 0) = 1/2$. This result, at odds with the experimentally observed phase-locking, is expected in systems without wave-function correlations. Indeed, in a disordered quantum dot, where the spatial correlations vanish beyond the elastic mean-free-path on a length scale smaller than the dot size, numerical simulations yielded an equal probability for the two signs of $D_\nu$ (Levy Yeyati, 2000-a). On the contrary, ballistic cavities with classically underlying chaotic dynamics exhibit wave-function correlations (given by (77) when the points $\mathbf{r}$ and $\mathbf{r}'$ are far away from the boundary).

Numerical calculations using asymmetric cavities and non-interacting electrons yield the $kL_\mathrm{d}$-dependence of the transmission amplitude $t$ exhibited in Figure 31. The change in $kL_\mathrm{d}$ (encoded in the color scale) results from the modulation of the Fermi energy $\varepsilon$ of the leads or from a rigid shift of the energy landscape in the dot induced by a gate voltage $V_\mathrm{g}$ (as in the setup of Figure 24). Analyzing $|t|$ it is possible to see that there exists relatively long sequences of peaks with exactly one zero in-between. In the complex plane this tendency can be seen by the fact that circles of a given color (nearby resonances) tend to turn in the same half-plane (indicating that there is a zero in between two of them). Occasionally, there is a "missing zero", signed by a switch of the half-plane in which the complex $t(\varepsilon)$ evolves.

# Emergence of phase-locking in the semiclassical limit

Identifying $D_\nu$ with its ensemble average $\langle D_\nu \rangle = \langle \gamma_\nu^{(1)} \gamma_\nu^{(2)} \rangle \langle \gamma_{\nu+1}^{(1)} \gamma_{\nu+1}^{(2)} \rangle$ and using the two-point correlation function (77) leads to $\overline{\mathcal{P}(\langle D_\nu \rangle < 0)} \simeq 1/k_\nu L_\mathrm{d}$ for $k_\nu L_\mathrm{d} \gg 1$, where the bar stands for the average over an interval of $\pi$ in the variable $k_\nu L_\mathrm{d}$ (Molina, 2012-a). In this approach the universal regime of systematic phase shifts between resonances does not appear abruptly, but it rather progressively emerges in the semiclassical limit. Thus, there is always a finite probability of finding out-of-phase peaks.

When the fluctuations of $D_\nu$ are taken into account by using the Gaussian probability density (76), with the parameters resulting from the two-point correlation function (77), the probability $\mathcal{P}(D_\nu < 0)$ of missing a $\pi$ phase-slip between resonances $\nu$ and $\nu + 1$ can be evaluated (Jalabert, 2014-a). This quantity has an overall decreasing behavior as a function of $k_\nu L_\mathrm{d}$ (red dashed line in Figure 32), with superimposed oscillations of quasi-period $\Delta k_\nu = \pi/L_\mathrm{d}$ (red dotted line).

Quantum numerical calculations yield a $k_\nu L_\mathrm{d}$ dependent distribution of $D_\nu$ that is well represented by a Gaussian density, but the characteristic parameters are not those resulting from Eq. (77), indicating the importance of the border corrections. The local value of $\mathcal{P}(D_\nu < 0)$ (thin blue line in Figure 32) oscillates with $k_\nu L_\mathrm{d}$. The numerically obtained smoothed $\overline{\mathcal{P}(\langle D_\nu \rangle < 0)}$ (thick blue line) decreases faster than the result of Eqs. (76) and (77), but slower than the $1/k_\nu L_\mathrm{d}$ dependence obtained by ignoring the fluctuations of $D_\nu$.

Despite their qualitative differences, the three above-described approaches agree in predicting that in the semiclassical limit of $k_\nu L_\mathrm{d} \gg 1$, it becomes less likely to find departures from phase-locking.

# Growing-rate of the accumulated transmission phase

In the numerics, like in the analytical approaches, there are always "missing zeros", though they are less likely to be observed when moving into the semiclassical limit of $k_\nu L_\mathrm{d} \gg 1$. Thus, there is an *emergence* of universality, but not a characteristic dot-size beyond which all peaks are always in phase. In order to characterize this emergent behavior, it is possible to track the number of resonances ($N_\mathrm{r}$, obtained by following the scattering phase) and of zeros ($N_\mathrm{z}$, obtained by passages through the origin in the complex plane) as a function of $V_\mathrm{g}$ (or $\varepsilon$), as presented in Figure 33. These two numbers tend to grow with the same rate in the semiclassical limit. As shown in the inset, the percent difference between the number of resonances and zeros in a given $k$-interval $\mathcal{P} = (\Delta N_\mathrm{r} - \Delta N_\mathrm{z})/\Delta N_\mathrm{r}$ (dots) follows a $1/kL_\mathrm{d}$ law (solid line).

The evolution of $t(\varepsilon)$ allows to define the *accumulated transmission phase* $\alpha_\mathrm{c}^{(-)}$, not restricted to the interval $[0, 2\pi)$, by taking a phase-slip of $-\pi$ at each transmission-zero. In Figure 33 the curve $\alpha_\mathrm{c}^{(-)}$ versus $V_\mathrm{g}$ shows that the regions where $N_\mathrm{z}$ lags $N_\mathrm{r}$ (indicated by grey spots) are separated approximately by $\pi$ in $kL_\mathrm{d}$. The overall flat behavior of $\alpha_\mathrm{c}^{(-)}$ between these regions is locally altered by a resonance followed by a transmission-zero (up excursions of $\alpha_\mathrm{c}^{(-)}$) or by pairs of successive zeros (down excursions of $\alpha_\mathrm{c}^{(-)}$).

The arbitrariness adopted in the definition of the accumulated transmission phase can be lifted by imposing a small magnetic field. In this case $t(\varepsilon)$ has a vanishing probability of passing by the origin of the complex plane, and there are no phase-slips of $\pi$, but a continuos evolution. The accumulated transmission phases $\alpha_\mathrm{c}$ for small positive and negative magnetic fields are shown by the thin brown lines of Figure 33, and exhibit the same growing rate of the scattering phase. Thus, the Friedel sum-rule is fulfilled on average

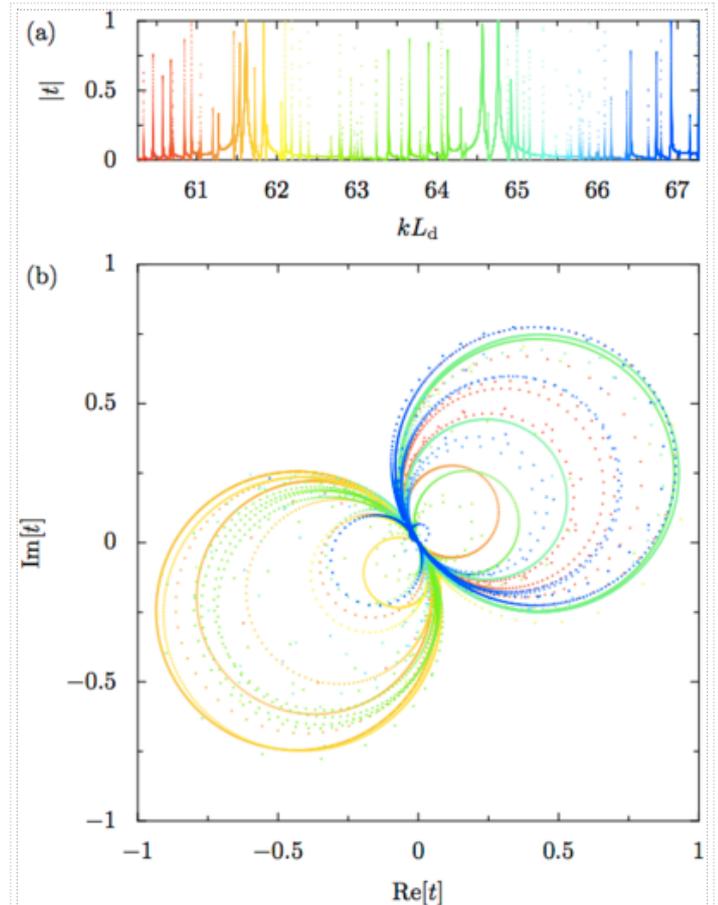

Figure 31: (a) Absolute value of the transmission amplitude for the setup of Figure 24, as a function of $kL_\mathrm{d}$. (b) The evolution of the transmission amplitude as a function of the electron energy (or $kL_\mathrm{d}$) is presented in the complex plane (encoded by the $kL_\mathrm{d}$ dependence common to the color plot of both panels). When there is a transmission-zero between two resonances, $t$ "turns" twice on the same half-plane. When there is no transmission-zero, the evolution

for the transmission phases once the ambiguity in the definition of the phase-slips at the transmission-zeros is removed.

The analytical and numerical results of the phase-behavior of Coulomb blockade resonances within the constant-interaction model, and using some drastic approximations, seem to indicate that there are sequences of in-phase peaks progressively larger when $kL_d$ increases. The sequences of in-phase resonance exhibited in the experiments of Refs. (Yacoby, 1995-a; Schuster, 1997-a; Avinum-Kalish, 2005-a) are relatively small. It is then expected that if larger sequences were measured, departures from the claimed universal behavior of a complete phase-locking would be observed.

# Concluding thoughts

The quantum chaos studies of mesoscopic transport deal with the signatures, on the quantum transmission, of the geometry through which a scattering process takes place. There are two distinct problems involved in this pursuit:

- The sensitivity on the underlying classical dynamics of the scatterer, considered as an ideal *clean* cavity.
- The experimentally measurable consequences of the designed shape of a *ballistic* microstructure on its low-temperature conductance.

The first problem is defined by the first three terms of the Hamiltonian (2). It has been the object of an intense theoretical endeavor in the domain of quantum chaotic scattering, and some conclusions have been clearly established.

- It is possible to differentiate classically chaotic and integrable clean cavities through the study of conductance fluctuations and the weak-localization effect.
- The above-mentioned differences are generally not evident at first sight, but only reveled by a quantitative analysis and performing appropriate averages: conductance fluctuations present larger weight of high harmonics in the case of integrable structures; the magnitude of the conductance fluctuations increases with energy in integrable geometries (2.7.2, 2.7.3) and it is energy-independent in the chaotic case (2.4.4); a Lorentzian weak-localization peak is obtained in the chaotic case versus a triangular line-shape in the integrable case (2.5.2).
- Unlike chaotic systems, integrable ones are not characterized by a generic behavior. Thus, their transport properties are very dependent on the particular system under consideration. In addition, in corresponding to each resonance is on a different half-plane. (Adapted from Ref. (Jalabert, 2014-a), copyright 2014, American Physical Society.)

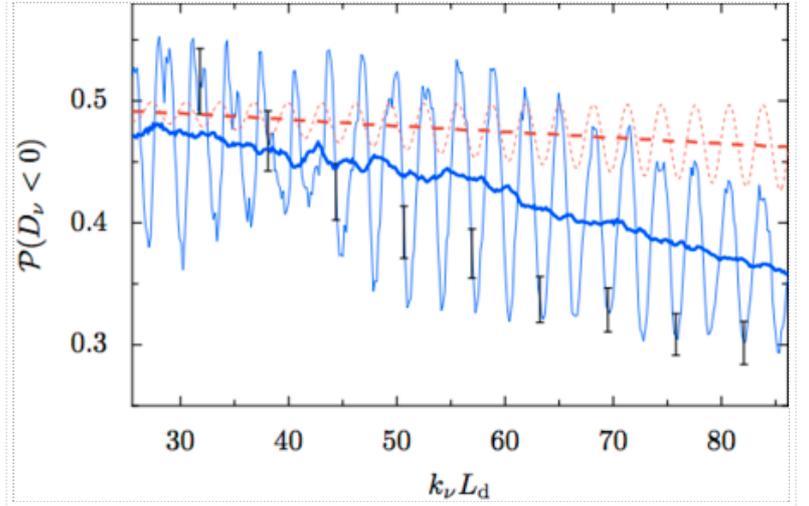

Figure 32: Numerically obtained $\mathcal{P}(D_\nu < 0)$ (thin blue line) and the result of its smoothing over an interval $k_\nu L_d = \pi$ (thick blue line). The error bars indicate the statistical errors. The red dotted line is obtained assuming a Gaussian distribution of the wave-functions (76), with a correlator given by Eq. (77). The red dashed line represents a smoothing of the previous curve. (Adapted from Ref. (Jalabert, 2014-a), copyright 2014, American Physical Society.)

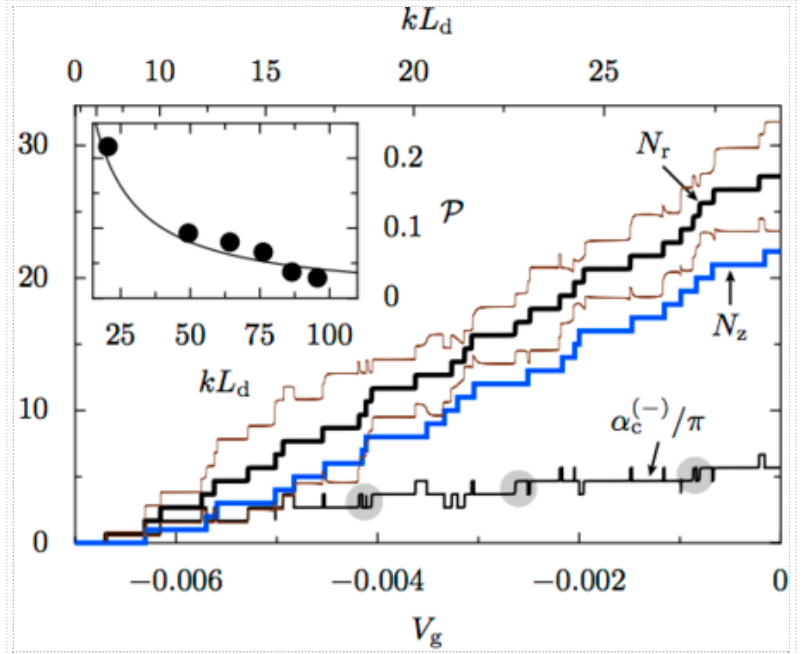

Figure 33: Accumulated transmission phase $\alpha_c^{(-)}$ (solid black), together with the number of resonances $N_r$ (thick solid black) and zeros $N_z$ (thick solid blue) as a function of $V_g$ (or $kL_d$) for the setup of Figure 24. The shaded spots along the curve of $\alpha_c^{(-)}$, approximately separated by $\pi$ in the variable $kL_d$, indicate the regions where $N_z$ lags $N_r$. Thin brown lines: $\alpha_c$ for small positive and negative magnetic field. Inset: $\mathcal{P} = (\Delta N_r - \Delta N_z)/\Delta N_r$, taken at various intervals of $V_g$, as a function of $kL_d$ (filled dots) and a guide-to-the-eye that decreases as $1/kL_d$ (solid). (Adapted from Ref. (Molina, 2012-a), copyright 2012, American Physical Society.)

- the case of integrable cavities the interplay between the conserved quantities of the leads and those of the system plays an important role.
- The limited time that classical trajectories spend inside the cavity is the cause for the differences between the chaotic and integrable cases to be more subtle than for closed systems. In closed systems trajectories of all lengths are relevant, while in the scattering setup trajectories may leave before exhibiting their chaotic character. Quantitative important effects then appear in the case of quantum scattering, like short trajectories (2.4.4), direct trajectories (2.5.2, 2.7.1), and Ehrenfest-time corrections (2.6.3).
- Chaotic cavities tend to be characterized by a single parameter governing the distribution of trajectory lengths and the energy-dependent conductance fluctuations. The area distribution, the magnetic field correlation-length, and the half-width of the weak-localization peak are related with this single parameter. Contrariwise, the length and area distributions of integrable systems are not characterized by a single parameter.
- The random-matrix theory of open systems yields universal predictions for the conductance fluctuations, the weak-localization effect, and the shot noise of chaotic cavities, provided that the effect of short trajectories has been eliminated (i.e. working with geometries like those of Figure 12 and Figure 14). The effect of short trajectories (3.7) and Ehrenfest-time (2.6.3) modify the universal predictions into sample-dependent results.
- Semiclassical expansions provide a valuable tool to the quantum scattering through a clean cavity when the particle wavelength is much smaller than the typical size of the structure ($ka \gg 1$) and the number of modes in the leads is large ($N \gg 1$). This approach allows to easily taking into account the time-symmetry breaking effect of an external magnetic field. It is then possible to incorporate sample-specific features when the expansions are evaluated as a trajectory sum, or generic information when the global character of the chaotic dynamics is considered.
- The diagonal semiclassical approximation of only of pairing symmetry-related trajectories does not respect the unitarity of the scattering coefficients resulting from particle conservation. Thus, the magnitude of the conductance fluctuations and the weak-localization effect are not accessible within this approach. Loop corrections go beyond the diagonal approximation and preserve unitarity. The corresponding calculation of of the conductance fluctuations and the weak-localization effect, within the ergodic hypothesis of the classical dynamics, leads to equivalent results as those of the random-matrix theory approach (which are not sample specific).
- The transmission through a cavity connected to the leads through tunnel barriers is determined by the values of the resonant wave-function close to the entry and exiting points. Assuming that the wave-functions are distributed according to the Gaussian ensembles provides a good description of the resonance widths. The evolution of the transmission phase between two resonances is described by wave-funtion correlations that go beyond the basic random-matrix theory description.

The second problem is of great interest, as it is directly related with laboratory measurements. Some relevant questions can be addressed in this context.

- Does the effective one-particle Hamiltonian (2) provide a good description for the low-temperature transport through ballistic systems? Ballistic mesoscopic structures typically belong to the remarkably large class of Condensed Matter systems where it is indeed possible to use a description of weakly interacting quasi-particles moving in a self-consistent field. The nature of the mean-field is generally difficult to determine, but it does not hinder the possibility of establishing quantitative predictions.
- Are classical trajectories relevant for quantum transport through high-quality samples? There is indeed ample evidence on the signature of classical trajectories in low-temperature conductance measurements. The conductance fluctuations and weak-localization of ballistic cavities have been shown to be distinct from those of a diffusive sample. The measured characteristic scales of these two phenomena are related as in the clean case, indicating their common quantum interference origin. The shape of a ballistic cavity affects its transmission; the quantum interference between pairs of trajectories can be monitored. An external magnetic field effectively suppresses the quantum interference between two time-reverse symmetric trajectories reducing the conductance fluctuations.
- Does the chaotic or integrable nature of the classical dynamics in the patterned microstructure have measurable consequences on its low-temperature conductance? The lack of consensus about answering this question stems from the above-mentioned subtle differences between chaotic and integrable scatterers, as well as from the complexity of ballistic mesoscopic systems (1.1.6). The finite coherence time and the effect of residual disorder suppress the relevance of long trajectories, which are the most sensitive ones to the chaotic nature of the dynamics. The difficulty to achieve an electrostatic confinement with sharp boundaries makes that nominally chaotic billiard microstructures (lithographically designed with specific shapes) yield a mixed classical dynamics for the electron motion in the resulting self-consistent electrostatic potential. These considerations should be taken into account when analyzing the examples of nominally integrable cavities presenting a weak-localization peak characteristic of the chaotic case, or the difficulties to separate the different classical dynamics from the conductance fluctuations. The proposal of analyzing the power spectrum of the conductance fluctuations aims to deal with these issues, by separating the length scales of the contributing trajectories and comparing them with the physical cutoffs of the experimental system (2.4.2, 2.4.3, 4.1.1, 4.3.4). Weak smooth disorder can substantially alter the long-time dynamics of highly symmetric structures (like a circular dot), but it is expected to have a smaller effect in stadium-shaped structures where the classical dynamics is already chaotic in the clean case. A weak smooth-disordered potential does not alter the global stability of a chaotic geometry, but it might erase the signatures of an integrable dynamics in the case in which long trajectories are relevant for determining the observed quantity. Indeed, some of the geometries used to test the universal results of random-matrix theory were nominally integrable (4.1.3). The small number of propagating modes used in some experiments is another factor pointing towards the departure of the measurements from the semiclassical predictions.

- Is it possible to pursue the study of the signatures of a chaotic dynamics in quantum transport? The continuous improvement in the fabrication of better samples has been a technological drive to this kind of studies, and should continue to motivate further developments. The high quality of samples and the improved fabrication methods now allow cleaner and better-controlled structures with considerably larger transport mean-free-paths, as compared with those employed in the early studies of ballistic transport.
- Are there other interesting questions, besides that of the signatures of the nature of the classical dynamics? The interest of quantum chaos in mesoscopic transport extends well beyond that question. A particularly important aspect is the characterization of inelastic effects and the determination of the phase-coherence length $L_\Phi$ from a random-matrix hypothesis (3.8, 4.1.3). The characterization of the residual disorder can also be done from the connection with the underlying classical dynamics (4.3.2, 4.3.3).
- Are nearly-closed microstructures good laboratories for quantum chaos studies? The random-matrix theory, together with the constant-interaction model, yields predictions in very good agreement with experimental measurements of peak-heights in the Coulomb blockade regime. The role of integrability has not been put in evidence in experimental studies, probably due to the effect of weak disorder. The peak-spacing distribution of the Coulomb blockade regime requires a description going beyond the constant-interaction model on which traditional quantum chaos studies are based. However, the random character of the one-particle wave-functions appears as an important ingredient in the estimation of the interaction effects.
- Is the phase-locking of successive Coulomb blockade peaks understood? The correlations of one-particle wave-functions, rather than the electronic correlations, have been shown to be a key ingredient of this phenomenon. The chaotic nature of the underlying classical dynamics results in the tendency towards the phase-locking of successive peaks in the semiclassical limit. The complete phase-locking beyond certain dot size is not supported by present models based on a Gaussian wave-function distribution. The theoretical study of higher moments of the wave-function distribution, and the experimental determination of the typical length of the sequences of in-phase peaks, should help to advance in the understanding of this important problem.

# Acronyms

- 2DEG - two-dimensional electron gas
- DC - direct current
- CB - Coulomb blockade
- CIM - constant interaction model
- COE - circular orthogonal ensemble
- CSE - circular symplectic ensemble
- CUE - circular unitary ensemble
- QPC - quantum point contact
- QD - quantum dot
- RMT - random matrix theory
- SCA - semiclassical approximation
- SGM - scanning gate microscopy

# Research articles (cited as "first_author, year-a")

# Further reading, mainly review articles (cited as "first_author, year-r")

# Further reading, mainly books and theme issues (cited as "first_author, year-b")

# Thesis that deserve to be read (cited as "author, year-t")

# Internal references (cited as "first_author, year-i")